\documentclass[superscriptaddress,showpacs,amsmath,amssymb,aps,prd,twocolumn,floatfix,nofootinbib]{revtex4-1}

\usepackage{graphicx}
\graphicspath{{figures/}}

\usepackage{dcolumn}
\usepackage{bm}
\usepackage{color}
\usepackage{amsmath}
\usepackage{amssymb}
\usepackage{bbm}
\usepackage{lineno}
\usepackage{hyperref}
\usepackage{todonotes}
\usepackage{hhline}
\usepackage[percent]{overpic}
\usepackage{layouts}
\hypersetup{
    colorlinks = true
}

\usepackage{natbib}
\usepackage{graphicx}
\usepackage{ amssymb }
\usepackage{siunitx}

\begin{document}

\title{Clustering of Electromagnetic Showers and Particle Interactions with\\Graph Neural Networks in Liquid Argon Time Projection Chambers Data}


\newcommand{\SLAC}{SLAC National Accelerator Laboratory, Menlo Park, CA, 94025, USA}
\affiliation{\SLAC}
\newcommand{\ICME}{ICME, Stanford University, Stanford, CA, 94305, USA}
\affiliation{\ICME}
\newcommand{\Stanford}{Stanford University, Stanford, CA, 94305, USA}
\affiliation{\Stanford}

\author{Fran\c cois~Drielsma} \email{drielsma@stanford.edu} \affiliation{\SLAC}
\author{Qing~Lin} \affiliation{\SLAC}
\author{Pierre~C\^ote~de~Soux} \affiliation{\ICME}
\author{Laura~Domin\'e} \affiliation{\Stanford}
\author{Ran~Itay} \affiliation{\SLAC}
\author{Dae~Heun~Koh} \affiliation{\Stanford}
\author{Bradley~J.~Nelson} \affiliation{\ICME}
\author{Kazuhiro~Terao} \affiliation{\SLAC}
\author{Ka~Vang~Tsang} \affiliation{\SLAC}
\author{Tracy~L.~Usher} \affiliation{\SLAC}

\collaboration{on behalf of the DeepLearnPhysics Collaboration}\noaffiliation

\begin{abstract}
    Liquid Argon Time Projection Chambers (LArTPCs) are a class of detectors that produce high resolution images of charged particles within their sensitive volume. In these images, the clustering of distinct particles into superstructures is of central importance to the current and future neutrino physics program. Electromagnetic (EM) activity typically exhibits spatially detached fragments of varying morphology and orientation that are challenging to efficiently assemble using traditional algorithms. Similarly, particles that are spatially removed from each other in the detector may originate from a common interaction. Graph Neural Networks (GNNs) were developed in recent years to find correlations between objects embedded in an arbitrary space. The Graph Particle Aggregator (GrapPA) first leverages GNNs to predict the adjacency matrix of EM shower fragments and to identify the origin of showers, i.e. primary fragments. On the PILArNet public LArTPC simulation dataset, the algorithm achieves a shower clustering accuracy characterized by a mean adjusted Rand index (ARI) of 97.8\,\% and a primary identification accuracy of 99.8\,\%. It yields a relative shower energy resolution of $(4.1+1.4/\sqrt{E (\text{GeV})})\,\%$ and a shower direction resolution of $(2.1/\sqrt{E(\text{GeV})})^{\circ}$. The optimized algorithm is then applied to the related task of clustering particle instances into interactions and yields a mean ARI of 99.2\,\% for an interaction density of $\mathcal{O}(1)$\,m$^{-3}$.
\end{abstract}


\maketitle

\section{Introduction}

\begin{figure*}[t]
    \centering
    \includegraphics[width=0.98\textwidth]{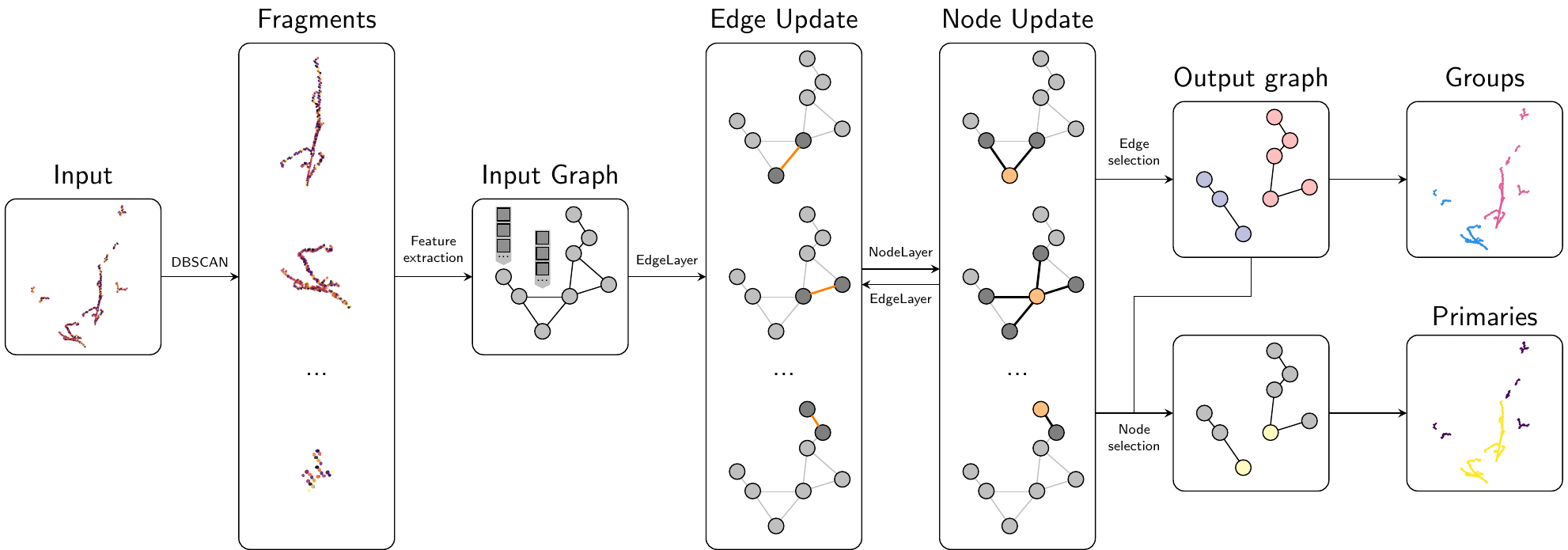}
    \caption{Architecture of the Graph Particle Aggregator (GrapPA) for shower clustering and primary identification. The input set of voxels associated with electromagnetic showers is passed through a density-based clustering algorithm that forms dense shower fragments. Each fragment is encoded into a set of node features in a graph connected by arbitrary edges carrying edge features. Edge and node features are updated through a series of message passing composed of edge and node updaters. The updated edge features are used to constrain the connectivity graph and the updated node features to identify primaries.}
    \label{fig:shower_gnn}
\end{figure*}

In recent years, accelerator-based neutrino oscillation experiments in the United States have been designed to use Liquid Argon Time Projection Chambers (LArTPCs) as their central neutrino detection technology~\cite{lartpc_us}. Charged particles that traverse these detectors ionize the noble liquid. The electrons so produced are drifted in a uniform electric field towards a readout plane. The location of the electrons collected on the anode, combined with their arrival time, offers mm-scale resolution images of charged particle interactions~\cite{rubbia_lartpc}. This level of tracking precision -- coupled to the detailed calorimetric information that a totally active detector provides -- is believed to be the key to resolving some of the ambiguities observed in previous experiments and to extending their energy reach to probe the MeV-scale physics sector~\cite{argoneut_electron, argoneut_mev_scale}.

The Short Baseline Neutrino Program (SBN)~\cite{sbn} aims to clarify an anomalous signal observed by the MiniBooNE experiment~\cite{miniboone}. It will eventually make use of three LArTPCs of varying sizes: the Short Baseline Near Detector (SBND, 112\,t), MicroBooNE (90\,t)~\cite{uboone} and ICARUS (600\,t)~\cite{icarus}. The DUNE experiment\,\cite{dune} will use the LArTPC technology to measure long-baseline neutrino oscillations with unprecedented precision. It will consist of a near detector (105\,t) and a far detector (40\,kt). The main signal for these physics endeavors is the unambiguous appearance of electron neutrinos -- which manifest themselves as electron showers -- in a beam of muon neutrinos. Their success thus centrally depends upon the accurate reconstruction of showers and specifically of their initial positions, directions, and energies. Both experiments also face the substantial challenge of assembling particles into complete neutrino interaction events, which are often accompanied by unrelated activity. Detectors located close to the surface, such as those of the SBN program, suffer from a high rate of cosmic rays, while the future DUNE near detector will observe a high rate of pileup events, with up to twenty neutrino interactions per beam pulse.

Electromagnetic (EM) showers exhibit an incoherent branching tree structure in LArTPCs. As an electron propagates through the dense detector medium, it loses energy through ionization and stochastically emits photons until it comes to a stop. The emitted photons propagate through the noble liquid with a mean free path of 15--30\,cm, for energies in the range 10--1000\,MeV, before they either produce an electron-positron pair or emit a Compton electron~\cite{uboone_pi0}. A single electron or photon creates a cascade of spatially distinct EM particles -- referred to as {\it fragments} in this study -- that may be far removed from one another in the image. Assembling these fragments into coherent shower objects has been a persistent challenge in LArTPCs that has not yet been fully resolved using traditional programming techniques.

Graph Neural Networks (GNNs) became popular in recent years as a way to leverage the concept of receptive field developed in the context of Convolutional Neural Networks and generalize it to arbitrary objects~\cite{gnns}. The receptive field is no longer exclusively determined by a square neighborhood of pixels in an image but rather defined by an adjacency matrix whose elements indicate whether objects are connected by an edge in a graph. This development is ideally suited to the clustering of EM showers in LArTPCs as they may each be represented by a collection of shower fragments (the graph nodes) that are connected by invisible photons (the graph edges). The task is then to identify the edges that connect fragments within a shower and to tag the fragments that initiate showers, i.e the primary fragments. 

In the case of interaction clustering, distinct sources of activity in the detector can be clustered into separate groups using a GNN by building a graph in which particles are nodes and edges represent correlations between particles. The task in then is to identify the edges that connect particles that belong to the same interaction.

The study presented in this paper is reproducible using a \texttt{singularity}~\cite{Singularity} software container \footnote{\href{https://singularity-hub.org/containers/11757}{https://singularity-hub.org/containers/11757}}, implementations available in the \texttt{lartpc\_mlreco}\footnote{\href{https://github.com/DeepLearnPhysics/lartpc\_mlreco3d}{https://github.com/DeepLearnPhysics/lartpc\_mlreco3d}} repository and a public simulation sample~\cite{pilarnet} made available by the DeepLearnPhysics collaboration\footnote{\href{https://osf.io/6gvf4/}{https://osf.io/6gvf4/}}.

Section~\ref{sec:architecture} presents the architecture of the reconstruction chain, as applied to the shower clustering task, from the LArTPC image input to the inference stage. Section~\ref{sec:optimization} outlines the studies that were conducted to optimize the chain. Section~\ref{sec:results} shows a detailed analysis of the shower reconstruction performance on this sample. Section~\ref{sec:inter_clustering} describes how the algorithm was adapted to the particle interaction clustering task and provides some performance metrics.

\begin{figure}[htb!]
    \centering
    \includegraphics[width=\linewidth, trim=1cm 0cm 1cm 3.5cm, clip]{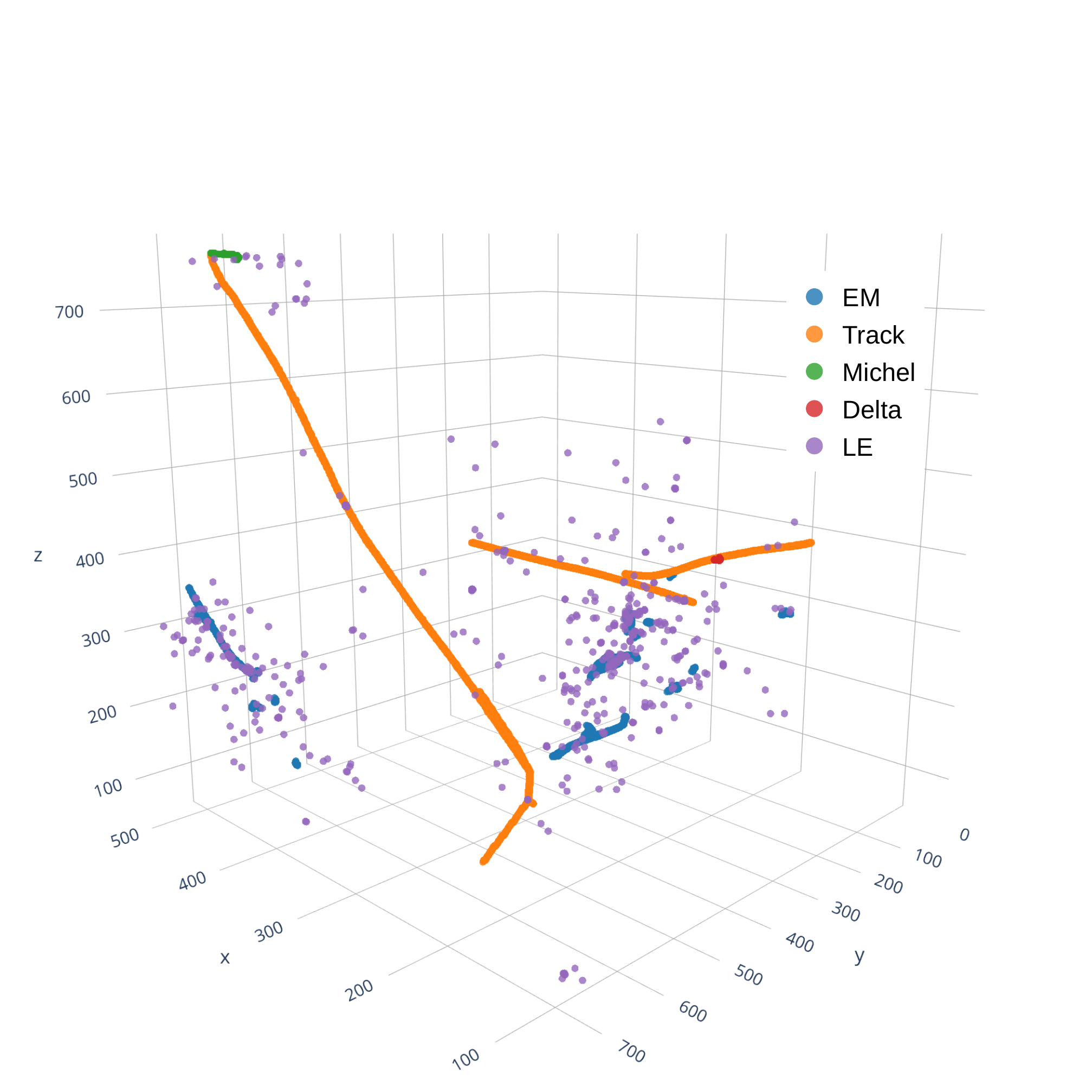}
    \caption{Example image from the simulated LArTPC input dataset. The colors correspond to the semantic type of the particle that deposited the energy: `EM' stands for electromagnetic, `track' for protons, pions and muons, `Michel' for muon decay electrons, `Delta' for delta electrons and `LE' for low energy scatters (low energy EM and nuclear activity).}
    \label{fig:input}
\end{figure}

\section{Reconstruction chain \label{sec:architecture}}

\subsection{Data}
The Graph Particle Aggregator (GrapPA) is schematically illustrated for the shower clustering task in figure~\ref{fig:shower_gnn}. The input LArTPC dataset, PILArNet~\cite{pilarnet}, consists of rasterized 3D energy deposition images of simulated ionizing particle interactions in liquid argon. An image corresponds to a $\sim 12\,$m$^3$ cubic volume of liquid argon with an edge length of 768 voxels ($1\,\text{voxel} = 3^3\,\text{mm}^3$). Each image includes a multiple-particle vertex, i.e. a set of particles originating from a common vertex, overlayed with randomly located track-like cosmic muon trajectories and shower-like instances. An example image is shown in figure~\ref{fig:input}. This image contains three tracks and two showers originating from the vertex in addition to two muon tracks and a detached shower.
 
\begin{figure}[htb!]
    \centering
    \includegraphics[width=\linewidth]{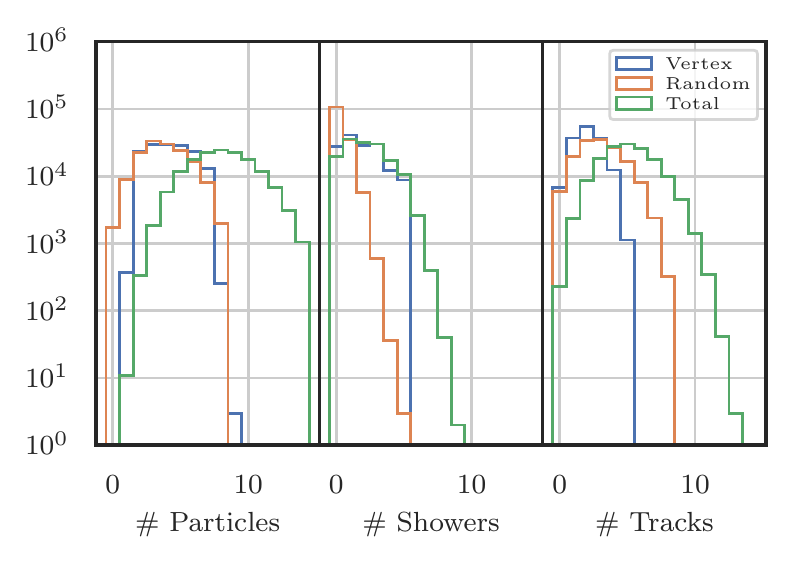}
    \caption{Distribution of the number of particles, showers and tracks in each image, originating from a common vertex (blue), randomly scattered (orange) and combined (green).}
    \label{fig:particle_count}
\end{figure}

Figure~\ref{fig:particle_count} shows the distributions of the number of particles coming from a common vertex in each image, the number of overlayed particles and their respective type compositions. These data emulate a particle density above that expected in an SBN program LArTPC detector. The data is split in two data samples: a training set of 125480 images and a validation set of 22439 images.
 
For the shower clustering task, the set of input voxels is constrained to those associated with electromagnetic (EM) activity. In the scope of this paper, the particle type of each voxel is assumed to be known, as it has been demonstrated that semantic segmentation neural network can identify EM voxels with a 99.4\,\% voxel-wise accuracy~\cite{uresnet_lartpc}. A similarly designed neural network has been shown to work on real data from the MicroBooNE LArTPC detector~\cite{UBPaper2}.
 
\subsection{Fragment clustering}
\begin{figure}[htb!]
    \centering
    \includegraphics[width=\linewidth]{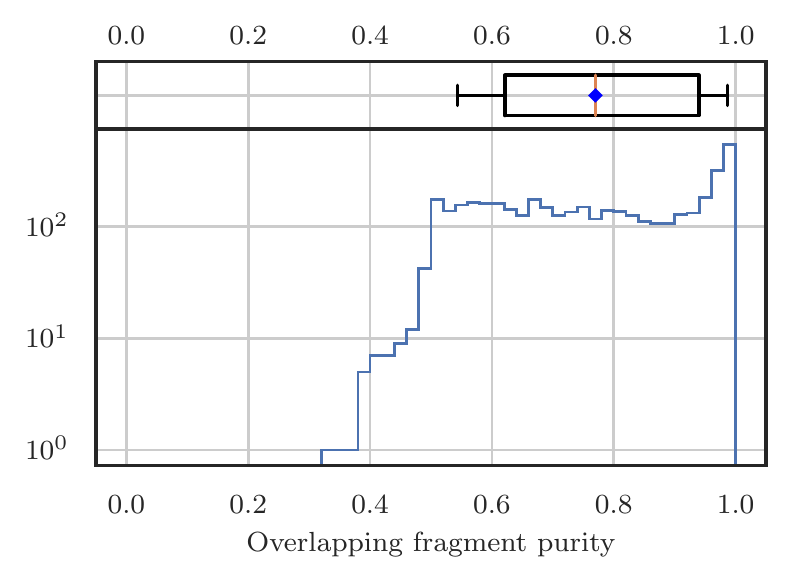}
    \caption{Distribution of the purity of overlapping shower fragments built using DBSCAN with a distance scale of 1.9. In the top box plot, the blue diamond represents the mean, the orange line the median, the box the IQR and the whiskers span from the $10^{\text{th}}$ to the $90^{\text{th}}$ percentiles.}
    \label{fig:dbscan_metrics}
\end{figure}

A shower object encompasses all energy deposition associated with a single primary\footnote{A primary electron or photon does not have an EM parent and is neither a delta ray, a Michel electron nor a deexcitation photon.} electron or photon and all its subsequent EM daughters. A shower fragment is defined as a spatially dense subset of voxels of a shower instance such that each voxel is in the Moore neighborhood of at least one other voxel in the fragment, i.e. at least `touches' it diagonally. As the ground truth\footnote{In the field of computer vision, ground truth refers to the data labels, i.e. a predefined target for the reconstruction.} fragments are not known a priori, EM voxels are clustered using the Density Based Spatial Clustering of Applications with Noise (DBSCAN) algorithm~\cite{dbscan} with a distance scale set to 1.9. DBSCAN cannot break ground-truth fragments, by definition, as any two touching voxels are merged by this algorithm. It can, however, merge two or more fragments that belong to separate shower instances into one. Purity is defined as the maximum fraction of voxels in a predicted fragment that belongs to a single instance. In this dataset, fragments reconstructed with DBSCAN contain more than one ground-truth label in 0.2\,\% of cases, which corresponds to $\sim2$\,\% of images. Figure~\ref{fig:dbscan_metrics} shows the purity distribution of the small fraction of fragments that do contain an overlap. It follows a close to uniform distribution between 0.5 and 1.
 
\begin{figure}[htb!]
    \centering
    \includegraphics[width=\linewidth]{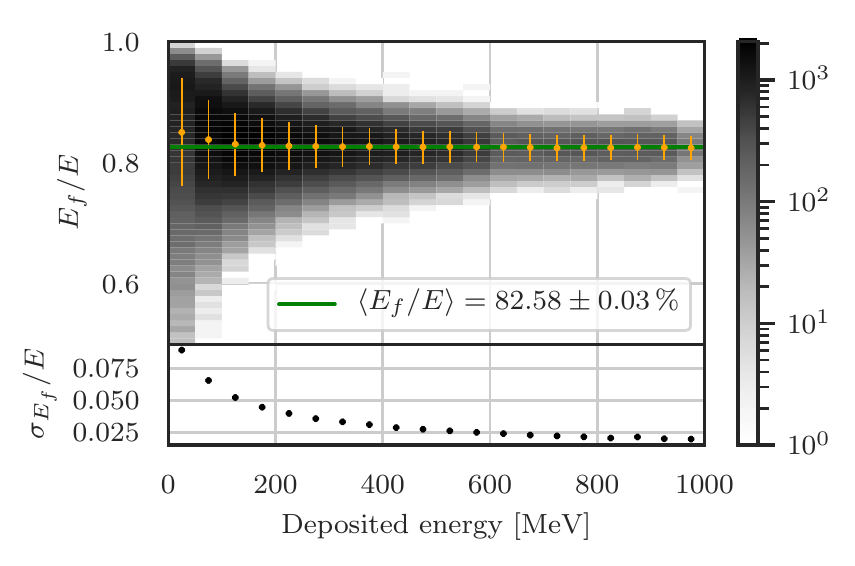}
    \caption{Fraction of the energy deposited by a shower in fragments of size 10 voxels and above. The orange markers on the top pad represent the mean and their error bars the RMS; the latter is shown on its own in the bottom pad. The green line is a constant fit to the markers above 100\,MeV.}
    \label{fig:shower_completeness}
\end{figure}

Fragments strictly smaller than 10\,voxels are not included in the input to the clustering task as they have no clear directionality. To characterize the impact of this selection on the shower energy resolution, primary showers that are at least 95\,\% contained in the image volume are selected. The fraction of the total shower energy deposited in fragments of size 10 and above is represented as a function of shower energy in figure~\ref{fig:shower_completeness}. The shower completeness is on average $\sim82.6$\,\%, for showers of 100\,MeV and above. For lower energy showers, the fraction of energy deposited in small fragments decreases slightly. This selection introduces an average relative uncertainty of 4.2\,\% on the final energy reconstruction.

\begin{figure}[htb!]
    \centering
    \includegraphics[width=\linewidth, trim=1cm 0cm 1cm 3.5cm, clip]{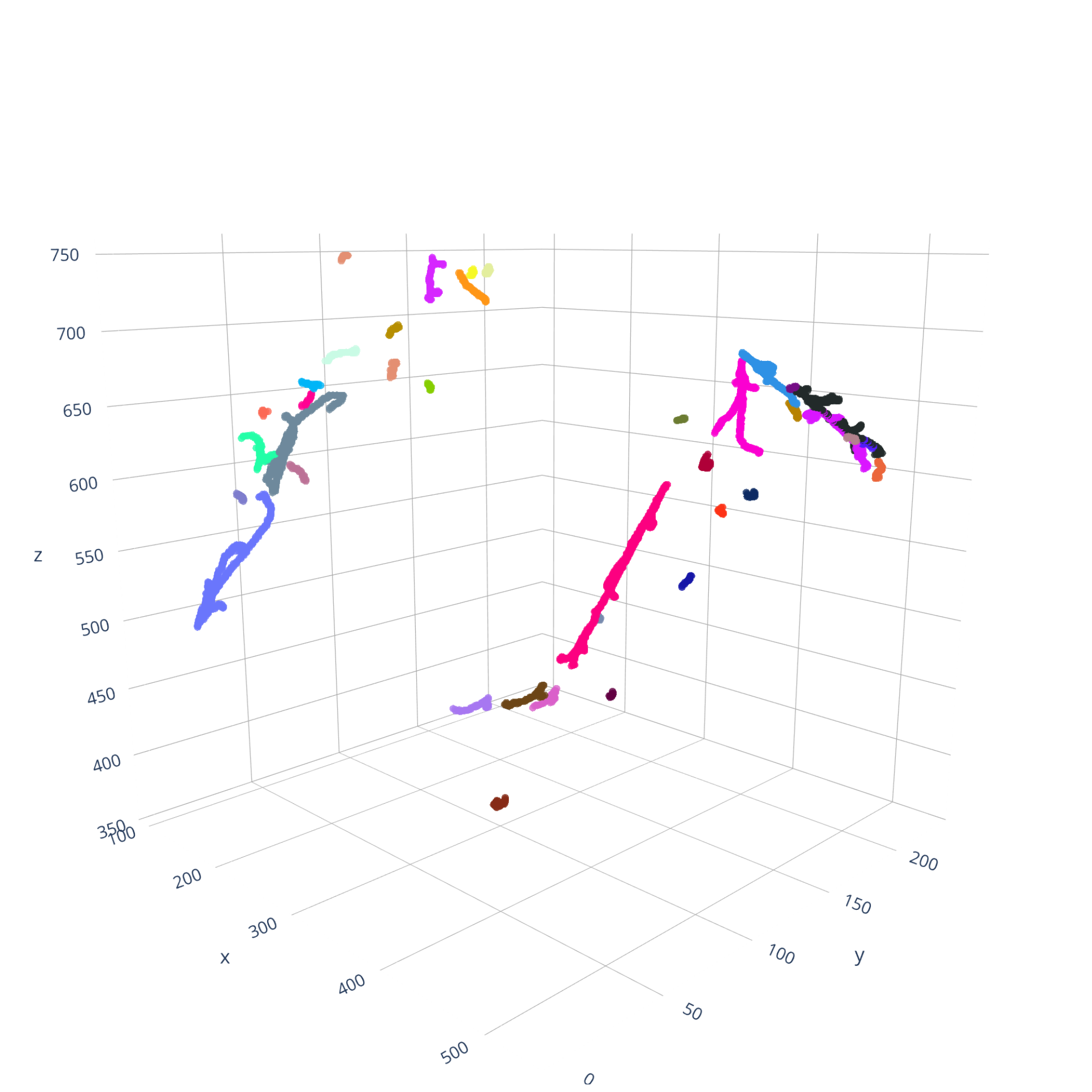}
    \caption{Image of the EM shower voxels with a color scale that represents the DBSCAN cluster ID.}
    \label{fig:fragments}
\end{figure}

Figure~\ref{fig:fragments} shows the fragments constructed upon the shower voxels of an image using the DBSCAN algorithm. The goal of the clustering algorithm is to cluster these fragments together into shower objects.

\subsection{Graph Representation}
A graph $G(V,E)$ is a collection of nodes $V$ and edges $E\subseteq V\times V$. Each shower fragment represents a node in a graph. There is an arbitrary number of ways to build a graph between the nodes. The optimal choice of input edges is discussed in section~\ref{sec:features}. Each node is encoded as a vector of $F_v$ {\em features} and each edge as a vector of $F_e$ features. Multiple ways of extracting these features are studied and optimized in section~\ref{sec:features}.

\subsection{Message Passing}
Message passing is used to communicate information within a graph~\cite{gnns, inductive_bias}. During the information propagation process, at step $s+1$, edge attributes are updated by combing the features coming from the nodes it connects together with its own through
\begin{equation}
    e_{ij}^{s+1} = \psi_{\bm{\Theta}}(\bm{x}_i^s, \,\bm{x}_j^s, \,\bm{e}_{ij}^s),
\end{equation}
with $\bm{x}_i$, $\bm{x}_j$ the node feature vectors associated with nodes $i$ and $j$, respectively, $e_{ij}$ the features of the edge connecting $i$ to $j$ and  $\psi$ a differentiable function such as a Multi Layer Perceptron (MLP)~\cite{neural_nets}. In order to update the node features, the message coming from node $j$ communicated to node $i$ at step $s+1$ is defined as
\begin{equation}
    \bm{m}_{ji}^{s+1} = \phi_{\bm{\Theta}} (\bm{x}_j^s, \,\bm{e}_{ji}^{s+1}),
    \label{eq:message}
\end{equation}
with $\phi_{\bm{\Theta}}$ a differentiable function such as an MLP. The messages coming from the neighborhood $\mathcal{N}(i)$\footnote{The neighborhood of node $i$, $\mathcal{N}(i)$, is the set of nodes which are adjacent to node $i$ in the input graph, i.e. that share an edge with node $i$.} of node $i$ are then aggregated with its own at each step to update its features following
\begin{equation}
    \bm{x}_i^{s+1} = \chi_{\bm{\Theta}}(\bm{x}_i^s,\,\Box_{\mathcal{N}(i)}\bm{m}_{ji}^{s+1}),
    \label{eq:node_update}
\end{equation}
with $\chi_{\bm{\Theta}}$ a differentiable function such as an MLP and $\Box$ an aggregation function such as sum, mean or max. The specific implementation of the differentiable functions and the number of message passing steps are studied and optimized in section~\ref{sec:mp_opt}.

\subsection{Loss definition}
Downstream of the message passing steps, two fully connected linear layers reduce the edge and node features separately to two channels each. The outputs are passed through the softmax function and the second channel is used to create a vector of $N_v$ primary scores for nodes, $\bm{s}^v$, and a vector of $N_e$ adjacency scores for edges, $\bm{s}^e$. The binary cross-entropy loss is then applied to node and edge scores alike as follows:
\begin{align}
    \mathcal{L}_v & = -\frac{1}{N_v}\sum_i y_i\ln(s^v_i)+(1-y_j)\ln(1-s^v_i),\\
    \mathcal{L}_e & = -\frac{1}{N_e}\sum_{(i,j)\in E}a_{ij}\ln(s^e_{ij})+(1-a_{ij})\ln(1-s^e_{ij}),
\end{align}
with $y_i$ the primary label of node $i$ and $a_{ij}$ the adjacency label of the edge connecting node $i$ to node $j$. The primary label, $y_i$, is 1 if the fragment initiated the shower, 0 otherwise. The definition of the target adjacency matrix, $\bm{A}$, which determines the adjacency labels, is discussed in section~\ref{sec:target}. The total loss is defined as $\mathcal{L} = \mathcal{L}_v+\mathcal{L}_e$.

\begin{figure*}[t]
    \centering
    \includegraphics[width=0.98\textwidth]{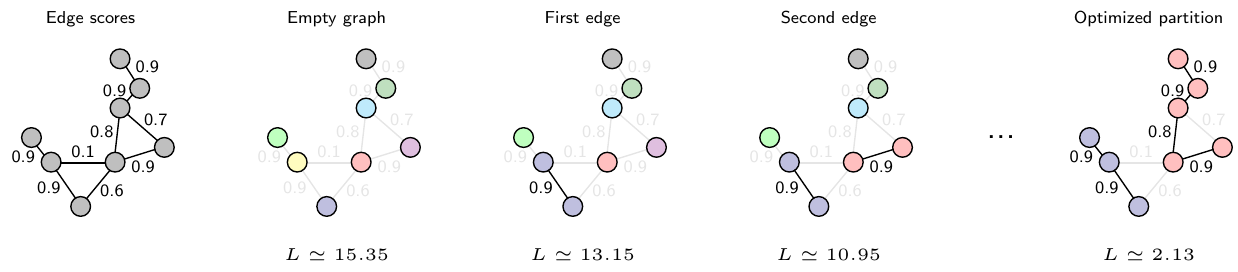}
    \caption{Schematics of the edge selection mechanism at the inference stage. The partition loss defined in equation\,\ref{eq:opt_loss} is first calculated for an empty graph in which each node forms its own group. Edges are sequentially added in order of decreasing score only if the new partition they form decreases the partition loss. The edge with score 0.6 is not added to the graph because it would put the nodes connected by the edge with score 0.1 in the same group and increase the partition loss.}
    \label{fig:inference}
\end{figure*}

\subsection{Inference}\label{sec:inference}
The network predicts an edge score matrix, $\bm{S}^e$, which tries to replicate the predefined ground-truth adjacency matrix, $\bm{A}$. In a graph partition problem, $\bm{A}$ should be designed such that, if $a_{ij}=1$, then nodes $i$ and $j$ belong to the same group. The converse statement does not have to hold, as nodes $i$ and $j$ may not be connected directly as long as they are linked through an indirect path.

Figure~\ref{fig:inference} schematically illustrates how the score matrix is converted into a node partition prediction. At the inference stage, one has to find the optimal node partition, $\hat{\bm{g}}$, such that, if $s^e_{ij}$ is close to 1, nodes $i$ and $j$ are encouraged to be put in the same true group. Mathematically, this corresponds to minimizing the partition cross-entropy loss defined as
\begin{align}
    L(\bm{S}^e|\bm{g}) = &-\frac{1}{N_e}\sum_{(i,j)\in E}\delta_{g_i,g_j}\ln(s^e_{ij})\nonumber\\
    & +(1-\delta_{g_i,g_j})\ln(1-s^e_{ij}),
    \label{eq:opt_loss}
\end{align}
for $\bm{g}$, i.e. $\hat{\bm{g}}=\min_{\bm{g}}L(\bm{S}^e|\bm{g})$, with $\delta$ the Kronecker delta. If an edge is not in the input graph, it does not contribute to the grouping optimization loss.

The cardinality of the set of all possible partitions of a set of $n$ nodes, $G$, corresponds to the Bell number $B_n$. This number grows quickly with the number of nodes to prohibitively large values that rule out brute-force optimization. Instead, edges are considered sequentially to be added to a predicted adjacency matrix, $\hat{\bm{A}}$, in order of decreasing edge score. At each step, the partition score is evaluated by running the Union-Find algorithm~\cite{algorithms} on the predicted matrix. The edge is permanently added to $\hat{\bm{A}}$ if the new partition improves the loss defined in equation~\ref{eq:opt_loss}. The optimizer stops when the next available edge has a score below 0.5.

Given the predicted partition of the graph nodes, $\hat{\bm{g}}$, the primary nodes are identified by picking those with the highest primary score in each group.

\subsection{Metrics}\label{sec:metrics}
Three clustering metrics are used to systematically characterize the performance of the clustering algorithms in this paper: efficiency, purity and adjusted Rand index (ARI)~\cite{clustering_metrics}. The efficiency and purity are defined as:
\begin{align}
    \text{Efficiency} = \frac{1}{N}\sum_{i=1}^{N_t}\max_j |c_j\cap t_i|, \\
    \text{Purity} = \frac{1}{N}\sum_{i=1}^{N_p}\max_j |c_i\cap t_j|,
\end{align}
with $N_t$ the true number of showers, $N_p$ the predicted number of showers, $N$ the total number of voxels in the image, $t_k$ the $k^{\text{th}}$ true cluster and $c_k$ the $k^{\text{th}}$ predicted cluster. The Rand index (RI) is defined as the accuracy on binary edge classification between any two pair of voxels; the ARI ajdusts for random chance by shifting this measure with respect to the average RI obtained for all possible permutations of the predicted labels, i.e.
\begin{equation}
    \text{ARI} = \frac{\text{RI}-E(\text{RI})}{\max(\text{RI}) - E(\text{RI})},
\end{equation}
with $E$ the expectation value. Note that if one of the partitions contains a single cluster, a single voxel mistake yields an ARI of 0, as permutations do not affect RI.

\section{Optimization\label{sec:optimization}}

\subsection{Training regiment}
In this section, the reconstruction steps are optimized to maximize the clustering accuracy in a model that uses ground-truth shower fragments as an input and does not attempt to predict shower primaries. Variations are studied with respect to a baseline model in which
\begin{itemize}
    \setlength\itemsep{0.2em}
    \item the input node and edge features are geometric;
    \item the input graph is a complete graph;
    \item the number of message passings is 3;
    \item the ground-truth adjacency matrix corresponds to a cluster graph built upon groups;
    \item the batch size is 128;
    \item the Adam optimizer~\cite{adam} is used with a learning rate is 0.0025.
\end{itemize}
The reconstruction chain is trained for 25\,epochs of the training set for each configuration under study. The edge classification accuracy and cross-entropy loss are cross-validated with 10000\,events from the validation set every $\sim1$\,epoch to check for overtraining.

\subsection{Feature extraction\label{sec:features}}
Each shower fragment has to be encoded into a set of node features and each edge in the input graph can be provided with a set of features of its own. Two methods of feature extraction have been considered in the context of this paper and are presented and compared in this subsection: Geometric and CNN.

\begin{figure}[htb!]
    \centering
    \includegraphics[width=\linewidth]{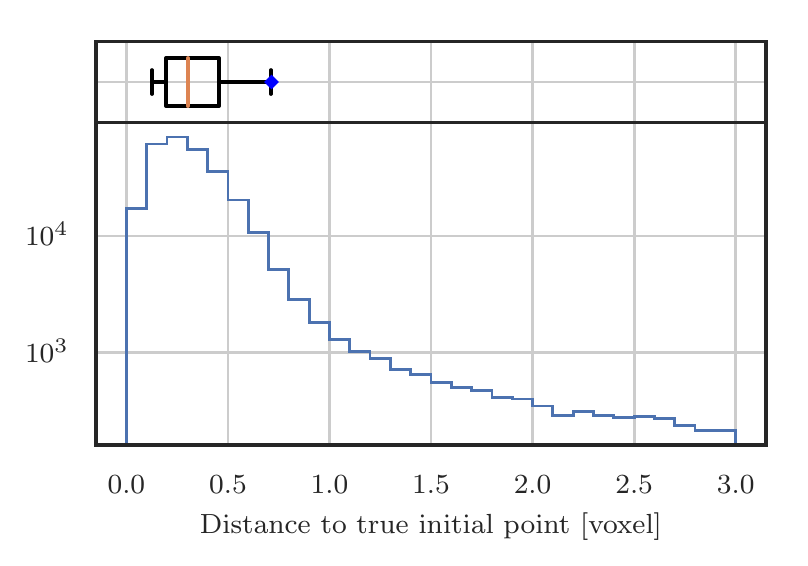}
    \caption{Distance between the point with the highest PPN score within a fragment and the true initial point of the fragment. In the top box plot, the blue diamond represents the mean, the orange line the median, the box the IQR and the the whiskers span from the $10^{\text{th}}$ to the $90^{\text{th}}$ percentiles.}
    \label{fig:ppn_perf}
\end{figure}

Geometric features are a list of summary statistics of the distribution of fragment voxels in Euclidean space. It includes the following 22 features:
\begin{itemize}
    \setlength\itemsep{0.2em}
    \item normalized covariance matrix (9 features);
    \item normalized principal axis (3 features);
    \item centroid (3 features);
    \item number of voxels (1 feature);
    \item initial point (3 features);
    \item normalized initial direction (3 features).
\end{itemize}
In this study, the initial point of a fragment is acquired by picking the center of the voxel, $\bm{v}_s$, closest the true simulated particle first energy deposition. In a realistic setting, it will be reconstructed using the Point Proposal Network (PPN)~\cite{ppn_paper}. Figure~\ref{fig:ppn_perf} shows the distribution of the distance between the point with the highest PPN score in a given fragment and the true initial fragment point. The point is closer than 3\,voxels from the true point in 96.3\,\% of cases.

\begin{figure}[htb!]
    \centering
    \includegraphics[width=\linewidth]{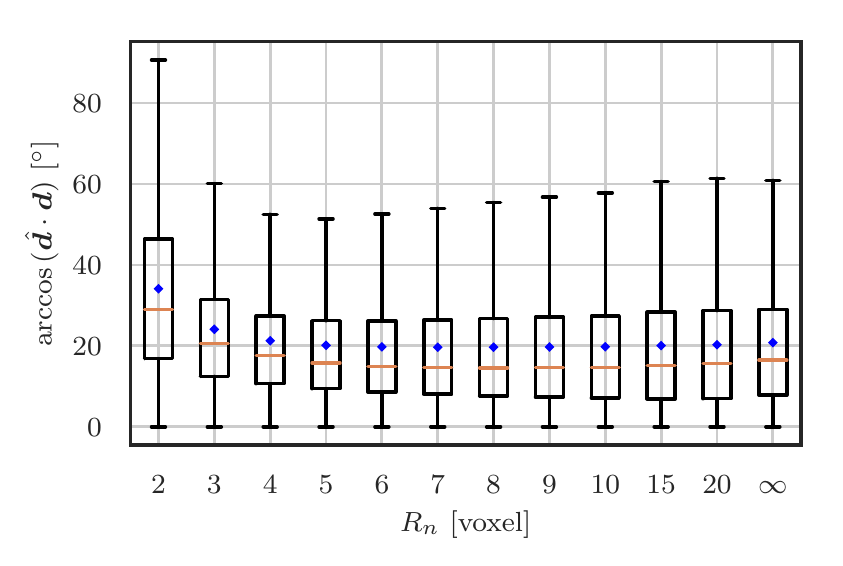}
    \caption{Boxplot of the angle between the reconstructed direction of a shower fragment, $\hat{\bm{d}}$, and the normalized true particle momentum, $\bm{d}$, as a function of the neighborhood cut, $R_n$. The blue diamonds represent the means, the orange lines the medians, the boxes the IQRs and the whiskers extend at most 150\,\% of the IQR on either side of the box.}
    \label{fig:dir_radius}
\end{figure}

The initial direction, $\hat{\bm{d}}$, is estimated by calculating the normalized mean direction from the initial point, $\bm{v}_s$, to all the other fragment voxels within a neighborhood distance $R_n$ of the initial point, i.e. 
\begin{equation}
 \hat{\bm{d}} = \langle \bm{v}\rangle/|\langle \bm{v}\rangle|,\quad \langle \bm{v}\rangle = \frac{1}{N_n}\sum_{\{i|d_{si}\leq R_n\}} (\bm{v}_i -\bm{v}_s),
 \label{eq:dir}
\end{equation}
with $d_{si}$ the Euclidean distance between $\bm{v}_i$ and $\bm{v}_s$ and $N_n=\#\{i|d_{si}\leq R_n\}$ the number of voxels in the neighborhood. The radius was optimized to $R_n=5$ by minimizing the spread of the angle between this direction estimate and the true normalized particle momentum, $\bm{d}=\bm{p}/|\bm{p}|$, as shown for multiple radial cut values in figure~\ref{fig:dir_radius}. In the following, the importance of the initial point and direction for the clustering task is studied by training the model without them (referred to as NI).

The geometric edge features include 19 components:
\begin{itemize}
    \setlength\itemsep{0.2em}
    \item closest points of approach (CPAs) (6 features);
    \item displacement between CPAs (3 features);
    \item outer product of displacement (9 features);
    \item length of displacement (1 feature).
\end{itemize}

The CNN feature extractor treats each fragment as an individual node image -- by masking out voxels that are not associated with it in the image -- and each pair of fragments as an edge image. Images are passed through a Convolutional Neural Network (CNN) which consists of alternating ResNet blocks~\cite{resnet} and strided convolutions which progressively reduce the spatial size of each image while increasing the number of features in each channel. Sparse convolutions are used to efficiently handle mostly empty images~\cite{scn}. In this study, the kernel size is set to 5, the number of strided convolutions to 8 and the input number of filters to 32. The features of the most spatially compressed $3^3$ voxels image are average pooled to form a vector of 64 features per node and 64 features per edge.

\begin{figure}[htb!]
    \centering
    \includegraphics[width=\linewidth]{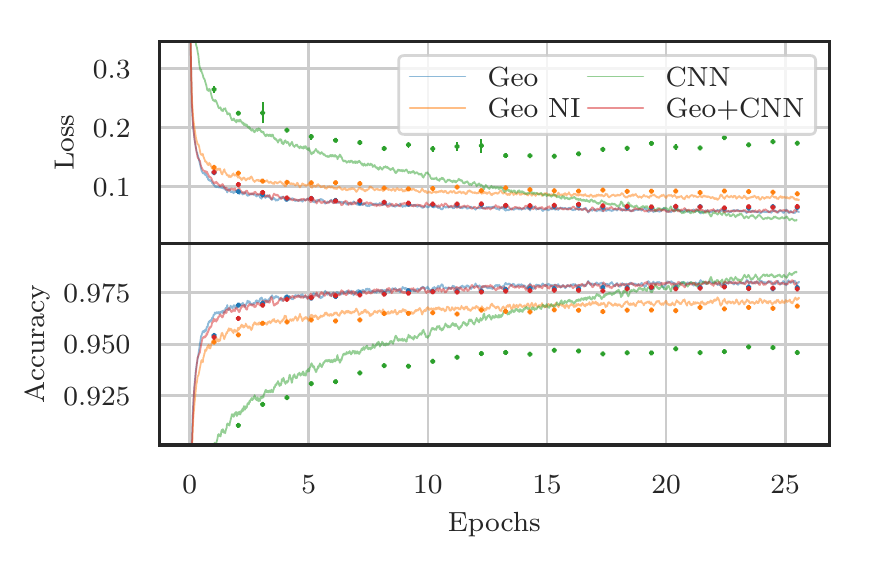}
    \caption{Edge score loss and edge prediction accuracy for the different encoders under considerations. The training curves are represented as lines and the validation points as round markers with statistical error bars.}
    \label{fig:encoders_train}
\end{figure}

Figure~\ref{fig:encoders_train} shows the training and validation curves for each type of feature extractor, produced using the training and validation datasets, respectively. Removing the initial point and initial direction (NI) from the geometric features reduces the edge prediction accuracy by $\sim 1\,\%$. The CNN as a standalone encoder quickly and dramatically overfits the training set. The addition of the CNN features to the baseline geometric features does not measurably improve the global edge classification loss.

\subsection{Input graph}
The input set of fragments is partitioned into groups based on an adjacency score matrix. An edge is given a score only if it appears in the input graph. Several graph construction methods were studied to find the optimal receptive field for nodes:
\begin{itemize}
    \setlength\itemsep{0.2em}
    \item complete graph (all possible pairwise edges);
    \item Delaunay graph (edges in the spatial Delaunay triangulation of the input voxels only);
    \item MST graph (edges in the spatial minimum spanning tree of the input voxels only);
    \item 5NN graph (edges connecting each node with its 5 nearest neighbors only).
\end{itemize}
These graphs are all defined undirected, so that if a message path exists from node $i$ to node $j$, its reciprocal path exists as well. The network is trained to activate both reciprocal paths if two nodes belong to the same group.

\begin{figure}[htb!]
    \centering
    \includegraphics[width=\linewidth]{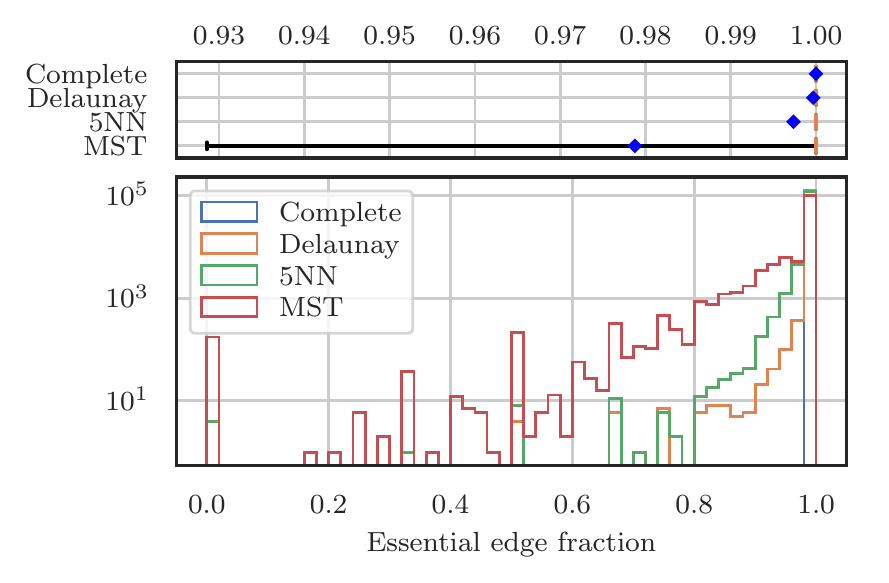}
    \caption{Fraction of the edges necessary to make perfect clustering predictions that appears in an image for the different input graphs under study. In the top box plot, the blue diamonds represent the means, the orange lines the medians and the whiskers span from the $10^{\text{th}}$ to the $90^{\text{th}}$ percentiles.}
    \label{fig:input_graph_eff}
\end{figure}

Restricting the number of edges in the input graph can potentially simplify the clustering task by allowing for the nodes to only focus on messages coming from nodes that are adjacent in the input graph. To be suitable, an input graph must include at least one {\em essential} path from each node to at least one other node that belongs to the same group, without passing through a node that does not. Figure~\ref{fig:input_graph_eff} shows the fraction of essential edges that appear in the input graph, i.e. the fraction of nodes that are reachable through an essential path. It shows that the MST graph is the most inefficient proposed network, on average missing a prohibitively large $\sim2.3\,\%$ of the edges necessary to make correct predictions. All the other graphs are viable options, although their accuracy will be limited in images missing essential edges.

\begin{figure}[htb!]
    \centering
    \includegraphics[width=\linewidth]{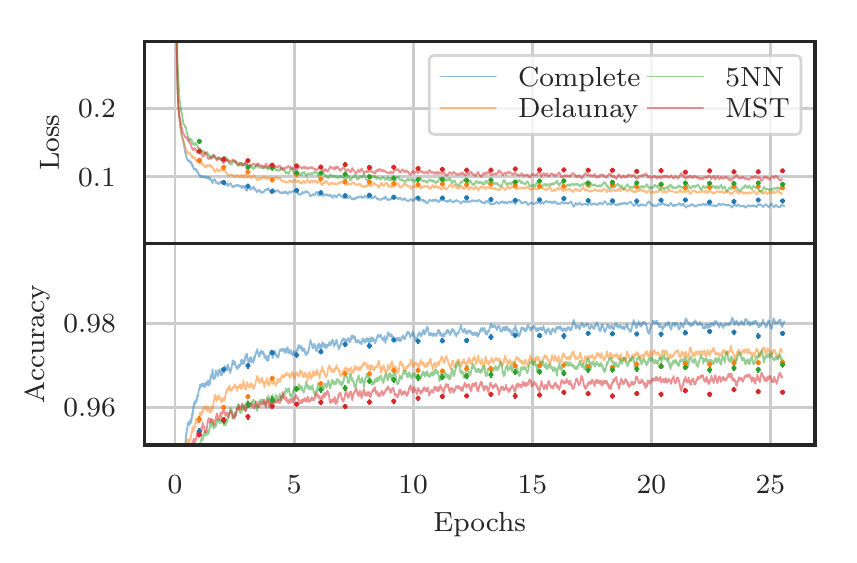}
    \caption{Edge score loss and edge prediction accuracy for the different input graphs under consideration. The training curves are represented as lines and the validation points as round markers with statistical error bars.}
    \label{fig:input_graph_train}
\end{figure}

\begin{figure}[htb!]
    \centering
    \includegraphics[width=\linewidth]{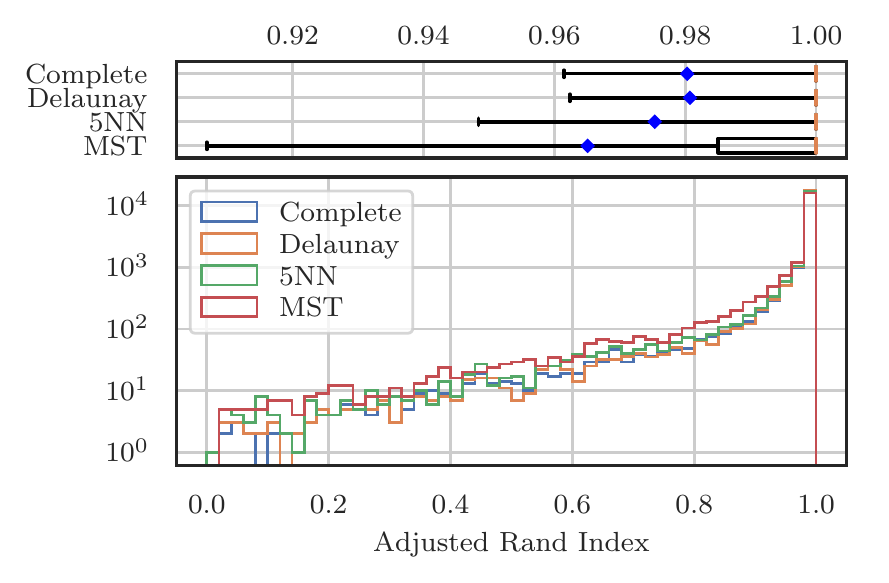}
    \caption{Adjusted Rand index (ARI) distributions on the test dataset for the four input graphs under consideration. In the top box plot, the blue diamonds represent the means, the orange lines the medians, the boxes the IQRs and the whiskers span from the $10^{\text{th}}$ to the $90^{\text{th}}$ percentiles.}
    \label{fig:input_graph_ari}
\end{figure}

Figure\,\ref{fig:input_graph_train} shows the training and validation curves for the aforementioned input graph structures. Figure\,\ref{fig:input_graph_ari} shows the adjusted Rand index clustering metric on the test set for each configuration. The complete graph, which is the one that includes all possible message passing routes, performs best. Delaunay graphs perform similarly -- at a much greater computational cost -- while other input graphs fail to yield a similar precision. This demonstrates the ability of the network to prioritize messages purely based on the features that it is provided with.

\subsection{Message passing\label{sec:mp_opt}}
At a message passing step $s+1$, the edge features are updated by an edge updater which maps $2F_v^{s}+F_e^s$ features coming from the nodes it connects and its own features to $F_e^{s+1}$ features, with $F_v^s$ the number of node features at step $s$ and $F_e^s$ the number of edge features at step $s$. In this study, regardless of the step number, $F_e^{s+1}$ is set to 64. The edge updater MLP consists of three linear layers, each preceded by a 1D batch normalization layer and followed by a LeakyRELU layer of leakiness $0.1$. The first linear layer brings the number of features to 64, the other two maintain that number.

The nodes are updated from $F_v^s$ features to $F_v^{s+1}$ features by a function of neighboring nodes and the edge attributes that connect them, as summarized by equations~\ref{eq:message} and \ref{eq:node_update}. Five \texttt{Pytorch Geometric}~\cite{pytorch_geo} {\em layers} were studied in the context of this paper: MetaLayer, NNConv, EdgeConv, GATConv and AGNNConv.

The MetaLayer~\cite{inductive_bias} uses two successive MLPs to combine the edge features, the node features and their neighbor features. The first MLP combines the $F_v^{s}$ source node features together with the $F_e^{s+1}$ edge features using three linear layers, each preceded by a 1D batch normalization and followed by a LeakyRELU layer of leakiness $0.1$. This produces one message per edge $e_{ij}$, $m_{ij}^{s+1}$, each containing $F_v^{s+1}$ features. The second MLP combines the $F_v^{s}$ target node features with the averaged $F_v^{s+1}$ message features using the same architecture as the first MLP. At each message passing step, $F_v^{s+1}$ is set to 64.

The NNConv~\cite{nnconv} layer is defined as
\begin{equation}
    \bm{x}_i^{s+1} = \bm{\Theta}\bm{x}_i^s + \sum_{j\in\mathcal{N}(i)}\bm{x}_j^{s}\cdot h_{\bm{\Theta}}(\bm{e}_{j,i}^{s+1}),
\end{equation}
with $\bm{\Theta}$ an $F_v^{s+1}\times F_v^{s}$ matrix of weights and $h_{\bm{\Theta}}$ an MLP that maps $F_e^{s+1}$ edge features to an $F_v^{s+1}\times F_v^{s}$ matrix. In this study, the MLP is composed of three layers of a batch normalization, a linear layer and a LeakyReLU function of leakiness 0.1. The first layer increases the number of features from $F_e$ to $F_v^{s+1}\times F_v^{s}$ and the following two keep it constant.

The EdgeConv~\cite{edgeconv} layer is defined as
\begin{equation}
    \bm{x}_i^{s+1} = \sum_{j\in\mathcal{N}(i)}h_{\bm{\Theta}}(\bm{x}_i^{s}||\bm{x}_j^{s}-\bm{x}_i^{s}),
\end{equation}
with $||$ the concatenation operator and $h_{\bm{\Theta}}$ an MLP that maps $2F_v^s$ concatenated features to $F_v^{s+1}$ features. The implementation of the MLP uses an identical implementation to that of the NNConv layer.

The GATConv~\cite{gatconv} layer uses the concept of attention:
\begin{align}
    \bm{x}_i^{s+1} & = \alpha_{ii}\bm{\Theta}\bm{x}_i^s + \sum_{j\in\mathcal{N}(i)}\alpha_{ij}\bm{\Theta}\bm{x}_j^{s},\nonumber\\
    \alpha_{ij}^{s} & = \frac{\exp(LR(\bm{a}^T[\bm{\Theta}\bm{x}_i^s||\bm{\Theta}\bm{x}_j^s]))}{\sum_{j\in\mathcal{N}(i)\cup\{i\}} \exp(LR(\bm{a}^T[\bm{\Theta}\bm{x}_i^s||\bm{\Theta}\bm{x}_j^s]))},
\end{align}
with $LR=\text{LeakyReLU}$ and $\bm{a}$ the attention weight vector of size $2F_v^s$. The weight of the message coming from each node in the neighborhood of $i$ is explicitly learned as a function of the node features.

The AGNNConv~\cite{agnnconv} node updater uses a slightly different attention mechanism:
\begin{align}
    \bm{X}^{s+1} & = \bm{P}^s\bm{X}^{s},\nonumber \\
    P_{ij}^s & = \frac{\exp(\beta\cos(\bm{x}_i^s,\bm{x}_j^s))}{\sum_{j\in\mathcal{N}(i)\cup\{i\}} \exp(\beta\cos(\bm{x}_i^s,\bm{x}_j^s))},
\end{align}
with $\beta$ a learnable parameter. Note that for both of the attention-based layers and the EConv layer, the node features do not take into account the edge features explicitly.

\begin{figure}[htb!]
    \centering
    \includegraphics[width=\linewidth]{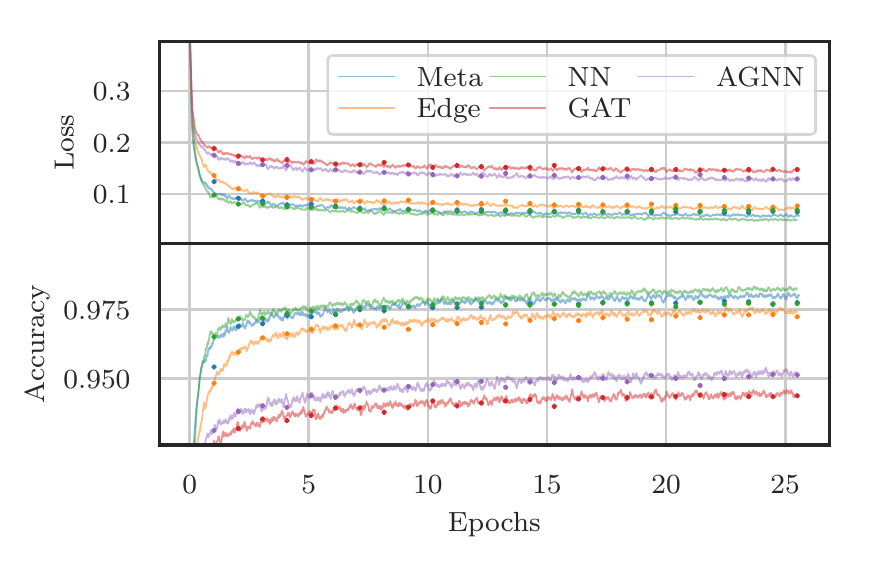}
    \caption{Edge score loss and edge prediction accuracy for the different node updater architecture. The training curves are represented as lines and the validation points as round markers with statistical error bars.}
    \label{fig:mp_archi_train}
\end{figure}

\begin{figure}[htb!]
    \centering
    \includegraphics[width=\linewidth]{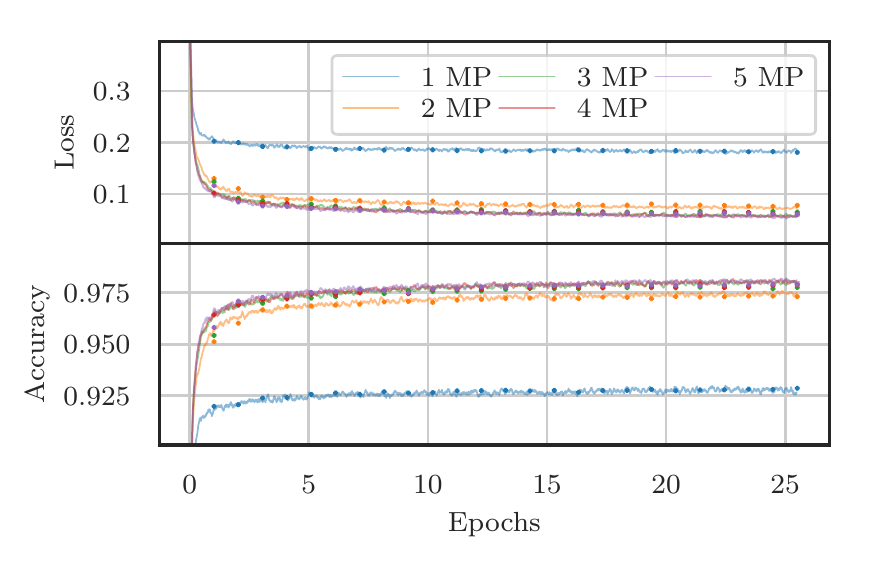}
    \caption{Edge score loss and edge prediction accuracy for the different number of message passing steps. The training curves are represented as lines and the validation points as round markers with statistical error bars.}
    \label{fig:mp_number_train}
\end{figure}

Figure\,\ref{fig:mp_archi_train} shows the training and validation curves for the functions under study and for three iterations of message passing. The MetaLayer node updater performs best, NNConv is comparable but overtrains faster at a large number of epochs. Note that, for this task, the added complexity through the use of an MLP is evidently useful. All layers that explicitly include the edge features -- or the difference between node features as a substitute in the EdgeConv function -- perform similarly. Figure\,\ref{fig:mp_number_train} shows the training and validation curves using the MetaLayer node updater with a number of message passing varying between 1 and 5. Adding more than three message passing steps does not measurably improve the edge classification accuracy.

\subsection{Ground-truth}\label{sec:target}
The ground-truth adjacency matrix may be defined in different ways. The only requirement is that it forms at least a tree within each group of nodes, so that the true partition can be predicted on an edge basis. One possibility is to set the value of edges that connect two nodes within the same group to 1 and all others to 0. This encourages the network to build a cluster graph, i.e a disjoint union of complete graphs.

The second possibility uses the score predictions to define a ground-truth. A tree of $n-1$ edges is built on each true group of $n$ nodes so as to maximize the sum of adjacency scores of the tree edges. The CE loss is then only applied to those edges in the trees and those separating different node groups. This allows the network freedom as to which edge to turn on, as long as it builds a forest, i.e. a disjoint union of trees. In the following discussion, the first ground-truth is referred to as the {\em cluster} target and the second as the {\em forest} target.

\begin{figure}[htb!]
    \centering
    \includegraphics[width=\linewidth]{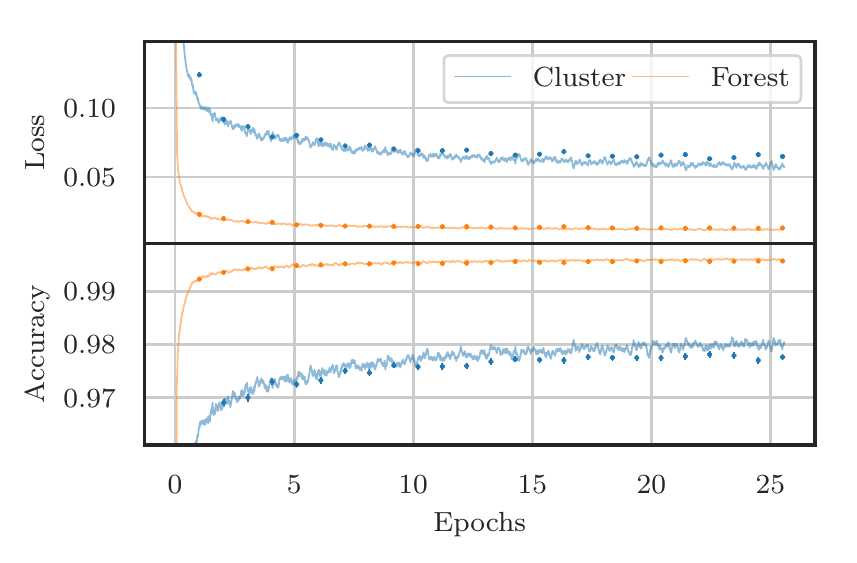}
    \caption{Edge score loss and edge prediction accuracy for the different ground-truths under consideration. The training curves are represented as lines and the validation points as round markers with statistical error bars.}
    \label{fig:target_graph_train}
\end{figure}

\begin{figure}[htb!]
    \centering
    \includegraphics[width=\linewidth]{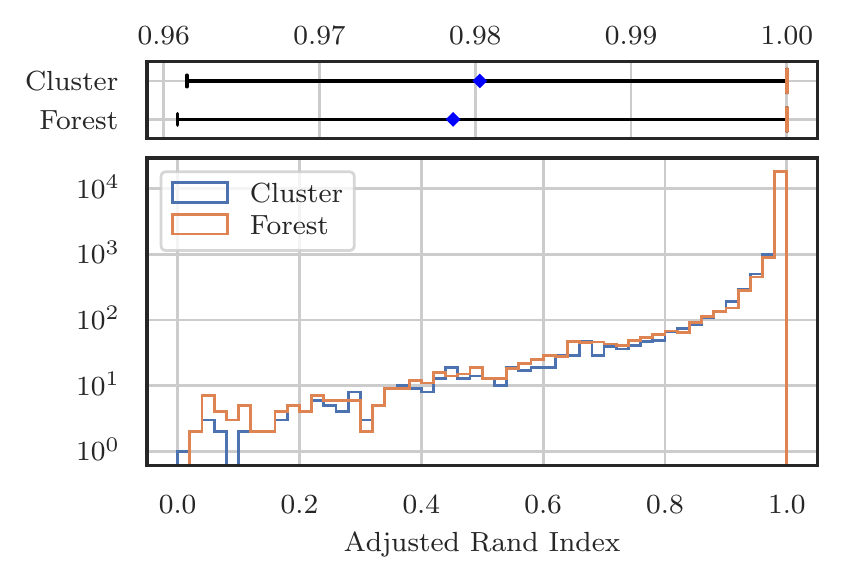}
    \caption{Adjusted Rand Index (ARI) distributions for the different ground-truths under consideration. In the top box plot, the blue diamonds represent the means, the orange lines the medians and the whiskers span from the $10^{\text{th}}$ to the $90^{\text{th}}$ percentiles.}
    \label{fig:target_graph_ari}
\end{figure}

Figure\,\ref{fig:target_graph_train} shows the training edge accuracy and edge loss for the two ground-truths defined above. Figure\,\ref{fig:target_graph_ari} shows their adjusted Rand index (ARI)~\cite{clustering_metrics} distribution on the test set. Both targets yield very similar results with a small lead for the baseline cluster target. Note that the forest target does not require the post-processing described in section~\ref{sec:inference}, as not all edges connecting nodes within a group are trained to be activated. This also means that a single edge incorrectly turned on in a forest introduces mistakes at the inference stage.

\section{Results \label{sec:results}}

\subsection{Training}\label{sec:full_train}

\begin{figure}[htb!]
    \centering
    \includegraphics[width=\linewidth]{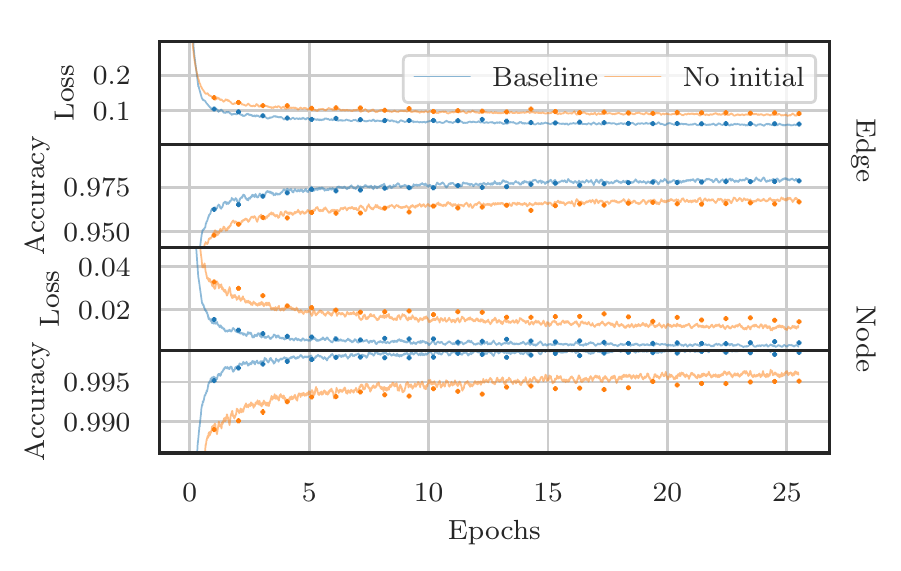}
    \caption{Cross-entropy loss and classification accuracy of edges and nodes for the full shower reconstruction model, with and without initial points. The training curves are represented as lines and the validation points as round markers with statistical error bars.}
    \label{fig:full_train}
\end{figure}

The baseline model is trained for 25\,epochs on the edge and node prediction tasks using the DBSCAN-formed shower fragments and achieves a loss and accuracy summarized in figure~\ref{fig:full_train}. The edge classification accuracy is not affected by the addition of the node classification task nor the use of DBSCAN. The primary classification accuracy is close to 1 when initial points are known.

\subsection{Shower grouping}\label{sec:shower_grouping}

\begin{figure*}[p]
    \centering
    \hfill
    \includegraphics[width=0.32\textwidth, trim=1cm 0cm 1cm 2cm, clip]{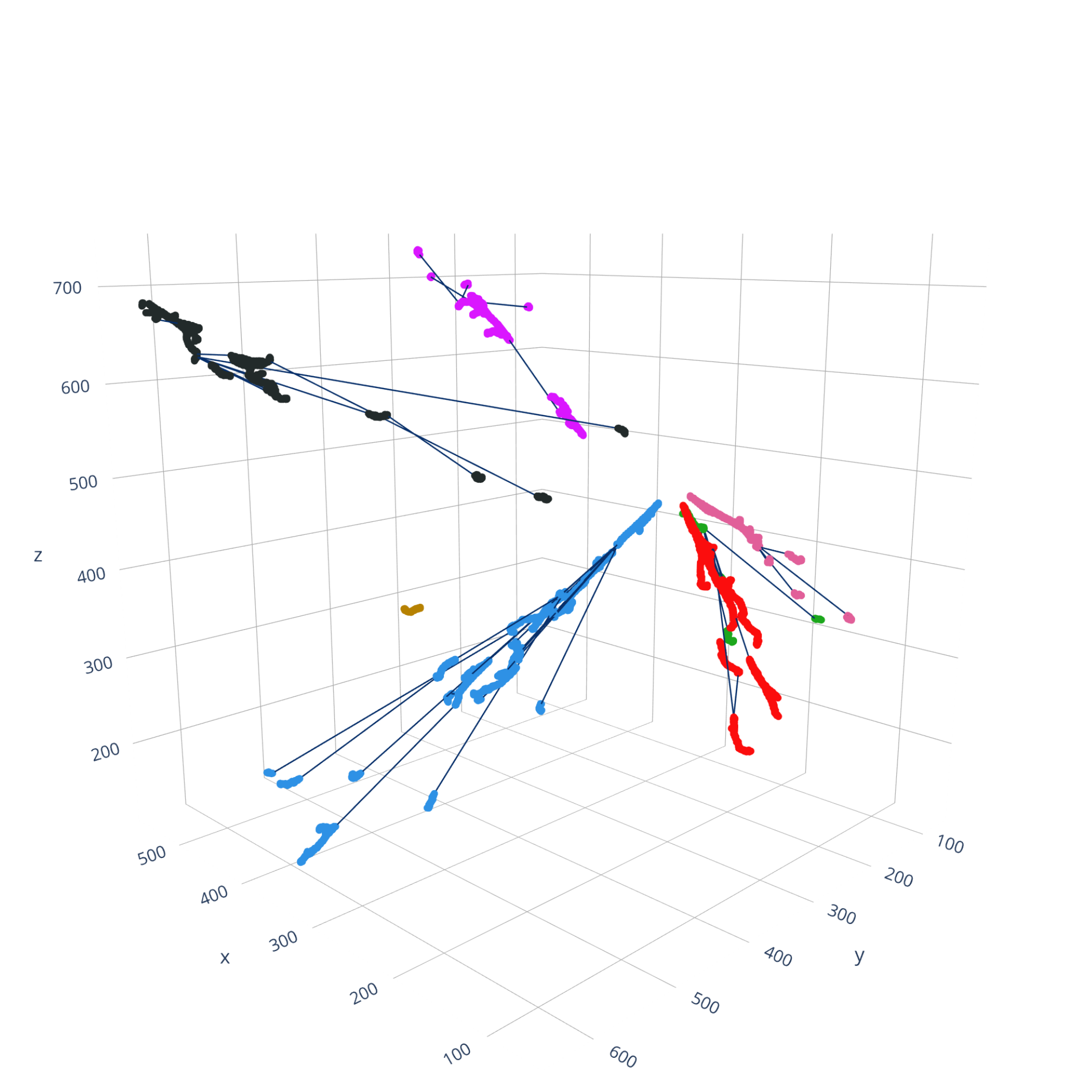}
    \hfill
    \includegraphics[width=0.32\textwidth, trim=1cm 0cm 1cm 2cm, clip]{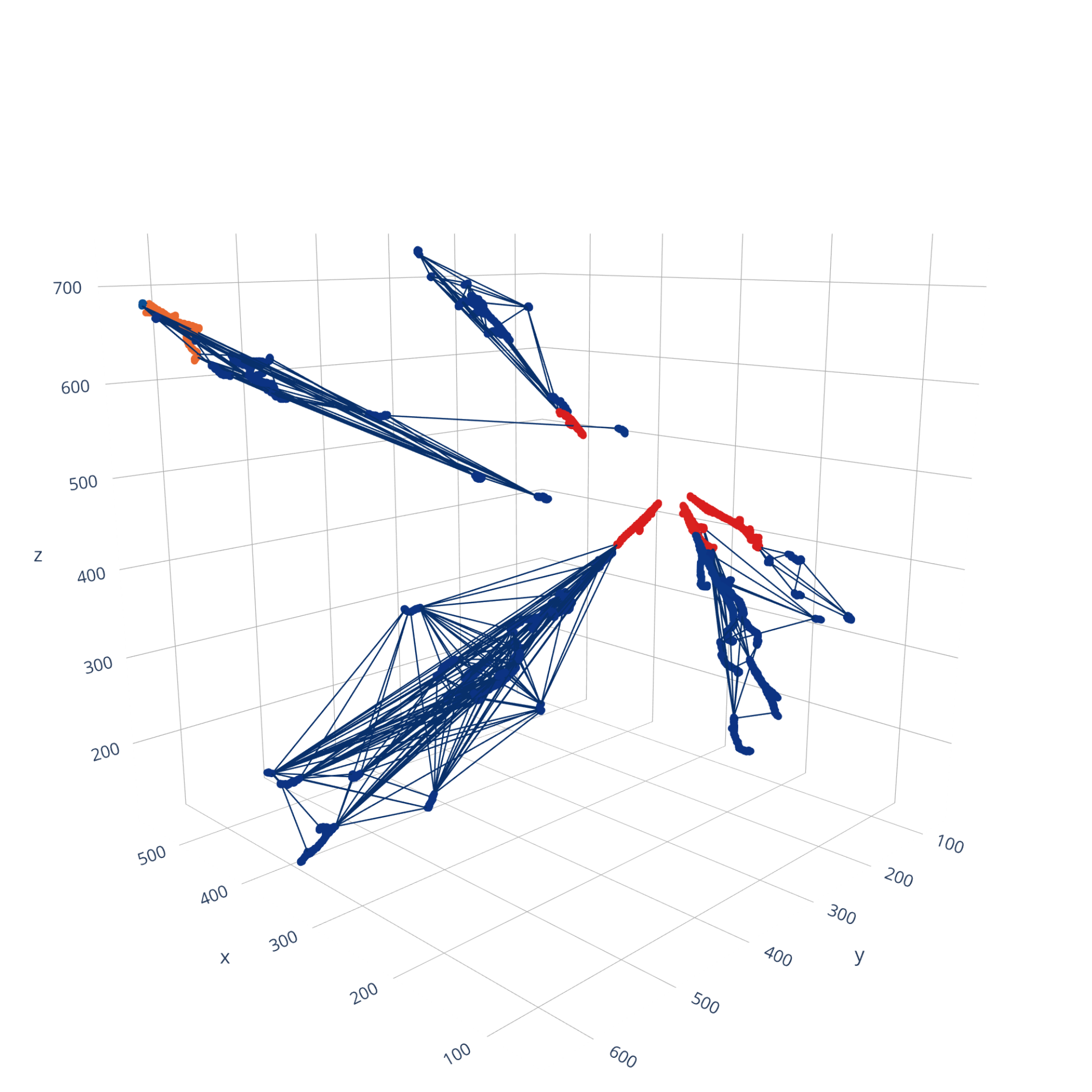}
    \hfill{}
    \begin{overpic}[width=0.32\textwidth, trim=1cm 0cm 1cm 2cm, clip]{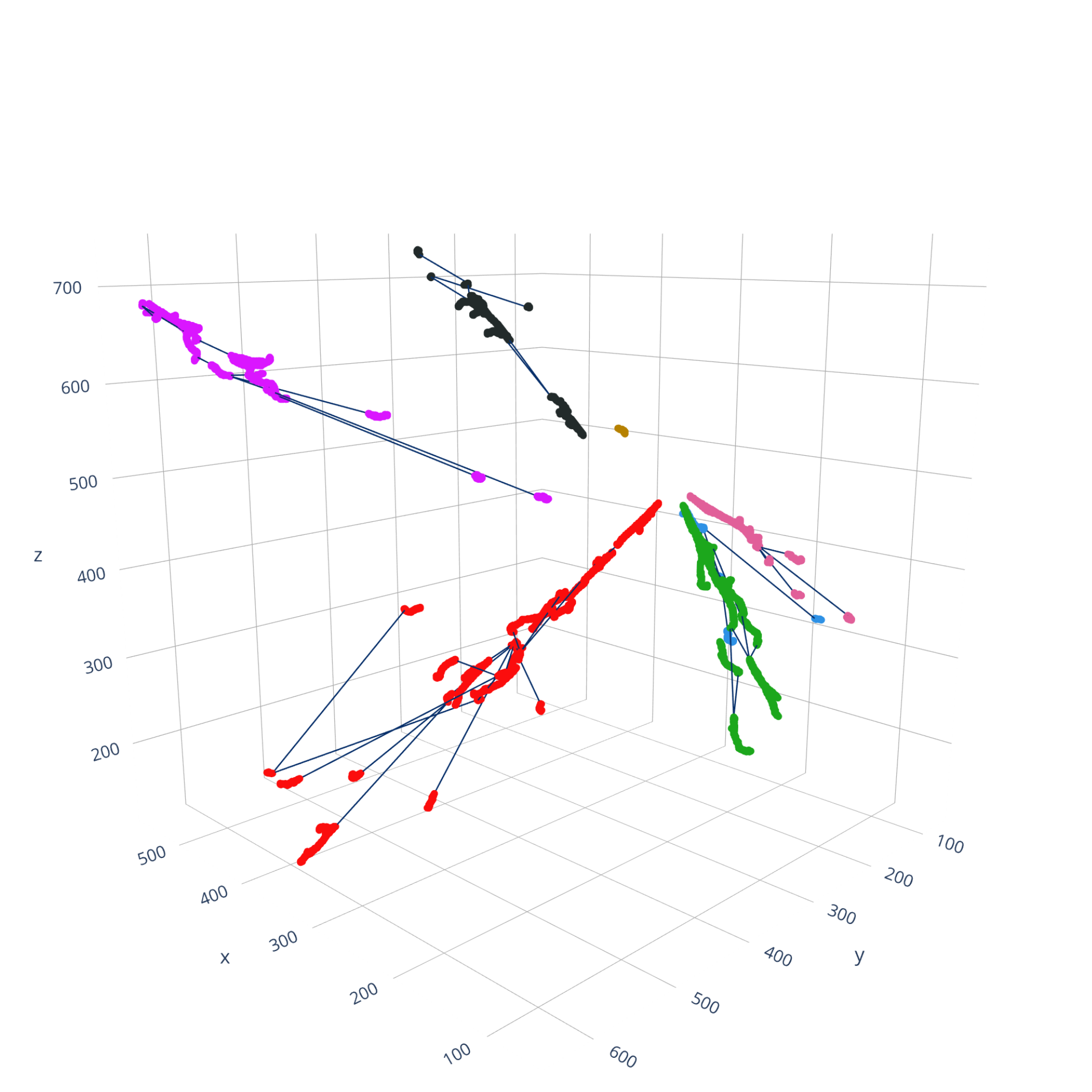}
        \put (35,15) {\fcolorbox{black}{white}{\footnotesize$\text{ARI}:\,98.1\,\%$}}
    \end{overpic}
    \hfill{}
    
    \hfill
    \includegraphics[width=0.32\textwidth, trim=1cm 0cm 1cm 2cm, clip]{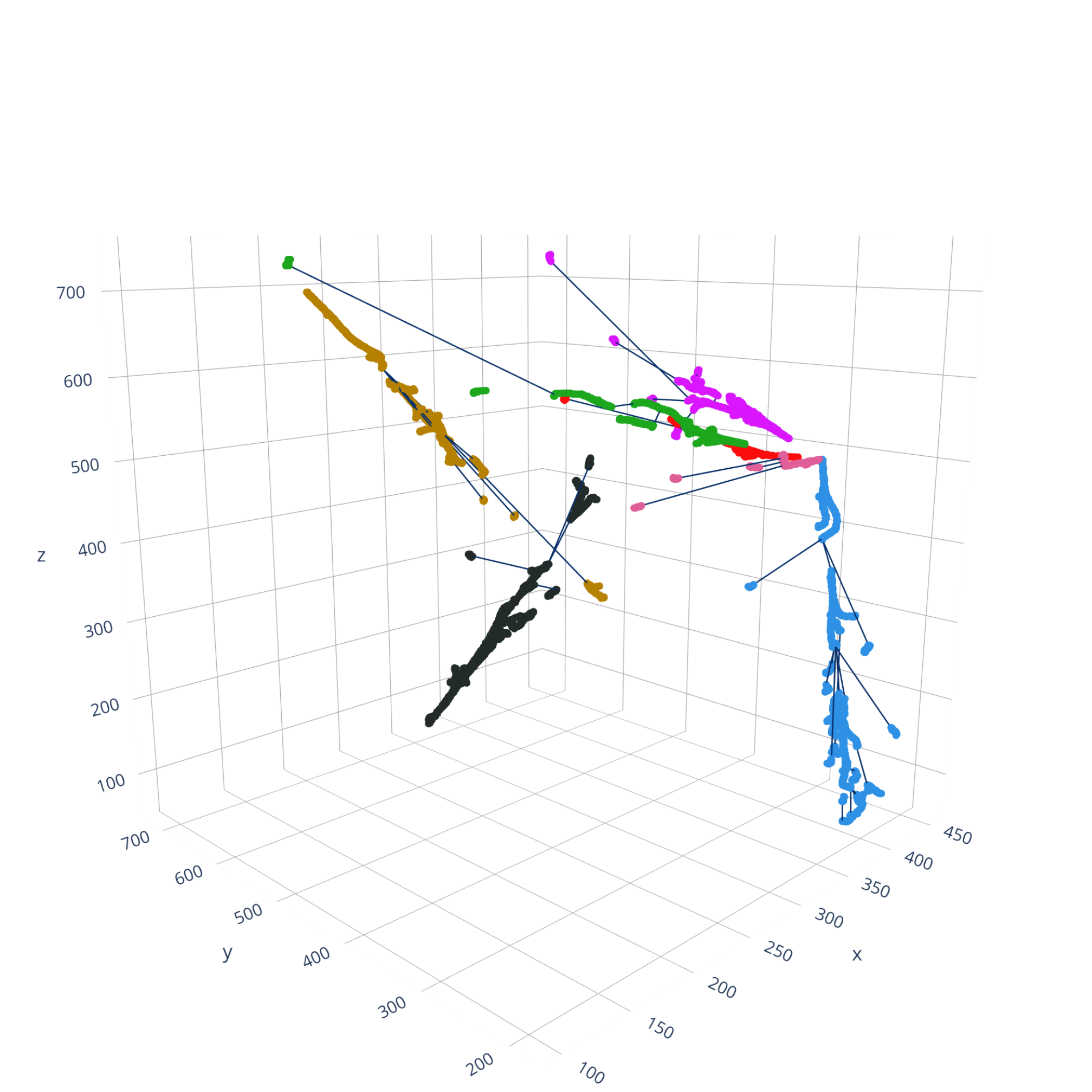}
    \hfill
    \includegraphics[width=0.32\textwidth, trim=1cm 0cm 1cm 2cm, clip]{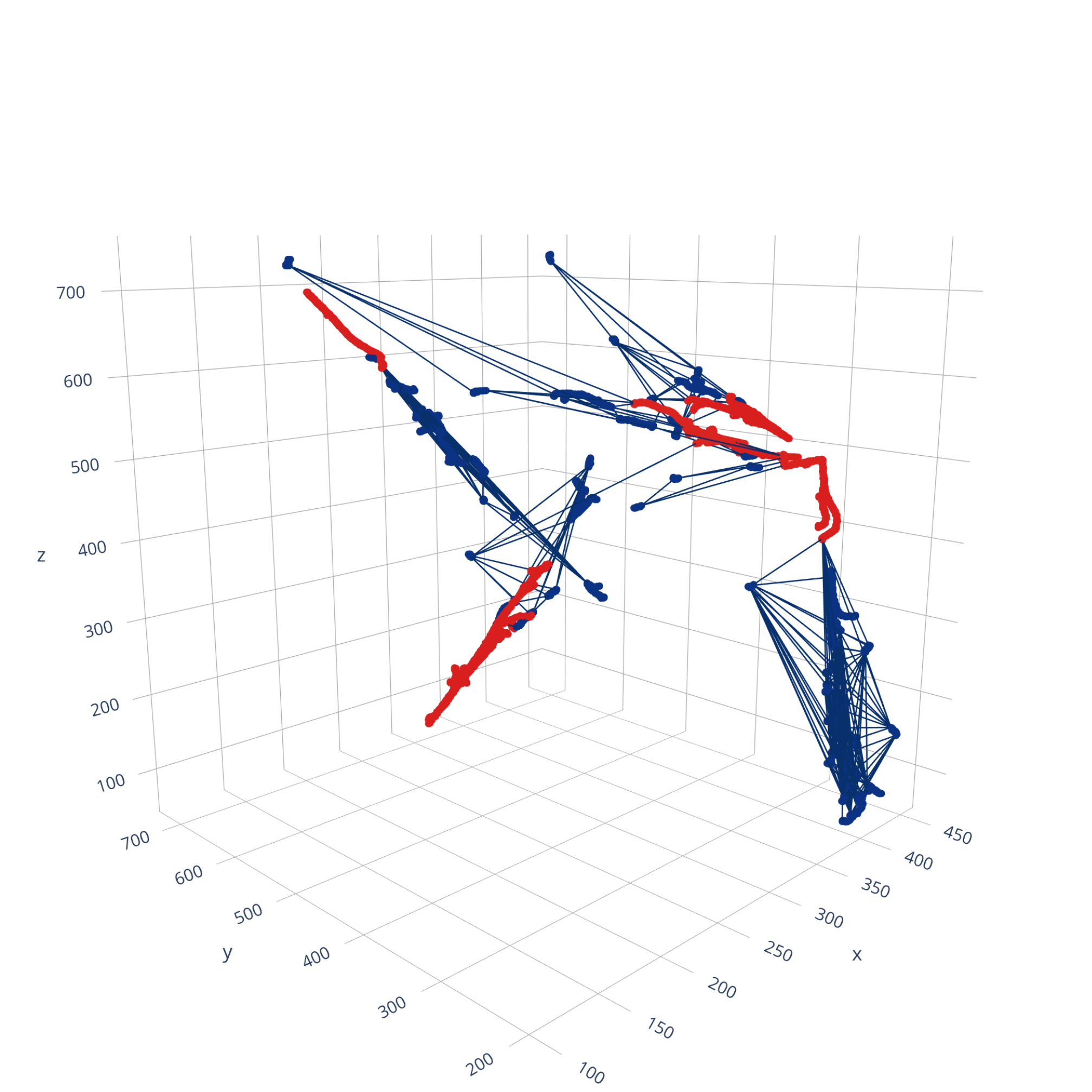}
    \hfill{}
    \begin{overpic}[width=0.32\textwidth, trim=1cm 0cm 1cm 2cm, clip]{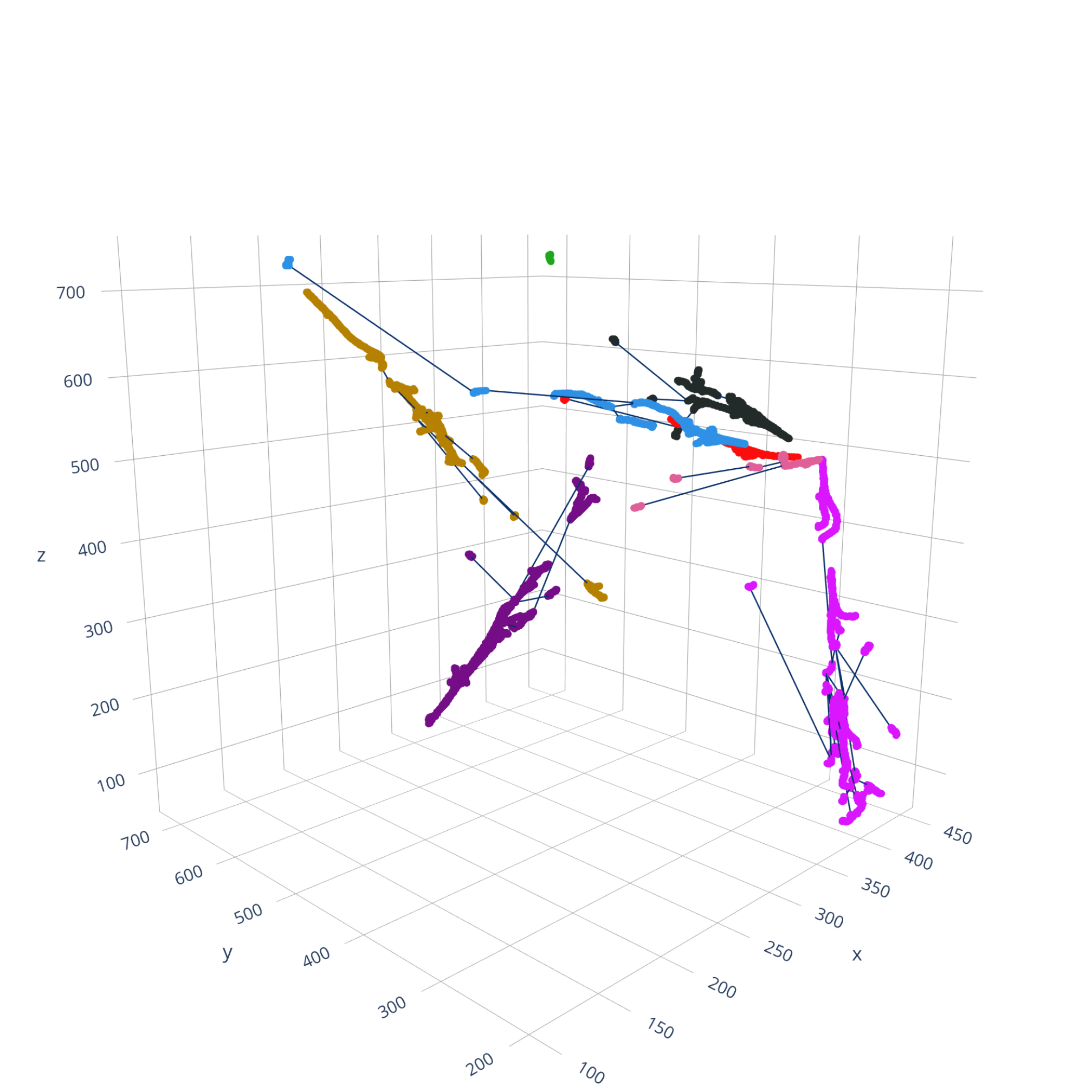}
        \put (35,15) {\fcolorbox{black}{white}{\footnotesize$\text{ARI}:\,99.7\,\%$}}
    \end{overpic}
    \hfill{}
    
    \hfill
    \includegraphics[width=0.32\textwidth, trim=1cm 0cm 1cm 2cm, clip]{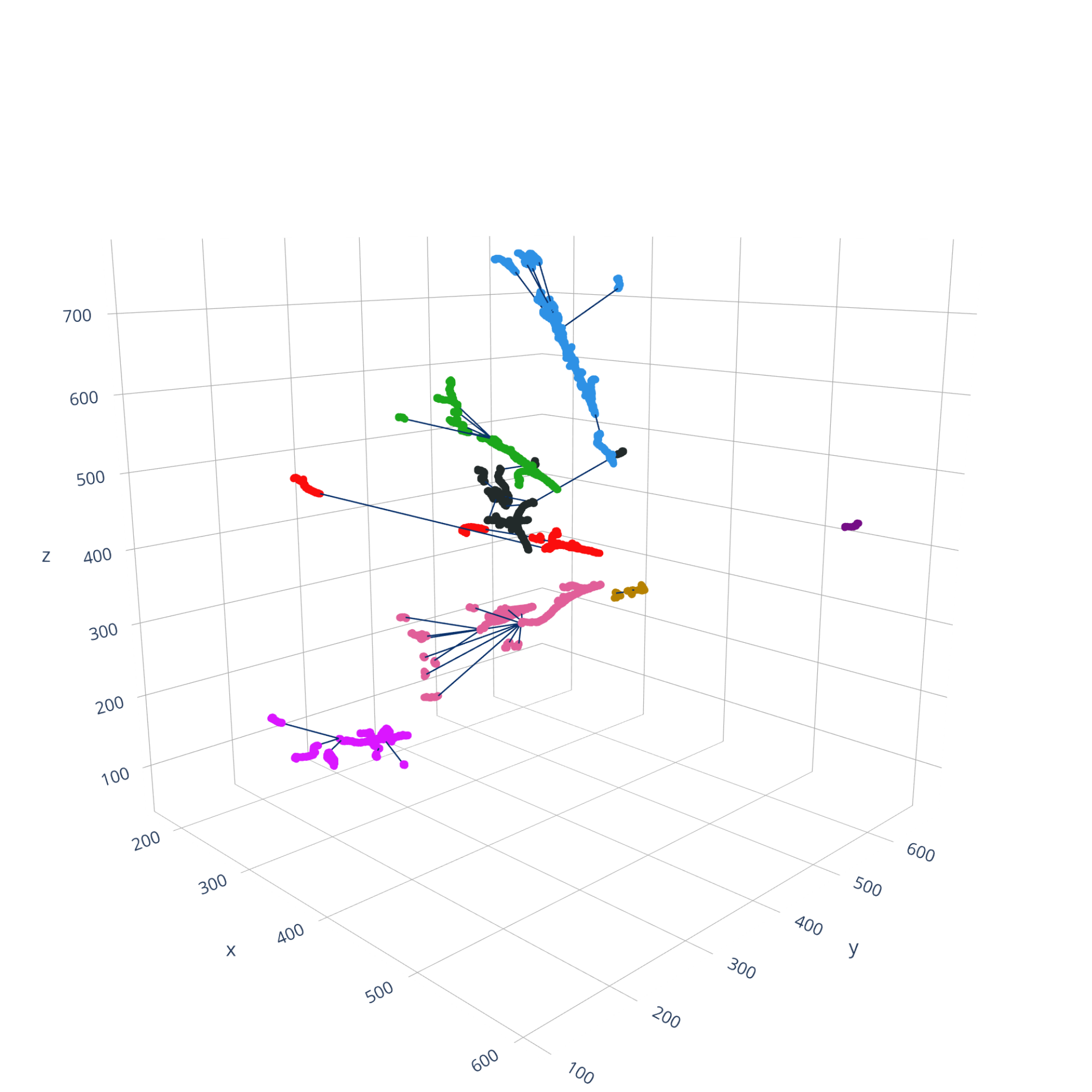}
    \hfill
    \includegraphics[width=0.32\textwidth, trim=1cm 0cm 1cm 2cm, clip]{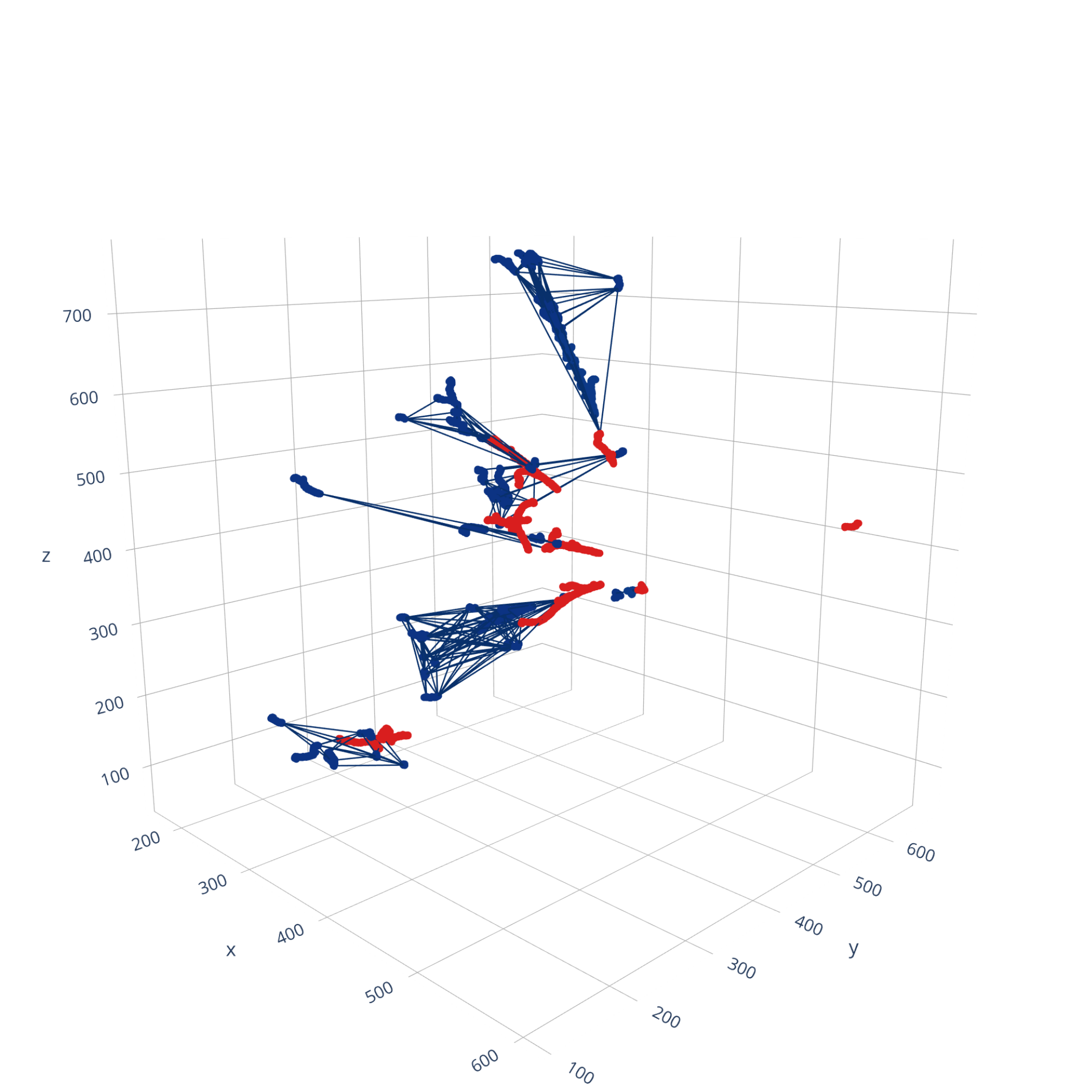}
    \hfill{}
    \begin{overpic}[width=0.32\textwidth, trim=1cm 0cm 1cm 2cm, clip]{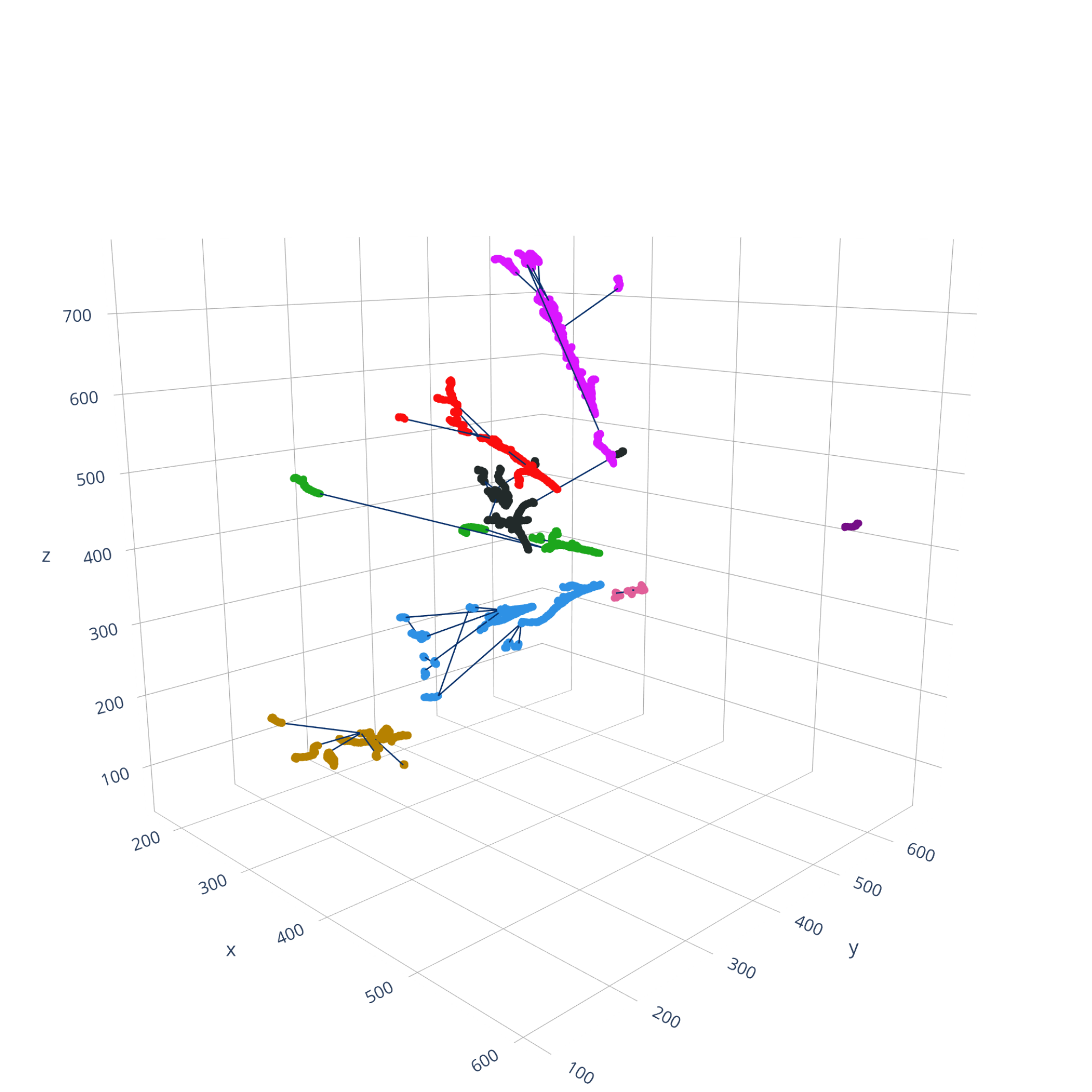}
        \put (35,15) {\fcolorbox{black}{white}{\footnotesize$\text{ARI}:\,100\,\%$}}
    \end{overpic}
    \hfill{}
    
    \hfill
    \includegraphics[width=0.32\textwidth, trim=1cm 0cm 1cm 2cm, clip]{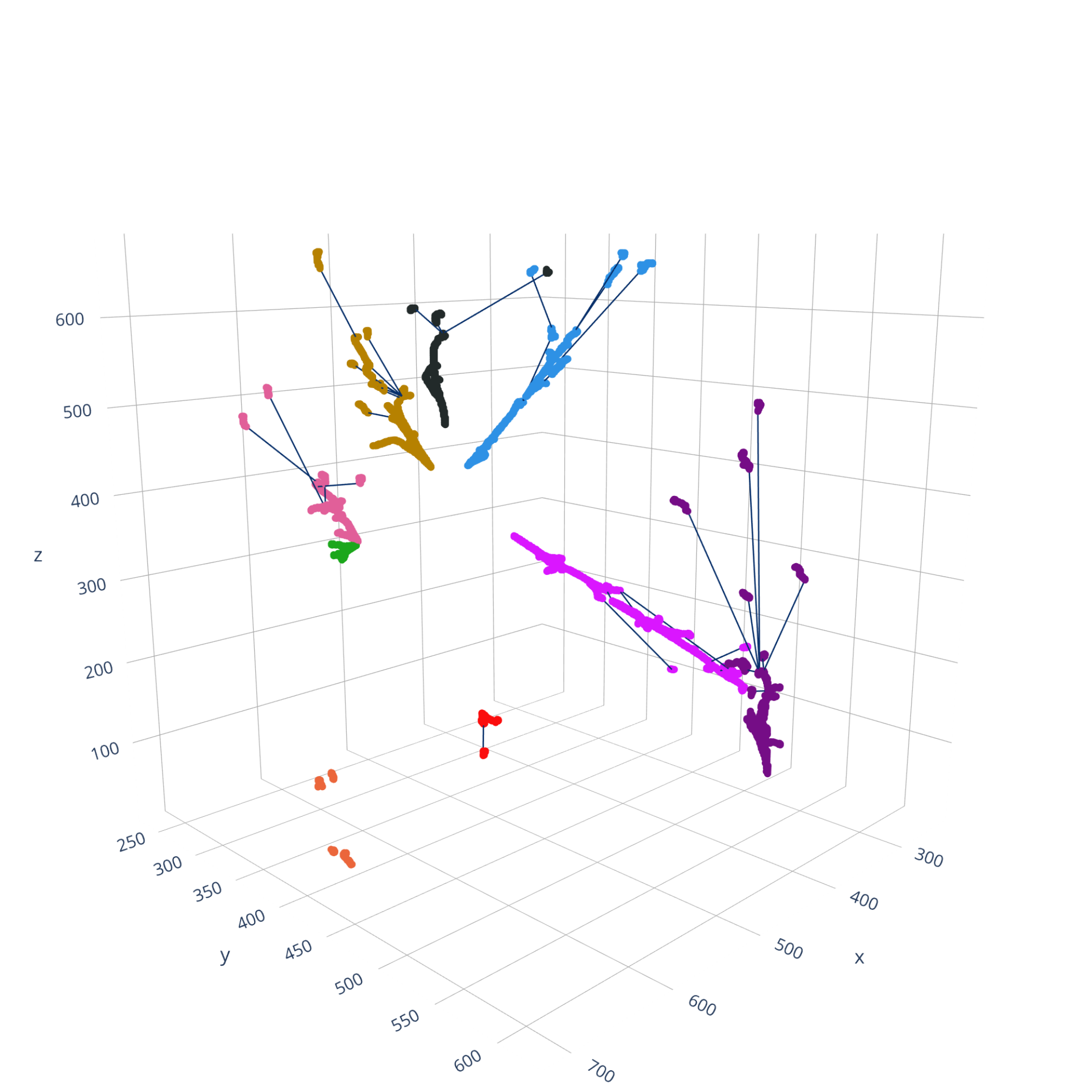}
    \hfill
    \includegraphics[width=0.32\textwidth, trim=1cm 0cm 1cm 2cm, clip]{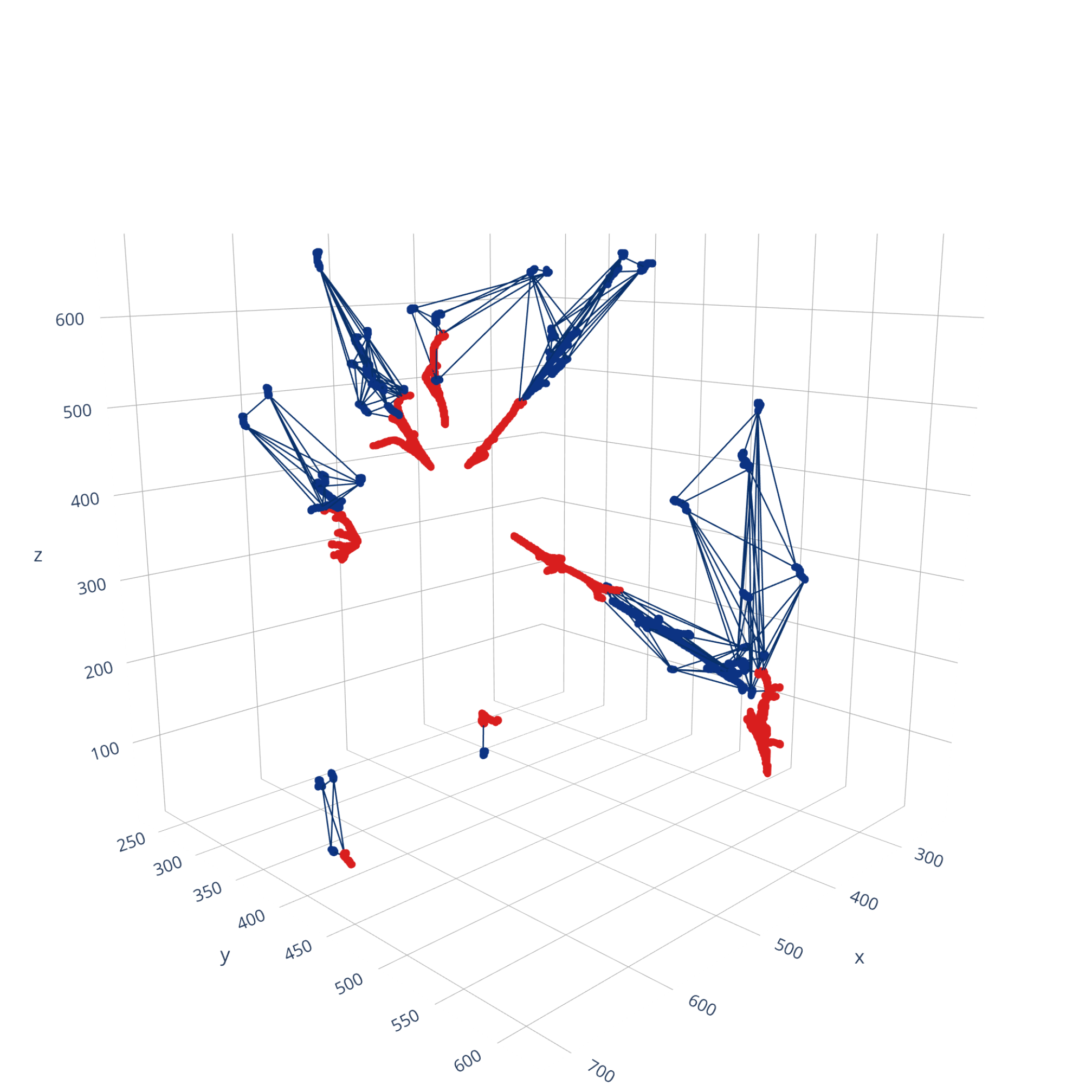}
    \hfill{}
    \begin{overpic}[width=0.32\textwidth, trim=1cm 0cm 1cm 2cm, clip]{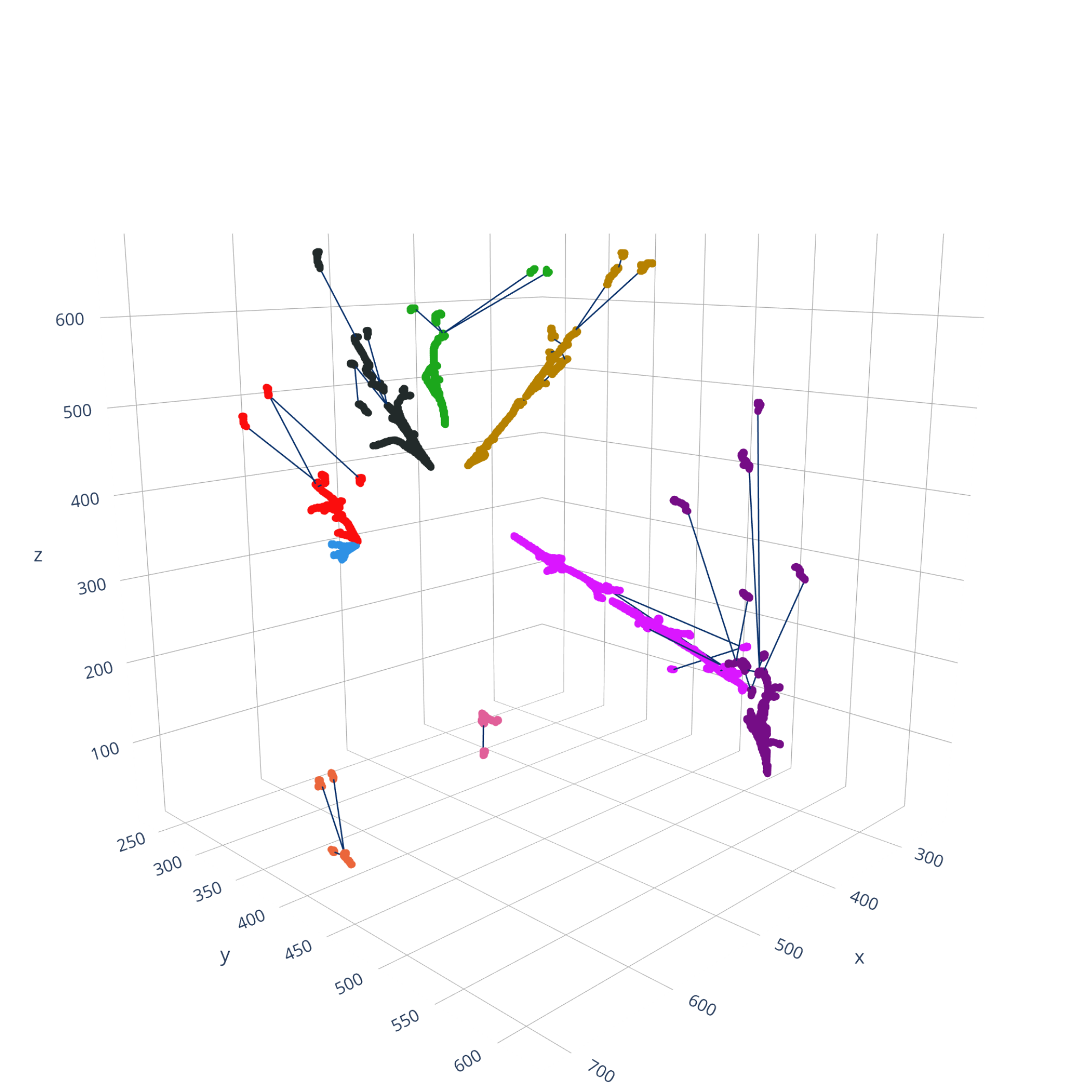}
        \put (35,15) {\fcolorbox{black}{white}{\footnotesize$\text{ARI}:\,99.4\,\%$}}
    \end{overpic}
    \hfill{}
    
    \caption{Shower clustering predictions for the four events with the highest number of shower fragments in the test dataset (one event per row). Left: ground-truth shower labels (color) and edges representing the true fragment parentage. Middle: primary node scores represented as a node color ranging from 0 (blue) to 1 (red) and edges with an adjacency score $>0.5$ (the closer to 1, the darker the edge). Right: inferred shower labels (color) and selected edges. }
    \label{fig:shower_clustering_examples}
\end{figure*}

Figure~\ref{fig:shower_clustering_examples} shows example outputs of the shower reconstruction algorithm for the four events that contain the most fragments in the test set. All four events have a high clustering accuracy, only missing or merging small fragments incorrectly. The top event highlights the importance of the partition loss optimization at the inference stage. Some edges are connecting two separate showers together but the loss minimization allows for the recovery of an almost perfect partition. The bottom left group of the fourth event originates from the same shower but the primary is out of volume. The network correctly associated them together but would not be penalized for making a mistake there. Also note that there is no primary identification mistake in these examples.

\begin{figure}[htb!]
    \centering
    \includegraphics[width=\linewidth]{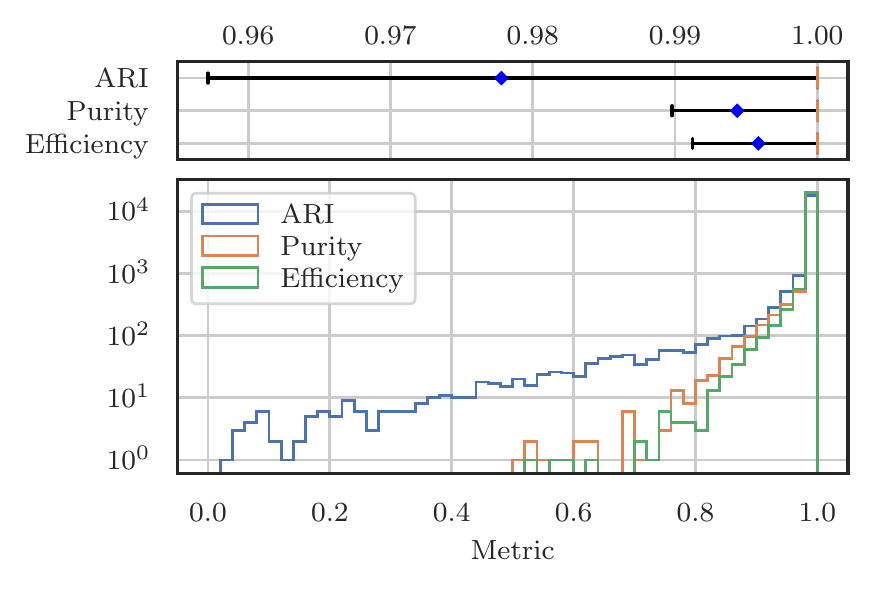}
    \caption{Distributions of the three clustering metrics for the shower clustering task. In the top box plot, the blue diamonds represent the means, the orange lines the medians and the whiskers span from the $10^{\text{th}}$ to the $90^{\text{th}}$ percentiles.}
    \label{fig:shower_clustering_metrics}
\end{figure}

\begin{figure}[htb!]
    \centering
    \includegraphics[width=\linewidth]{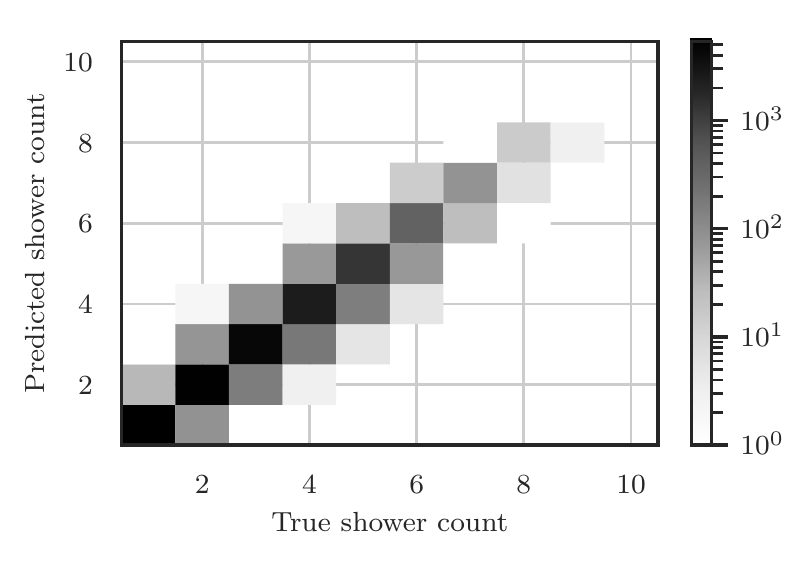}
    \caption{Number of predicted showers as a function of the number of true showers in an event. Only shower instances with over 100\,voxels are included.}
    \label{fig:shower_group_count}
\end{figure}

Figure~\ref{fig:shower_clustering_metrics} shows the clustering metrics associated with the baseline model applied to the test set. The cases with $\text{ARI}=0$ represent $\sim1\,\%$ of all events in this test set and are omitted from the ARI distribution. Figure~\ref{fig:shower_group_count} shows the number of reconstructed showers as a function of the number of true showers in a single image. To prevent small fragments that are either omitted or merged to affect the histogram content, only shower instances with more than 100\,voxels ($\sim60$\,MeV) are included. In 95.03\,\% of events, the estimated shower count is exact.

\subsection{Primary identification}
\begin{figure}[htb!]
    \centering
    \includegraphics[width=\linewidth]{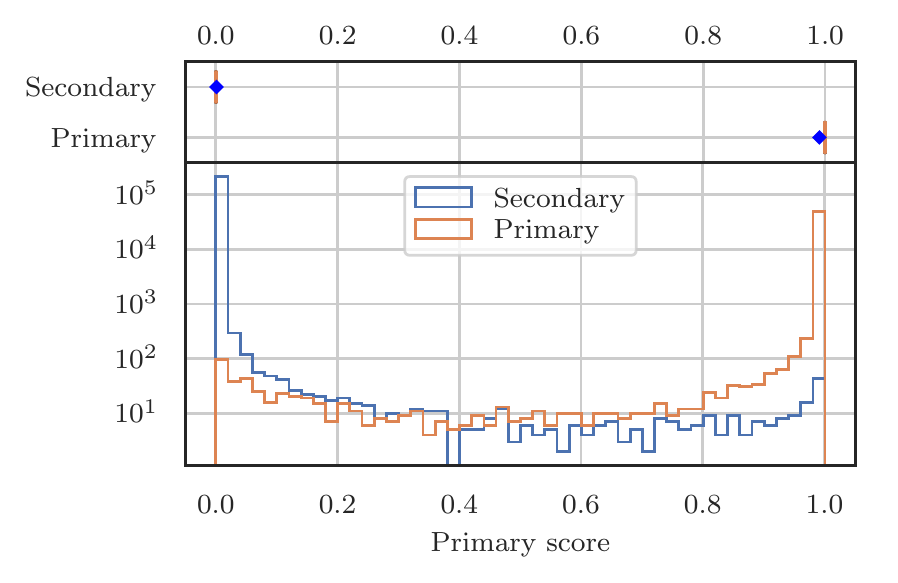}
    \caption{Fragment primary scores of ground-truth primary nodes and ground-truth secondary nodes. In the top box plot, the blue diamonds represent the mean scores and the orange lines the median scores.}
    \label{fig:shower_primary_prediction}
\end{figure}

Figure~\ref{fig:shower_primary_prediction} shows the distributions of primary scores for ground-truth primary and secondary nodes in the test set. This study shows that 0.08\,\% of secondary fragments have a score larger than $0.5$ and 0.84\,\% of primary fragments have a score below $0.5$.

Given the partition predicted by the network, a single primary fragment is assigned to each shower group by selecting the one with highest score. This scheme yields a group-wise primary identification accuracy of 99.77\,\% for showers consisting of two or more fragments. This task is relatively trivial given an understanding of the direction of travel of the shower. The prior knowledge of fragment initial points helps, but even without them, the primary identification accuracy is still at 98.94\,\%.

\subsection{Shower energy resolution}
\begin{figure}[htb!]
    \centering
    \includegraphics[width=\linewidth]{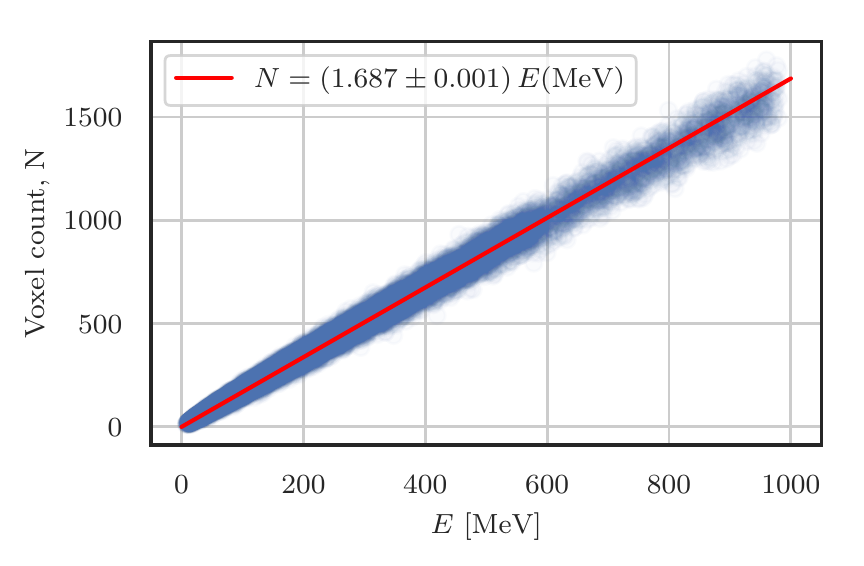}
    \caption{Number of shower voxels contained in shower fragments of size 10 and above as a function of the total energy deposited by the shower.}
    \label{fig:shower_energy_vs_count}
\end{figure}

For each ground-truth shower instance in the dataset, the total amount of energy that it deposits inside the image volume is integrated -- including the fragments smaller than size 10 that are not in the input to the reconstruction chain -- to form the ground-truth shower energy, $E$. For each true shower, the reconstructed cluster with the highest overlap is selected and the energy of its voxels is summed to form an energy estimate, $\hat{E}$. This estimate is multiplied by a fudge factor of 1.211 to compensate for the energy lost in small fragments measured in figure~\ref{fig:shower_completeness}. In order to assess the importance of the calorimetric information on the shower energy resolution, the energy is also estimated using the voxel count alone divided by a factor $1.69$, obtained by fitting the relation between the true number of EM voxels and the energy deposition, as shown in figure~\ref{fig:shower_energy_vs_count}.

\begin{figure}[htb!]
    \centering
    \includegraphics[width=\linewidth]{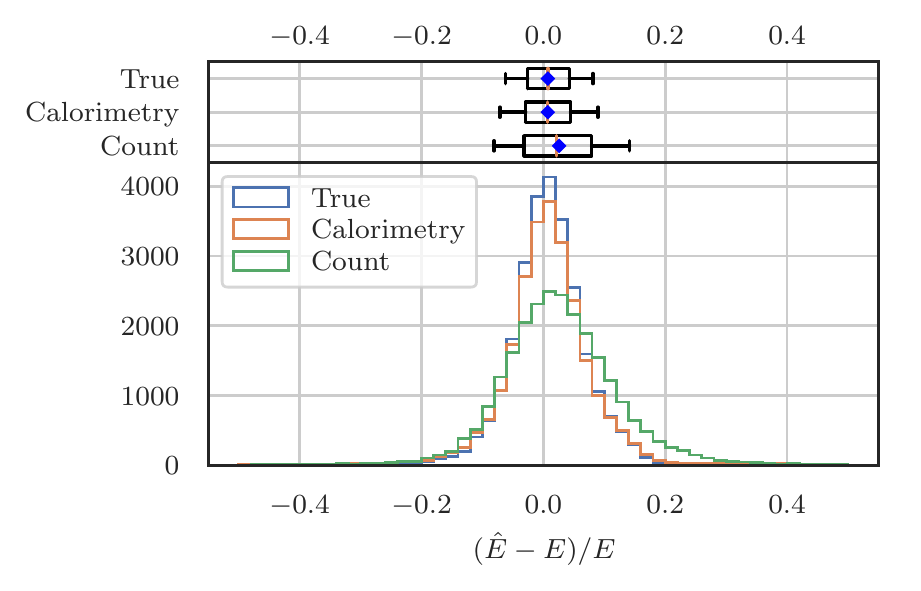}
    \caption{Reconstructed relative shower energy residual distribution. In the top box plot, the blue diamonds represent the means, orange lines the medians, the boxes the IQRs and the the whiskers span from the $10^{\text{th}}$ to the $90^{\text{th}}$ percentiles. `True' uses the true energy of true clusters, `Calorimetry' the true energy depositions of reconstructed clusters and `Count' a constant factor applied to the reconstructed voxel count.}
    \label{fig:shower_energy_resolution}
\end{figure}

\begin{figure}[htb!]
    \centering
    \includegraphics[width=\linewidth]{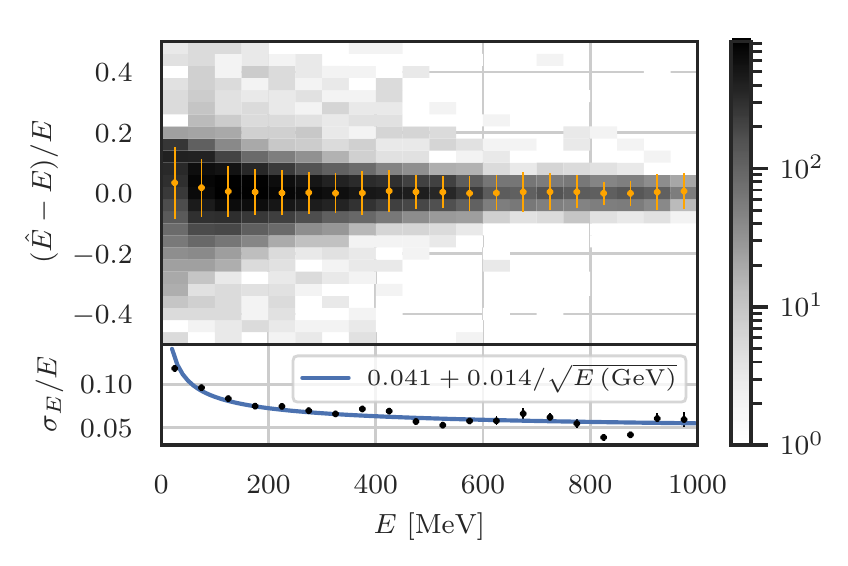}
    \caption{Reconstructed relative shower energy residual distribution as a function of the shower energy. The orange markers on the top pad represent the means and their error bars the RMS; the latter is shown on its own in the bottom pad.}
    \label{fig:shower_energy_resolution_vs_e}
\end{figure}

Figure\,\ref{fig:shower_energy_resolution} shows a distribution of the relative shower energy residuals for the showers that are at least 95\,\% contained inside the images of the test dataset. The residuals are provided with and without leveraging the calorimetric information. These results show that the uncertainty on the shower energy is mostly driven by the prior fragment size selection. Figure\,\ref{fig:shower_energy_resolution_vs_e} shows the $1\,\sigma$ energy resolution as a function of the shower energy. The uncertainty decreases as the shower energy increases and reaches an accuracy around 5\,\% at 1\,GeV.

\subsection{Shower angular resolution}
For each ground-truth shower in the dataset, its true direction is obtained by normalizing its primary momentum vector to its norm, $\bm{d}=\bm{p}/|\bm{p}|$. In practice, the fitted direction, $\hat{\bm{d}}$, is estimated by taking the mean direction from the predicted primary initial point to the primary points with a neighborhood radius $R_n$, as shown in equations\,\ref{eq:dir}. Setting $R_n$ to a constant is suboptimal because the geometry of primary fragments can vary significantly from one shower to another. Smaller radii are preferable for fragments that curve or branch out a lot, but larger radii are advantageous for mostly linear showers.

The radius is optimized in order to minimize the relative deviation of the points from a straight line. The primary fragment points are ordered from closest to farthest from the initial point, $\{\bm{v}\}_{i=1}^n$. The mean, $\bar{\bm{v}}_k$, and covariance matrix, $\bm{\Sigma}_k$, are first evaluated for the closest three points (as the covariance matrix is undefined for $k<3$). The mean and covariance matrix are then iteratively updated for $k={4,\ldots,n}$ following
\begin{equation}
    \bar{\bm{v}}_k = \frac{1}{k}\left((k-1)\bar{\bm{v}}_{k-1}+\bm{v}_k\right),
\end{equation}
\begin{equation}
    \bm{\Sigma}_k = \frac{k-1}{k}\bm{\Sigma}_{k-1}+\frac{1}{k-1}(\bm{v}_k-\bar{\bm{v}}_k)(\bm{v}_k-\bar{\bm{v}}_k)^T.
\end{equation}
For each combination of points, $k$, the ordered eigenvalues of the covariance matrix, $\{\lambda_{k,i}\}_{i=1}^3$, are evaluated. The optimal neighborhood of points minimizes the spread around the principal axis, i.e
\begin{equation}
    k^* = \min_k \frac{\lambda_{k,1}+\lambda_{k,2}}{\lambda_{k,3}}.
\end{equation}

\begin{figure}[htb!]
    \centering
    \includegraphics[width=\linewidth]{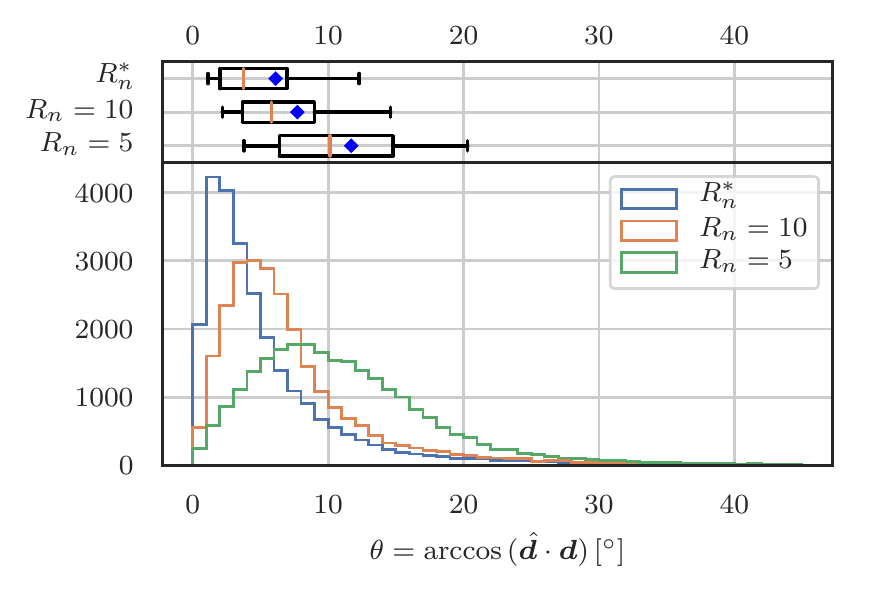}
    \caption{Reconstructed shower direction residual distribution. In the top box plot, the blue diamonds represents the means, the orange lines the medians, the boxes the IQRs and the the whiskers span from the $10^{\text{th}}$ to the $90^{\text{th}}$ percentiles.}
    \label{fig:shower_angle_resolution}
\end{figure}

\begin{figure}[htb!]
    \centering
    \includegraphics[width=\linewidth]{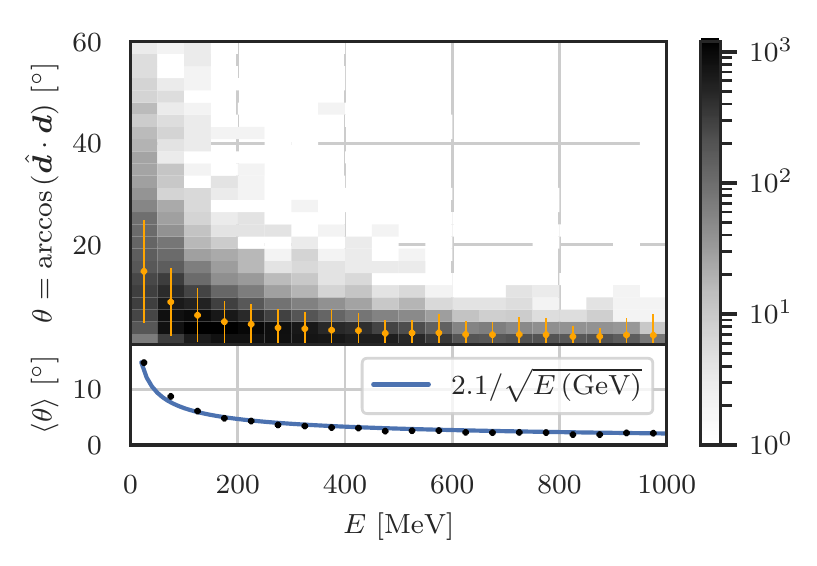}
    \caption{Reconstructed shower direction residual distribution as a function of the shower energy. The orange markers on the top pad represent the mean and their error bars the RMS; the former is shown on its own in the bottom pad.}
    \label{fig:shower_angle_resolution_vs_e}
\end{figure}

Figure\,\ref{fig:shower_angle_resolution} shows the angular distribution between the true direction and the estimate, $\theta=\arccos{(\hat{\bm{d}}\cdot\bm{d})}$, for all the true showers in the dataset. The residual angle distribution has a mode of $\sim2^{\circ}$, a mean of $6.1^{\circ}$ and a median of $3.8^{\circ}$, when using the optimized neighborhood, $R_n^*$. The distributions for fixed neighborhood radii perform significanlty worse. A radius of 5, while on average optimal for all fragments as shown in figure\,\ref{fig:dir_radius}, carries a large uncertainty when used for primary fragments. Figure\,\ref{fig:shower_angle_resolution_vs_e} shows the angular resolution as a function of the shower energy. It shows that the resolution improves siginificantly with energy to reach a mean as low as $\sim2.0^{\circ}$.

\subsection{Mistakes analysis}
\begin{figure*}[t]
    \centering
    \hfill{}
    \includegraphics[width=0.32\textwidth, trim=1cm 0cm 1cm 2cm, clip]{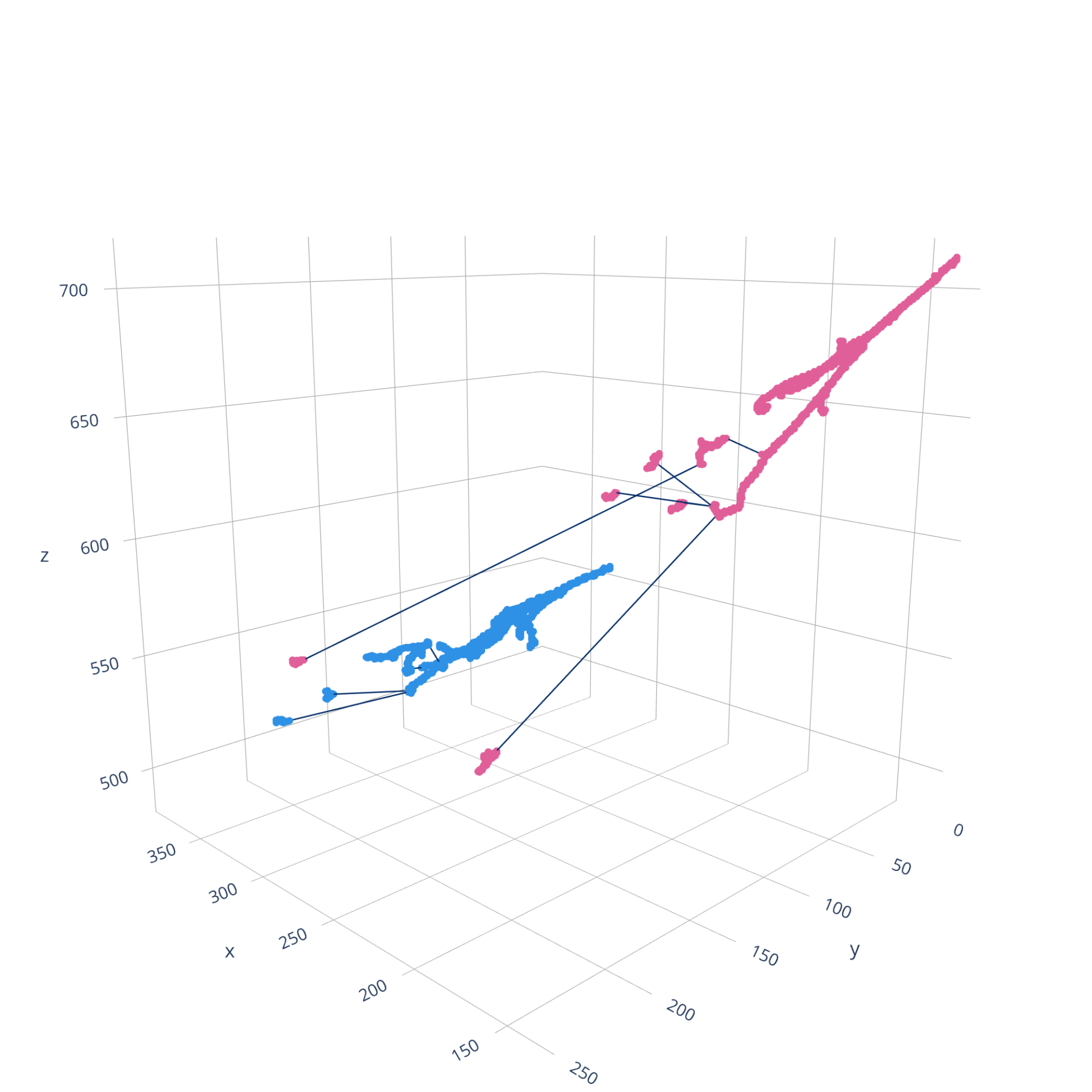}
    \hfill{}
    \includegraphics[width=0.32\textwidth, trim=1cm 0cm 1cm 2cm, clip]{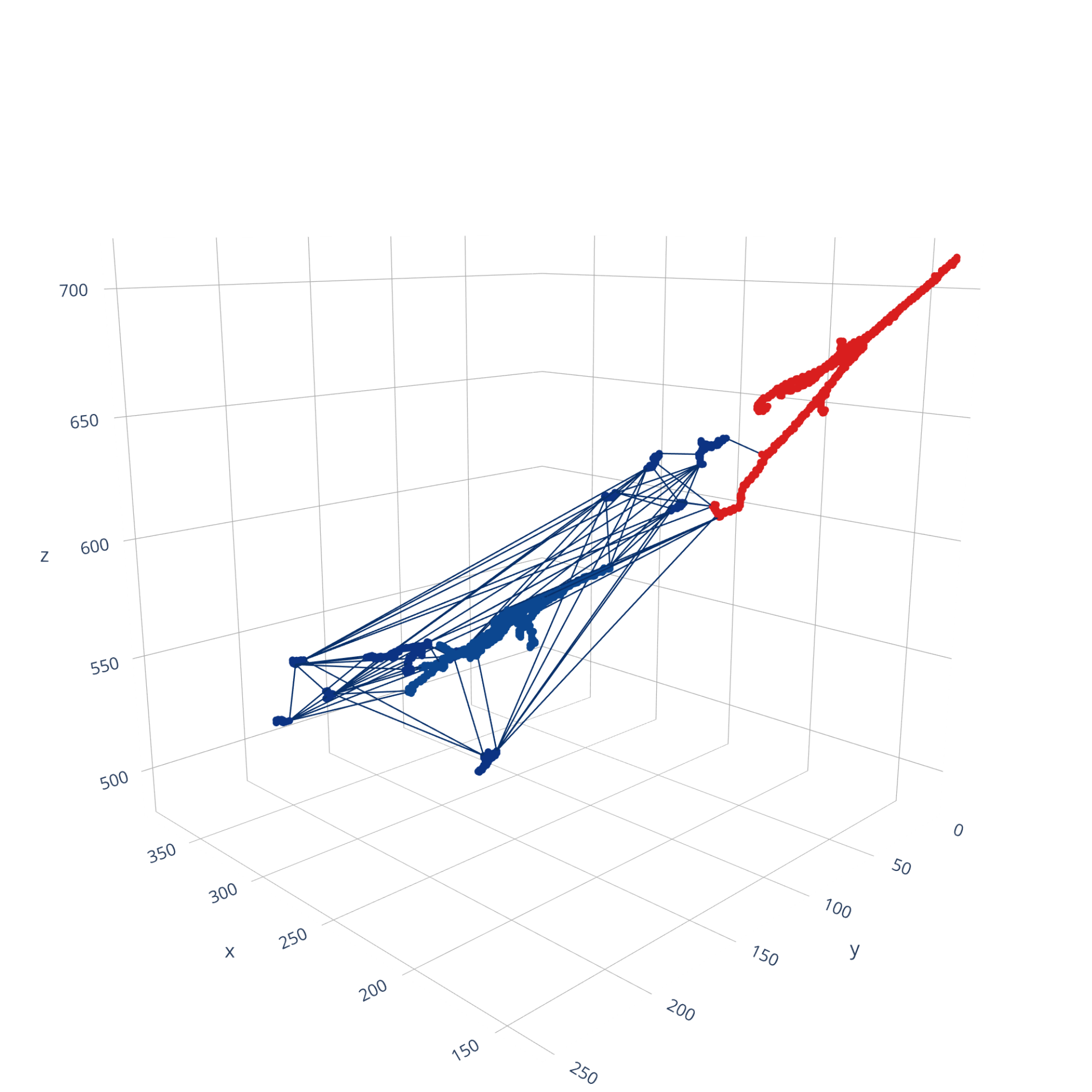}
    \hfill{}
    \begin{overpic}[width=0.32\textwidth, trim=1cm 0cm 1cm 2cm, clip]{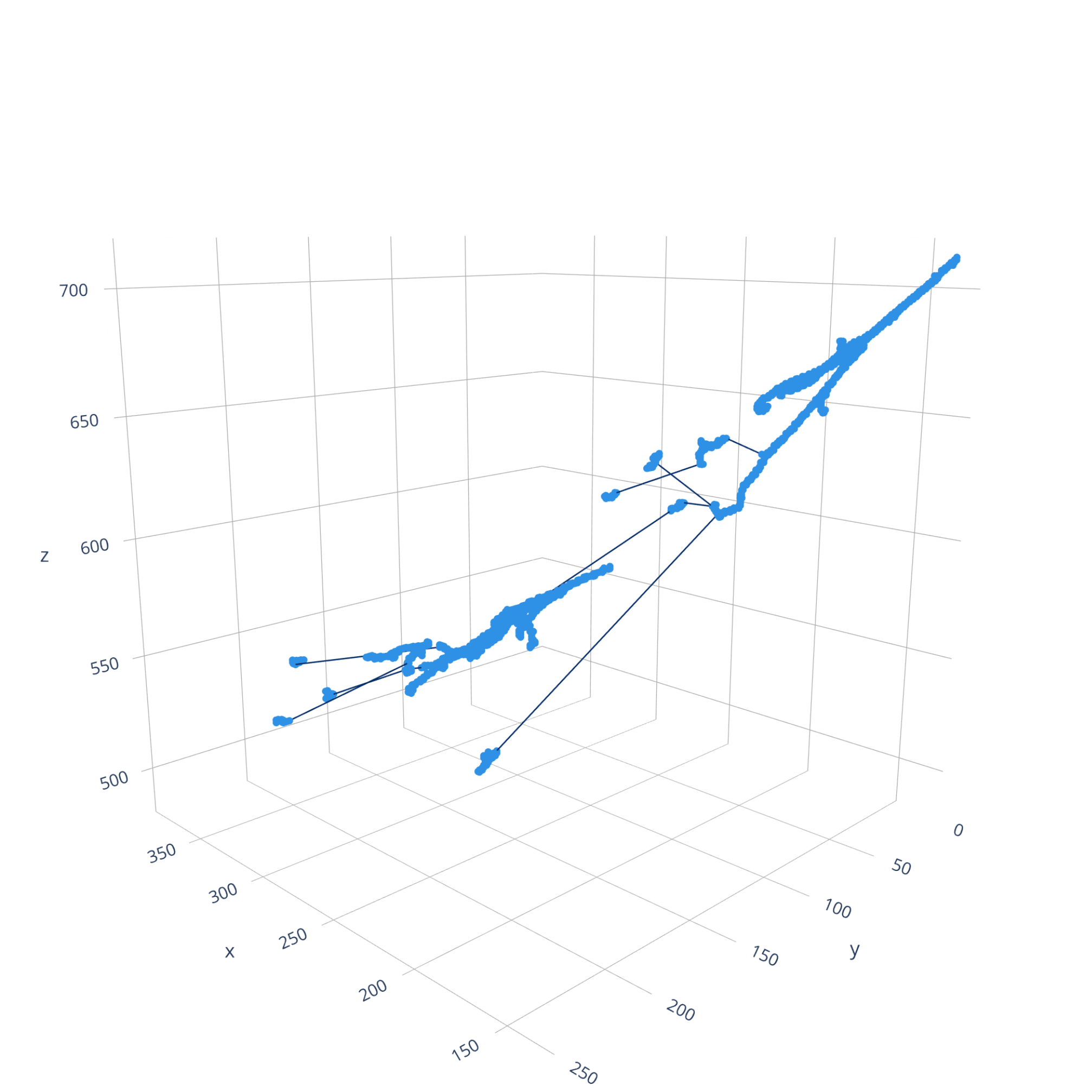}
        \put (35,15) {\fcolorbox{black}{white}{\footnotesize$\text{Pur.}:\,51.2\,\%$}}
    \end{overpic}
    \hfill{}
    
    \hfill{}
    \includegraphics[width=0.32\textwidth, trim=1cm 0cm 1cm 2cm, clip]{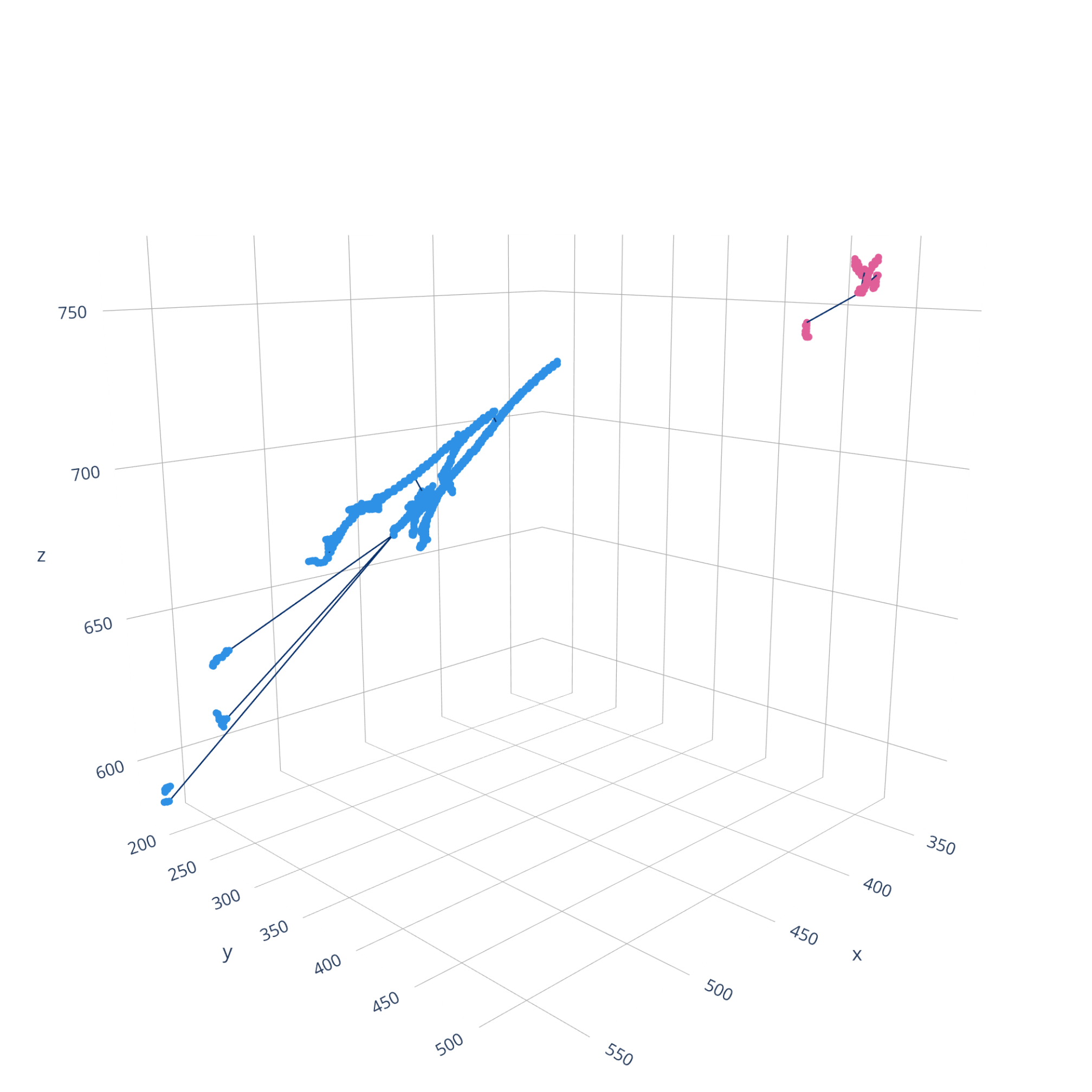}
    \hfill{}
    \includegraphics[width=0.32\textwidth, trim=1cm 0cm 1cm 2cm, clip]{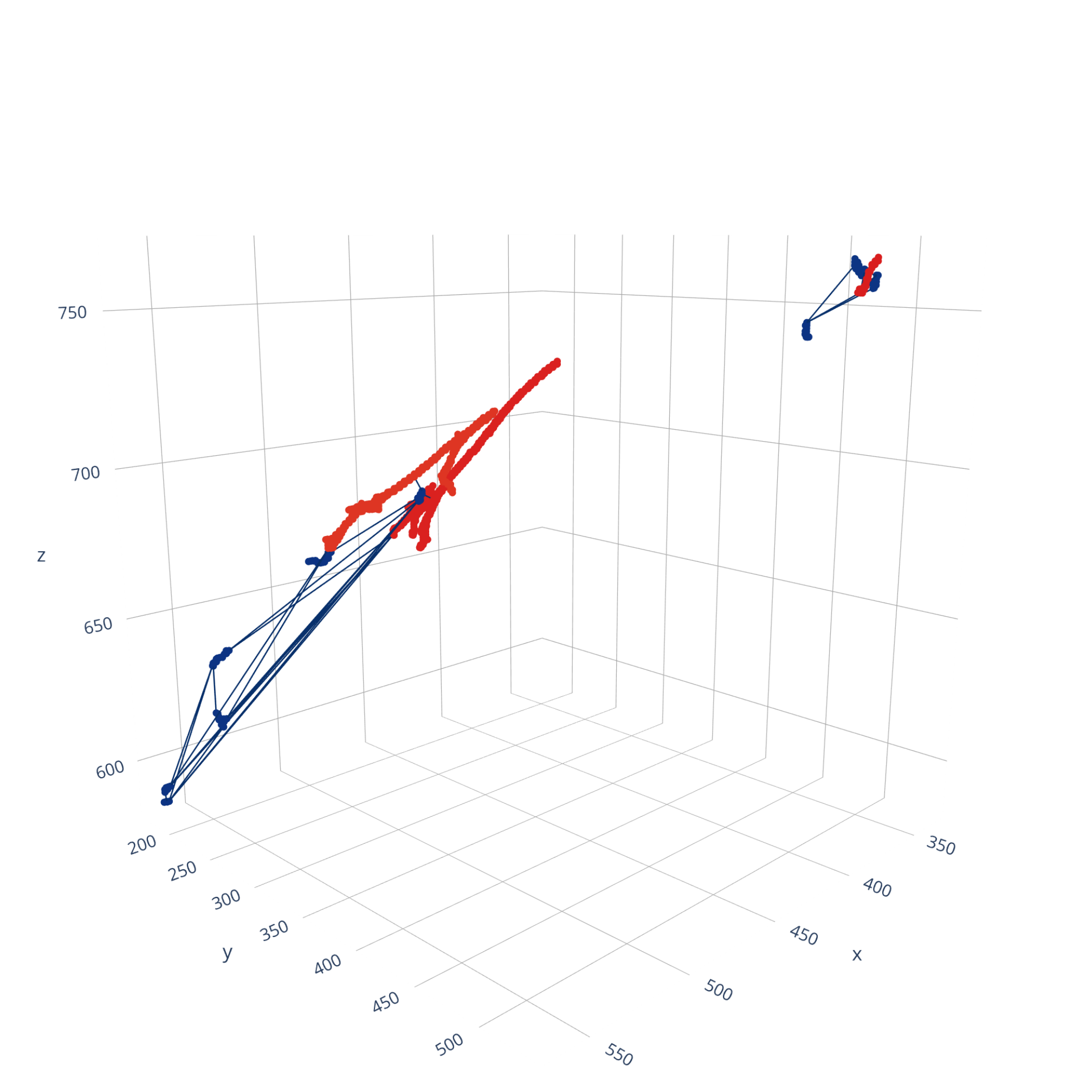}
    \hfill{}
    \begin{overpic}[width=0.32\textwidth, trim=1cm 0cm 1cm 2cm, clip]{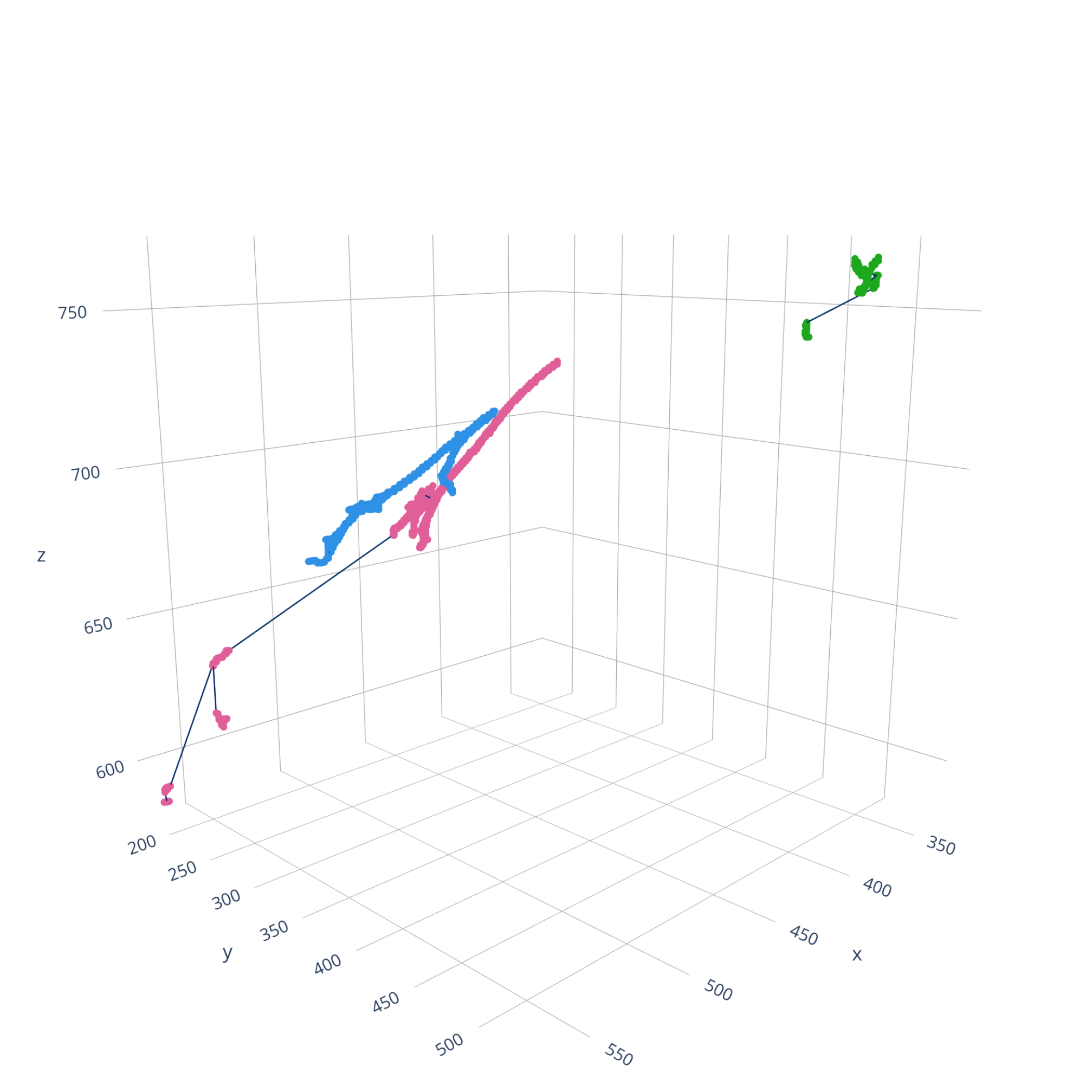}
        \put (35,15) {\fcolorbox{black}{white}{\footnotesize$\text{Eff.}:\,53.7\,\%$}}
    \end{overpic}
    \hfill{}
    
    \hfill{}
    \includegraphics[width=0.32\textwidth, trim=1cm 0cm 1cm 2cm, clip]{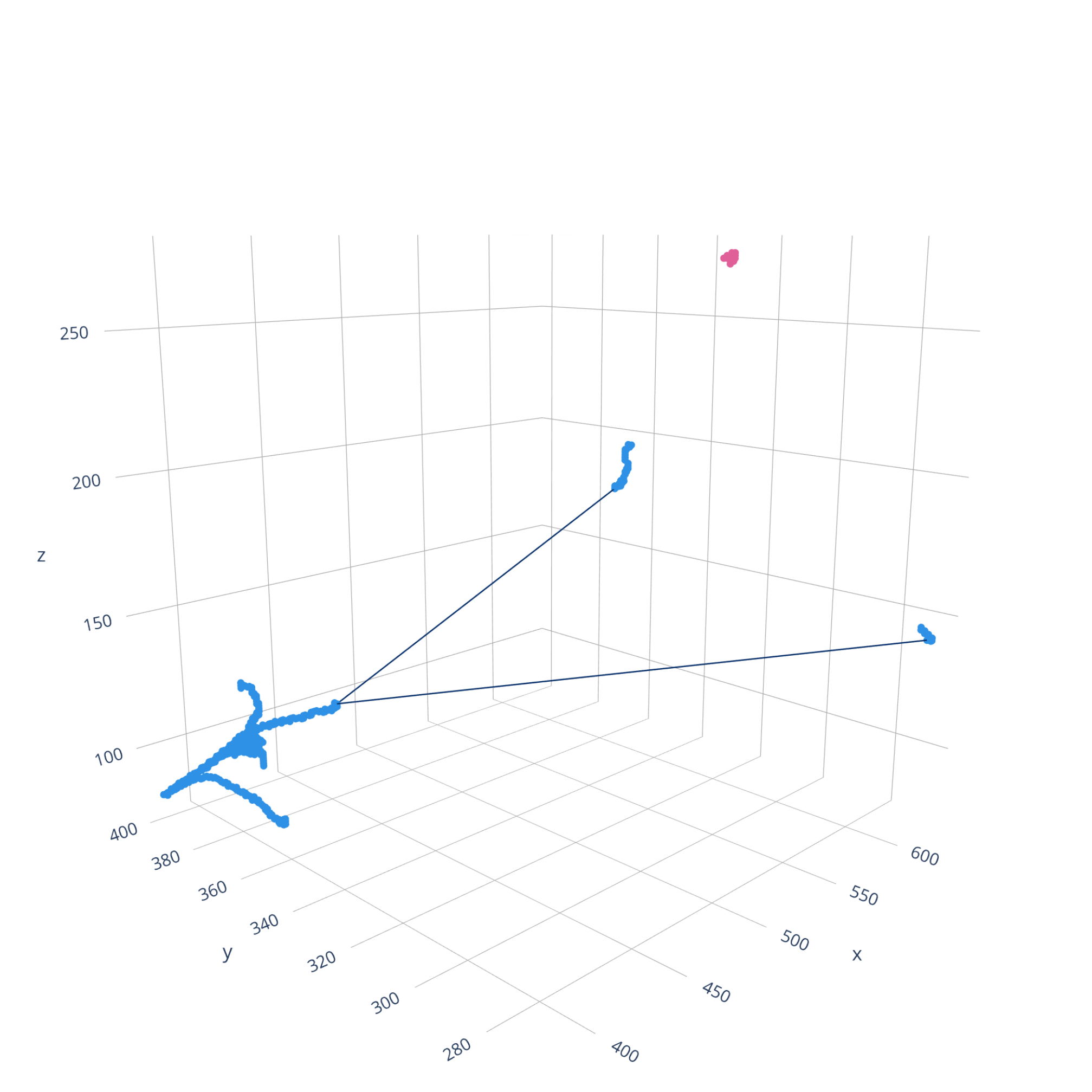}
    \hfill{}
    \includegraphics[width=0.32\textwidth, trim=1cm 0cm 1cm 2cm, clip]{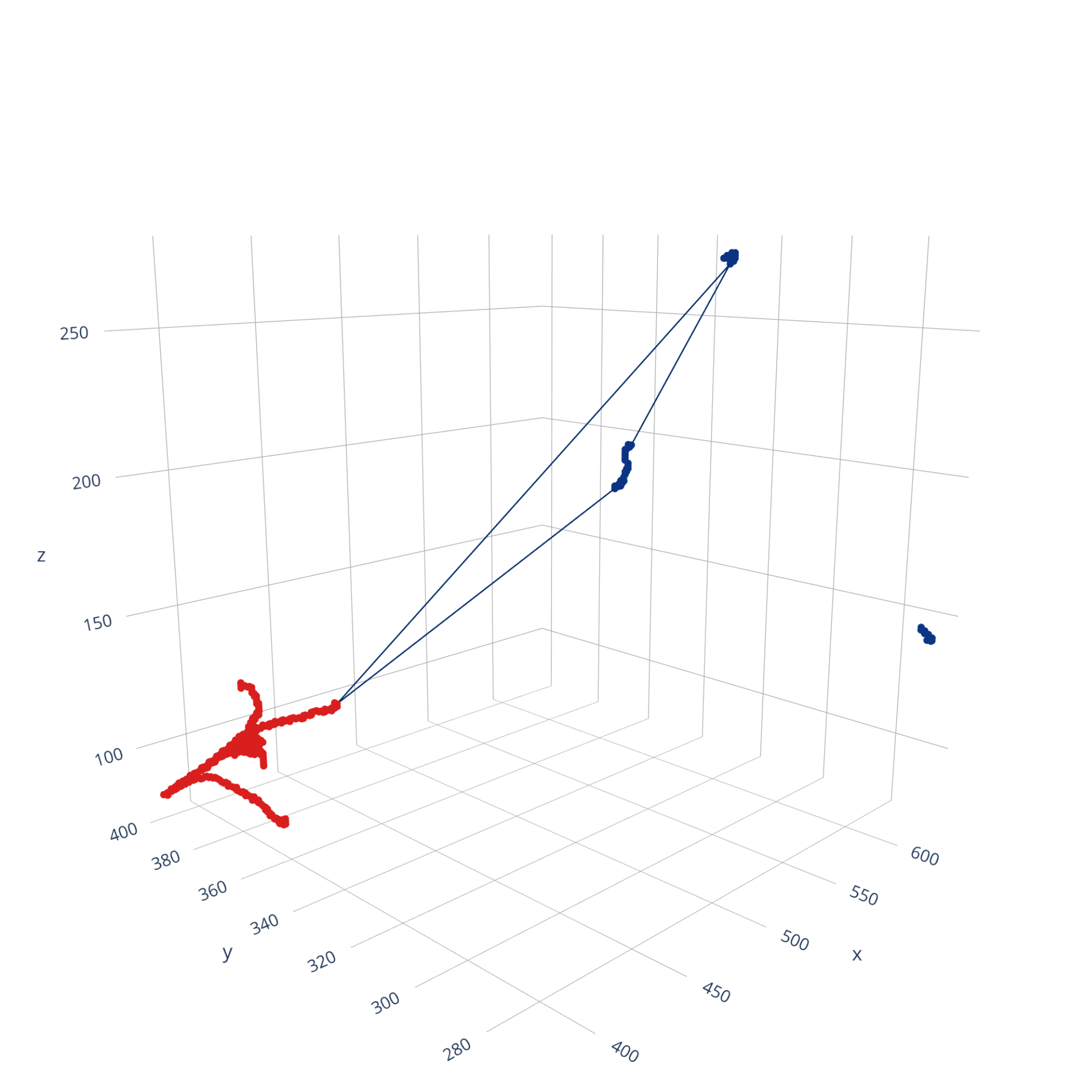}
    \hfill{}
    \begin{overpic}[width=0.32\textwidth, trim=1cm 0cm 1cm 2cm, clip]{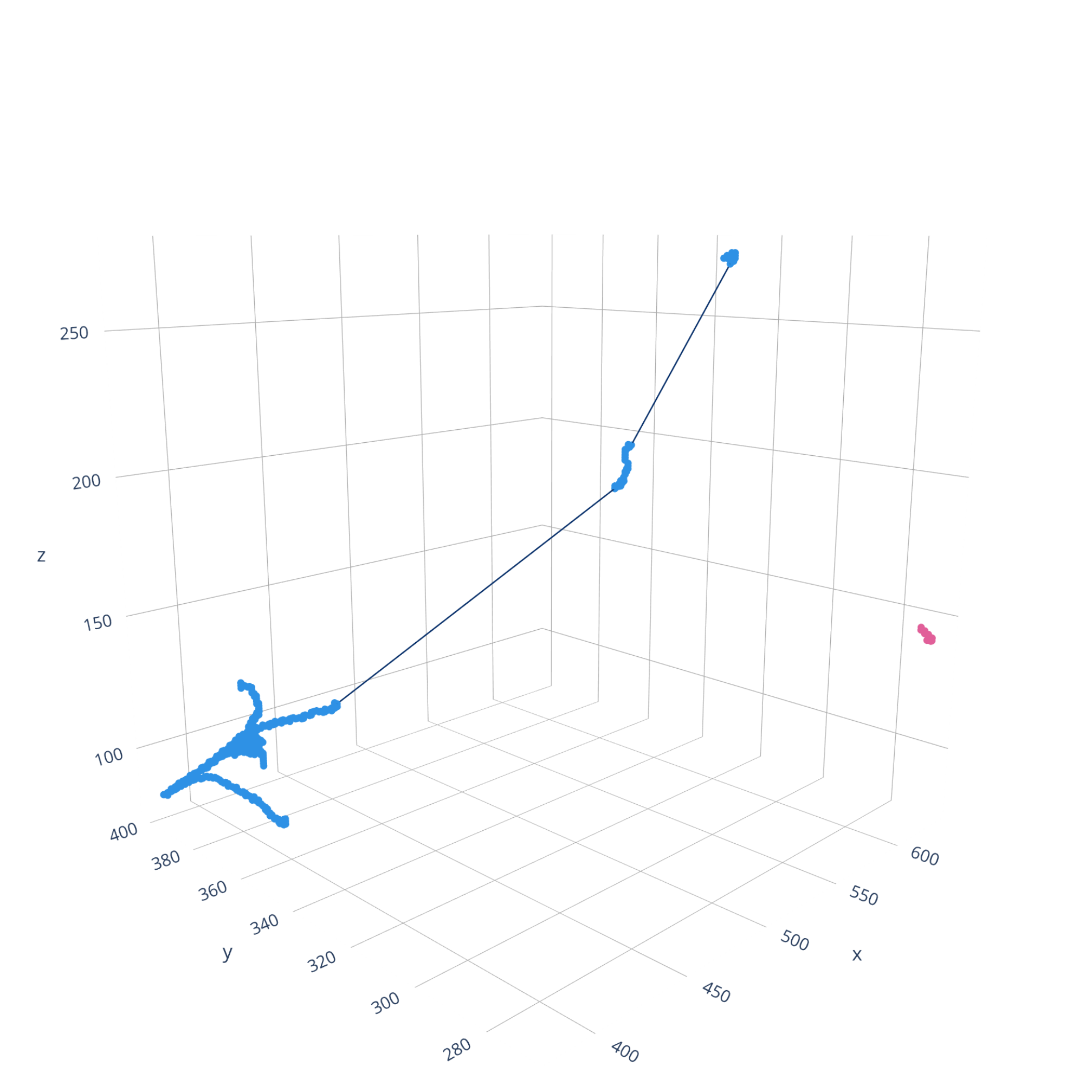}
        \put (35,15) {\fcolorbox{black}{white}{\footnotesize$\text{ARI}:\,-2.7\,\%$}}
    \end{overpic}
    \hfill{}
    
    \hfill{}
    \includegraphics[width=0.32\textwidth, trim=1cm 0cm 1cm 2cm, clip]{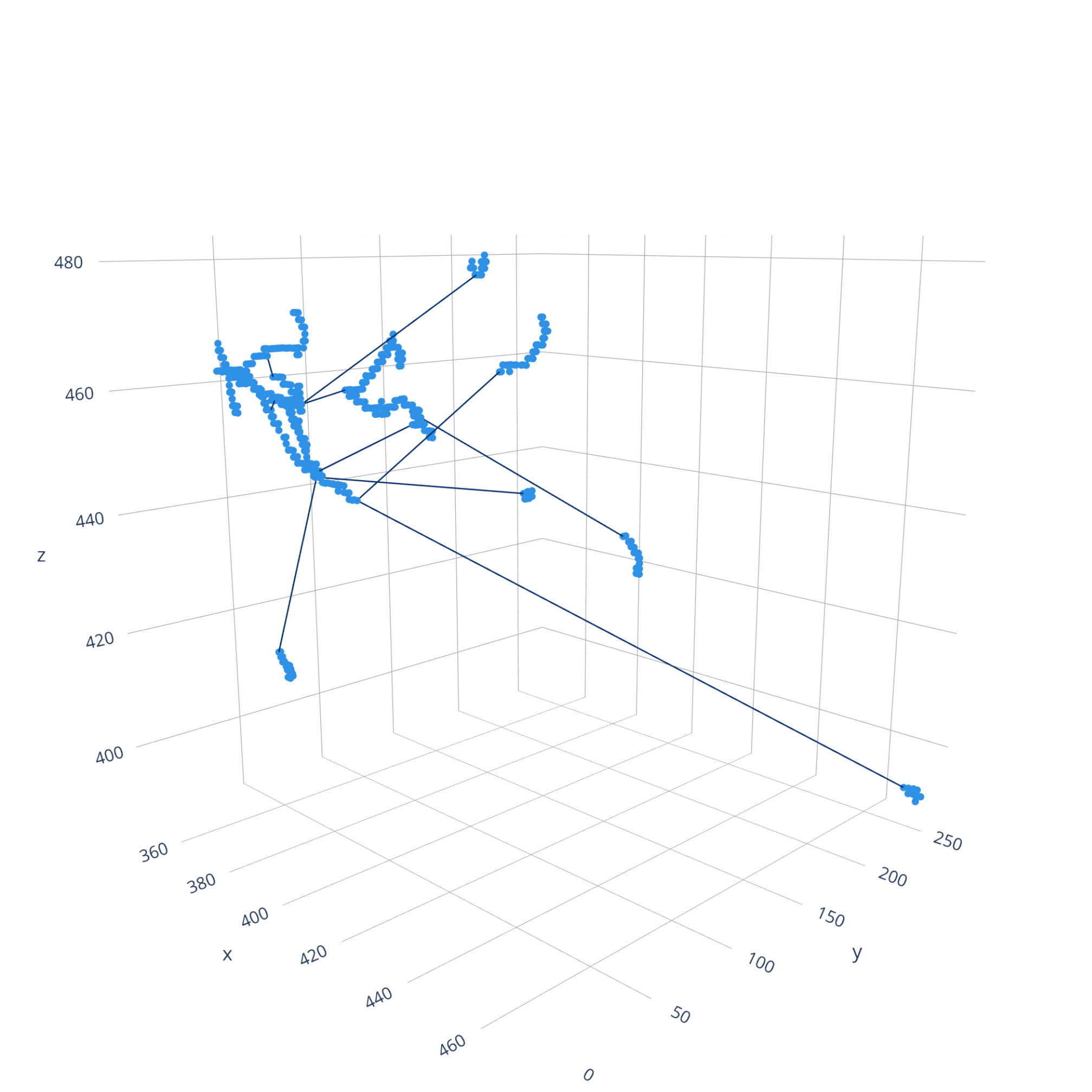}
    \hfill{}
    \includegraphics[width=0.32\textwidth, trim=1cm 0cm 1cm 2cm, clip]{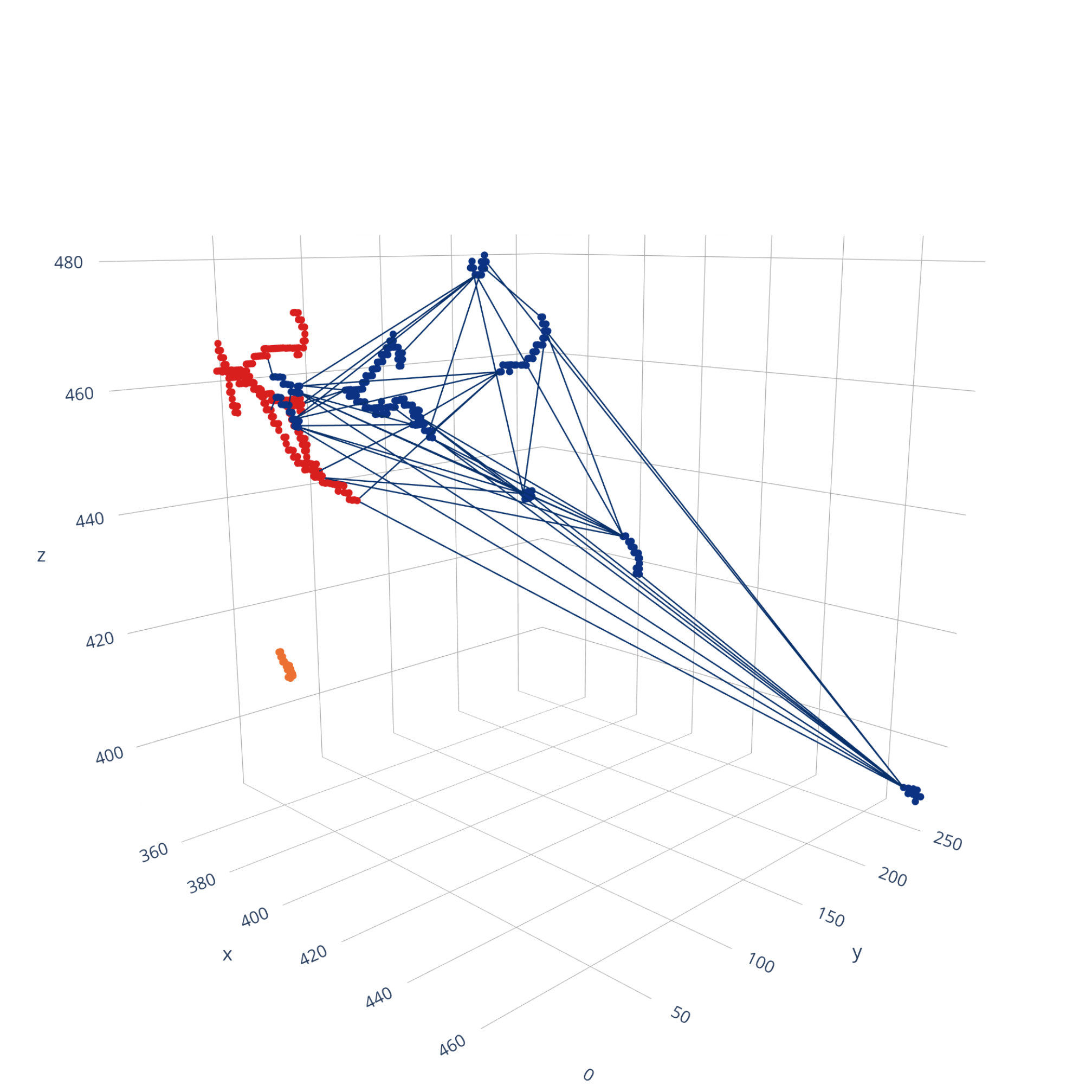}
    \hfill{}
    \begin{overpic}[width=0.32\textwidth, trim=1cm 0cm 1cm 2cm, clip]{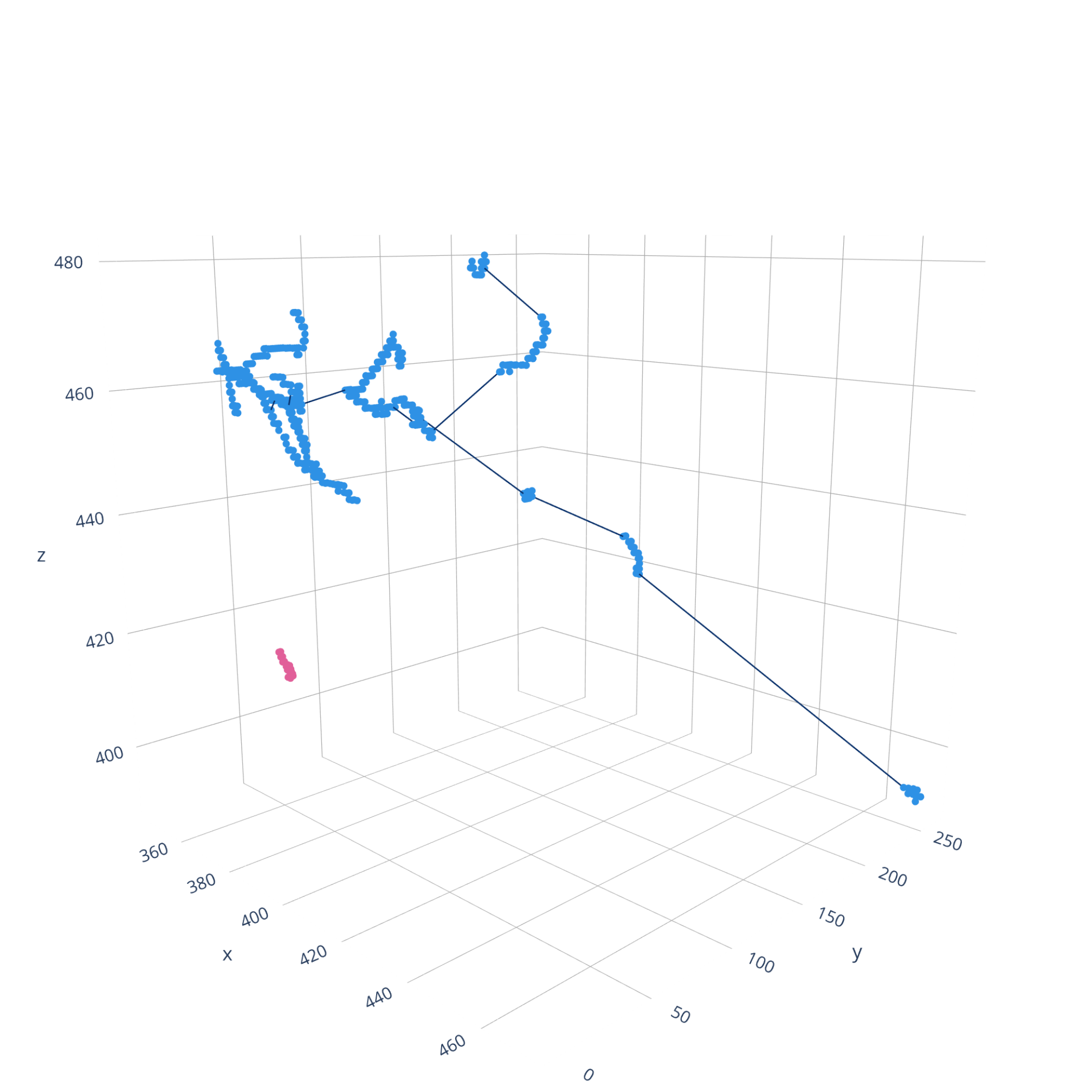}
        \put (35,15) {\fcolorbox{black}{white}{\footnotesize$\text{ARI}:\,0\,\%$}}
    \end{overpic}
    \hfill{}
    
    \caption{Shower clustering predictions with the largest mistakes in three categories and one with an ARI of 0 (one event per row). Left: ground-truth shower labels (color) and edges representing the true fragment parentage. Middle: primary node scores represented as a node color ranging from 0 (blue) to 1 (red) and edges with an adjacency score $>0.5$ (the closer to 1, the darker the edge). Right: inferred shower labels (color) and selected edges. }
    \label{fig:shower_clustering_mistakes}
\end{figure*}

A study of the events with low clustering purity reveals that the algorithm occasionally merges showers when the direction vector of one of its fragments can clearly be back-propagated to another shower fragment from a distinct instance. This may stem from an inconsistency between the true photon momentum and the local direction estimate of the fragment. The top row of figure~\ref{fig:shower_clustering_mistakes} shows an event with a purity of $0.51$ in the test set. Events with a purity $<0.9$ represent $\sim1$\,\% of this dataset. The middle row shows an event with an efficiency of $0.54$ in test set. Events with an efficiency $<0.9$ represent $\sim0.7$\,\% of this dataset. This event showcases the difficulty to choose whether to merge large colinear fragments or to separate them. The third row shows the event with the lowest ARI and the fourth with an ARI of 0. The last example highlights that a clustering may have an ARI of 0 but be mostly accurate, if the number of true or predicted showers is one.

\begin{figure}[htb!]
    \centering
    \includegraphics[width=.9\linewidth]{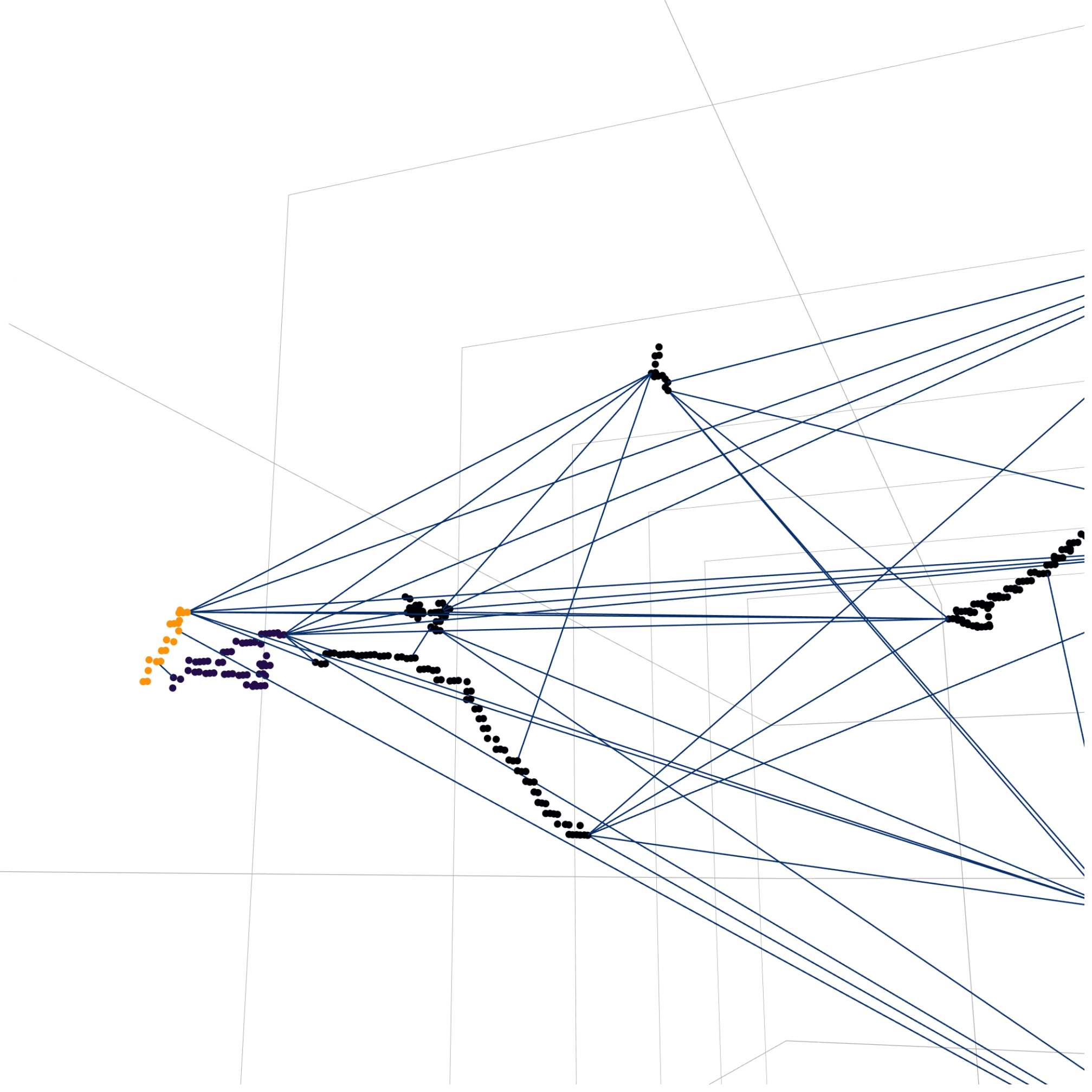}
    \caption{Primary scores for an ambiguous shower start with a color scale that ranges from 0 (black) to bright yellow (1).}
    \label{fig:primary_mistake}
\end{figure}

There is a total of 119 showers in the whole test set that have a misidentified primary. The majority of those mistakes stem from an incorrect partition of the nodes. The remaining mistakes can be attributed to ambiguous shower starts that do not have a fragment clearly upstream of the others. An example is provided in figure~\ref{fig:primary_mistake}. The network picks the leftmost fragment but the one directly to its right is the labeled shower start. The network shows its uncertainty by giving scores of 0.76 and 0.13 to the left and right fragments, respectively.

\section{Interaction clustering\label{sec:inter_clustering}}

\subsection{Modifications}
Interaction clustering is defined as the association of particle instances together into groups that share a common particle ancestor. In this study, the individual particle instances are assumed to be known a priori from the previous reconstruction steps. In the full reconstruction chain, the shower instances will be provided by the reconstruction algorithm described in the previous sections while tracks, Michel and Deltas are to be clustered by a separate algorithm such as DBSCAN~\cite{dbscan} or a CNN-based dense clustering algorithm~\cite{cnn_clustering}.

\begin{figure}[htb!]
    \centering
    \includegraphics[width=\linewidth, trim=1cm 0cm 1cm 3.5cm, clip]{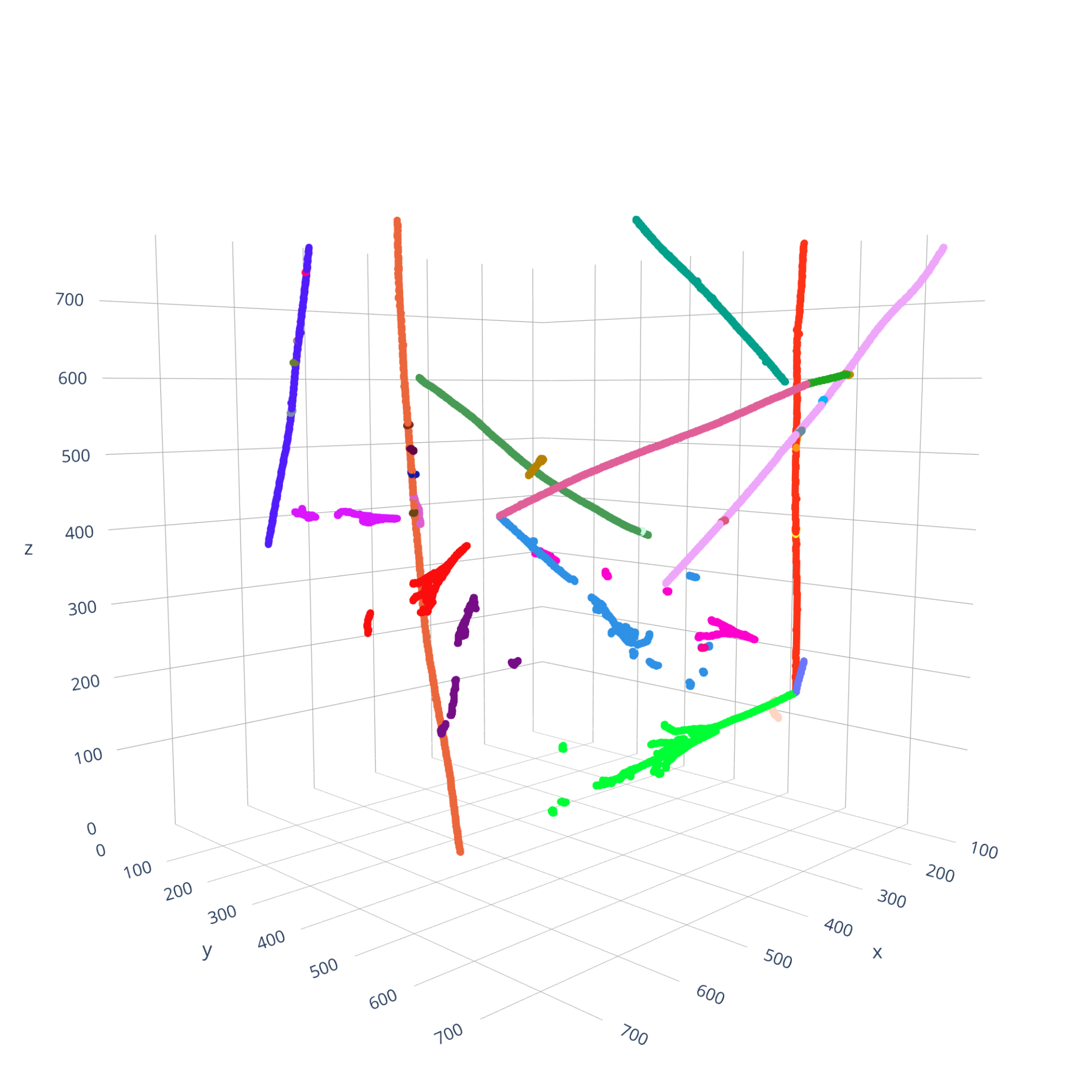}
    \caption{Particle labels of a superimposition of two events in the test set.}
    \label{fig:inter_input}
\end{figure}

Images simulated for this dataset only contain a single interaction vertex and multiple stray tracks and showers. In order to teach the network to separate multiple interaction vertices, several of these images may be stacked together. At the training stage, the number of images that are stacked together before being fed to the network follows a Poisson distribution of mean 2. Figure~\ref{fig:inter_input} shows an example of two stacked images and their particle labels.

This task utilizes an identical reconstruction chain to that used for the shower clustering. The input to the chain consists of particle instances instead of shower fragments and the target is interaction instances. The edge features are identically defined while node features are extended by adding the number-encoded particle class (0--4 corresponding to shower, track, Michel and delta rays, respectively), the mean and RMS energy deposition and the terminal point of tracks (other classes do not have well defined end points and are given a duplicate of the initial point instead).

\begin{figure}[htb!]
    \centering
    \includegraphics[width=\linewidth]{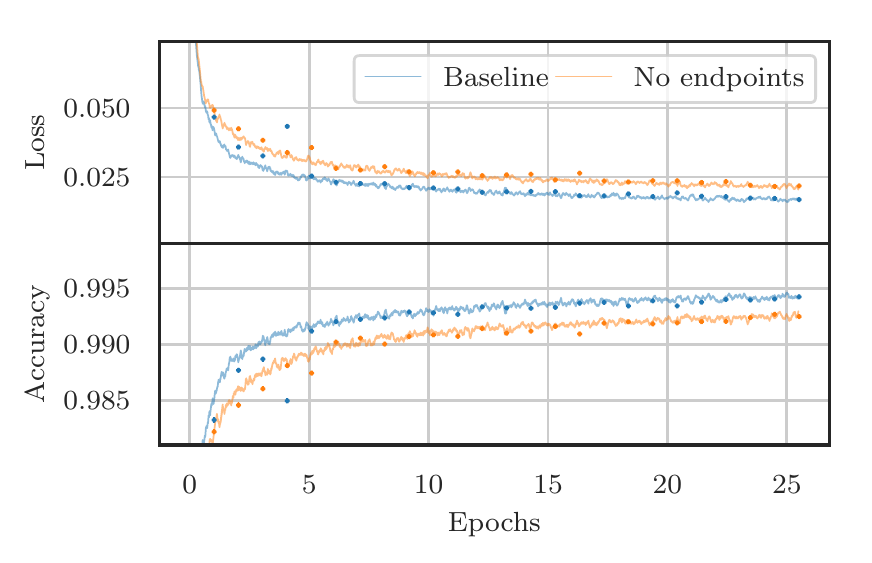}
    \caption{Edge score loss and edge prediction accuracy for the different interaction clustering models. The training curves are represented as lines and the validation points as round markers with an error bar.}
    \label{fig:inter_train}
\end{figure}

Downstream of the message passing stage, the updated node features are not explicitly used to make any prediction. The edge features are the basis for an adjacency matrix prediction, while the groups are extracted by using the method described in section~\ref{sec:inference}. Figure~\ref{fig:inter_train} shows the training and validation edge classification loss and accuracy for the interaction clustering task. These metrics are evaluated with and without the end point information; adding the endpoints to the particle instances increases the edge classification accuracy by $\sim0.2\,$\%.

\subsection{Performance}
\begin{figure*}[t]
    \centering
    \hfill{}
    \includegraphics[width=0.32\textwidth, trim=1cm 0cm 1cm 2cm, clip]{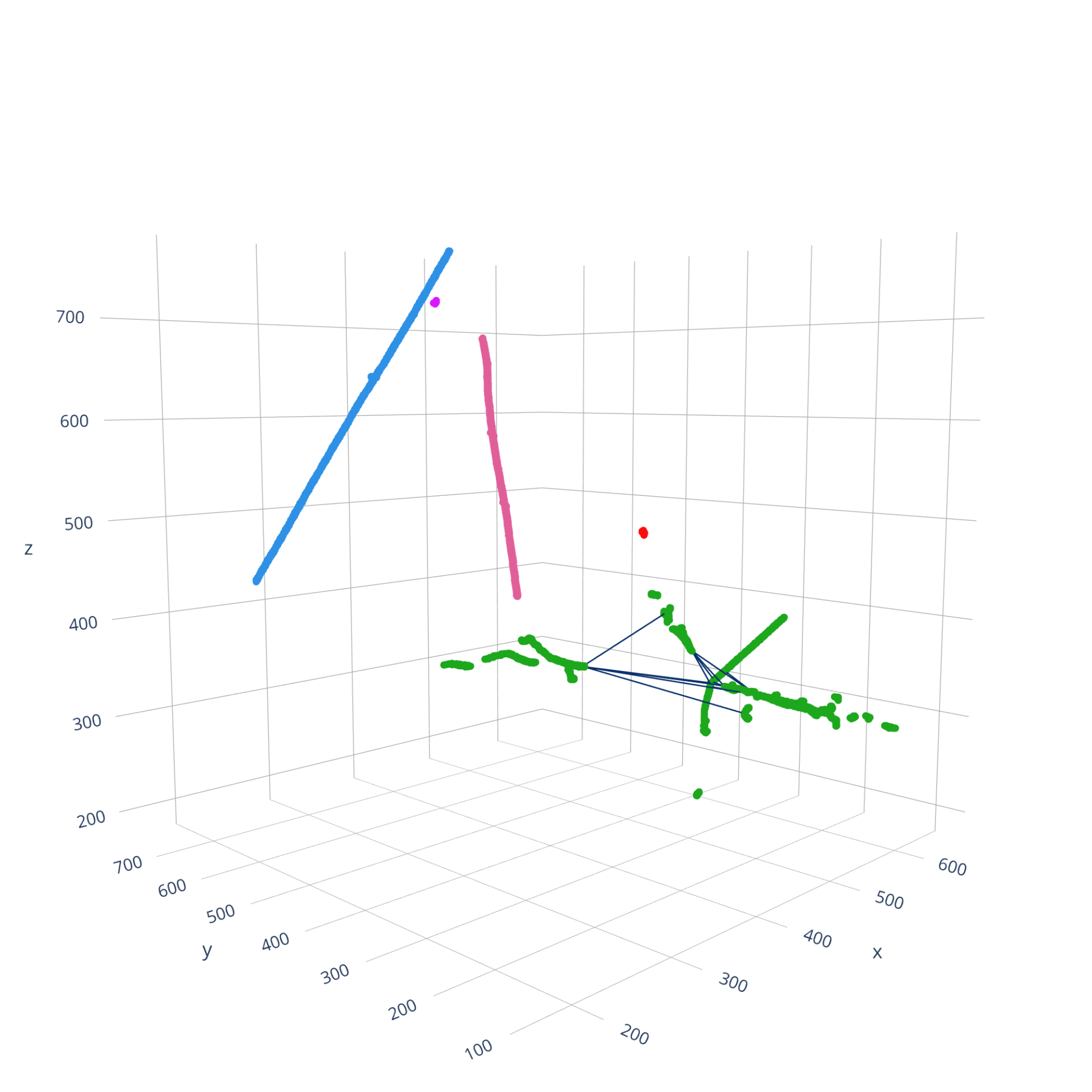}
    \hfill{}
    \includegraphics[width=0.32\textwidth, trim=1cm 0cm 1cm 2cm, clip]{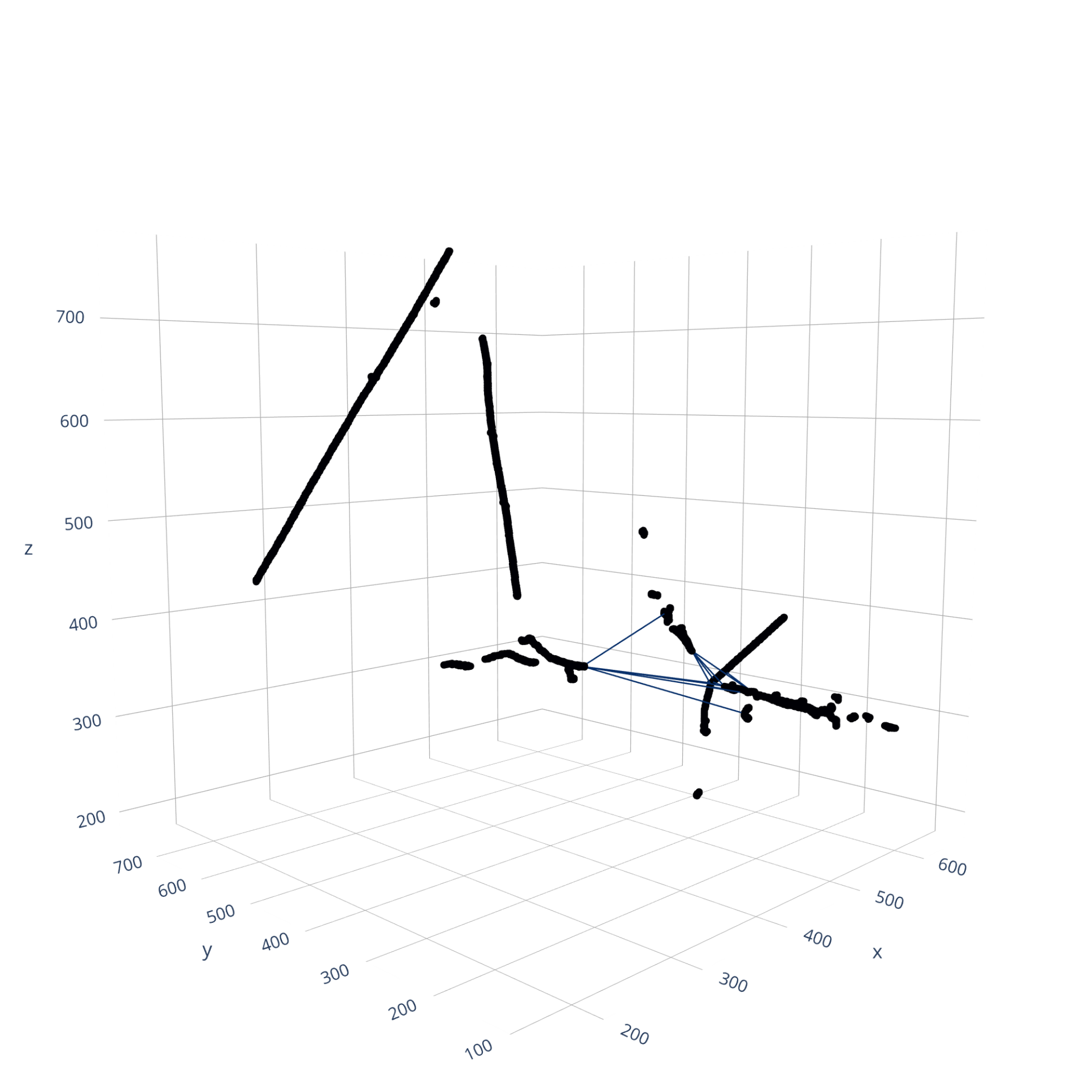}
    \hfill{}
    \begin{overpic}[width=0.32\textwidth, trim=1cm 0cm 1cm 2cm, clip]{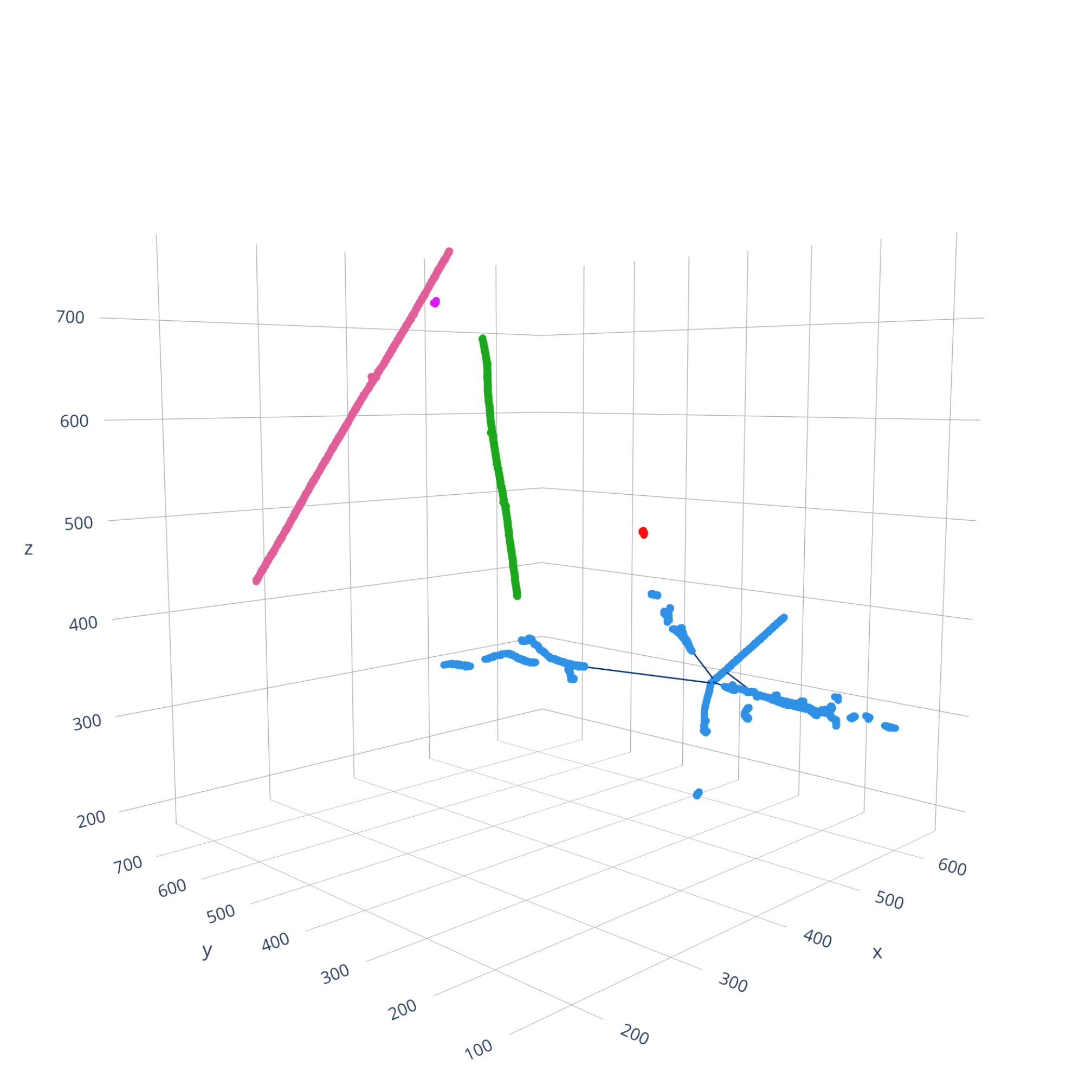}
        \put (35,15) {\fcolorbox{black}{white}{\footnotesize$\text{ARI}:\,100\,\%$}}
    \end{overpic}
    \hfill{}
    
    \hfill{}
    \includegraphics[width=0.32\textwidth, trim=1cm 0cm 1cm 2cm, clip]{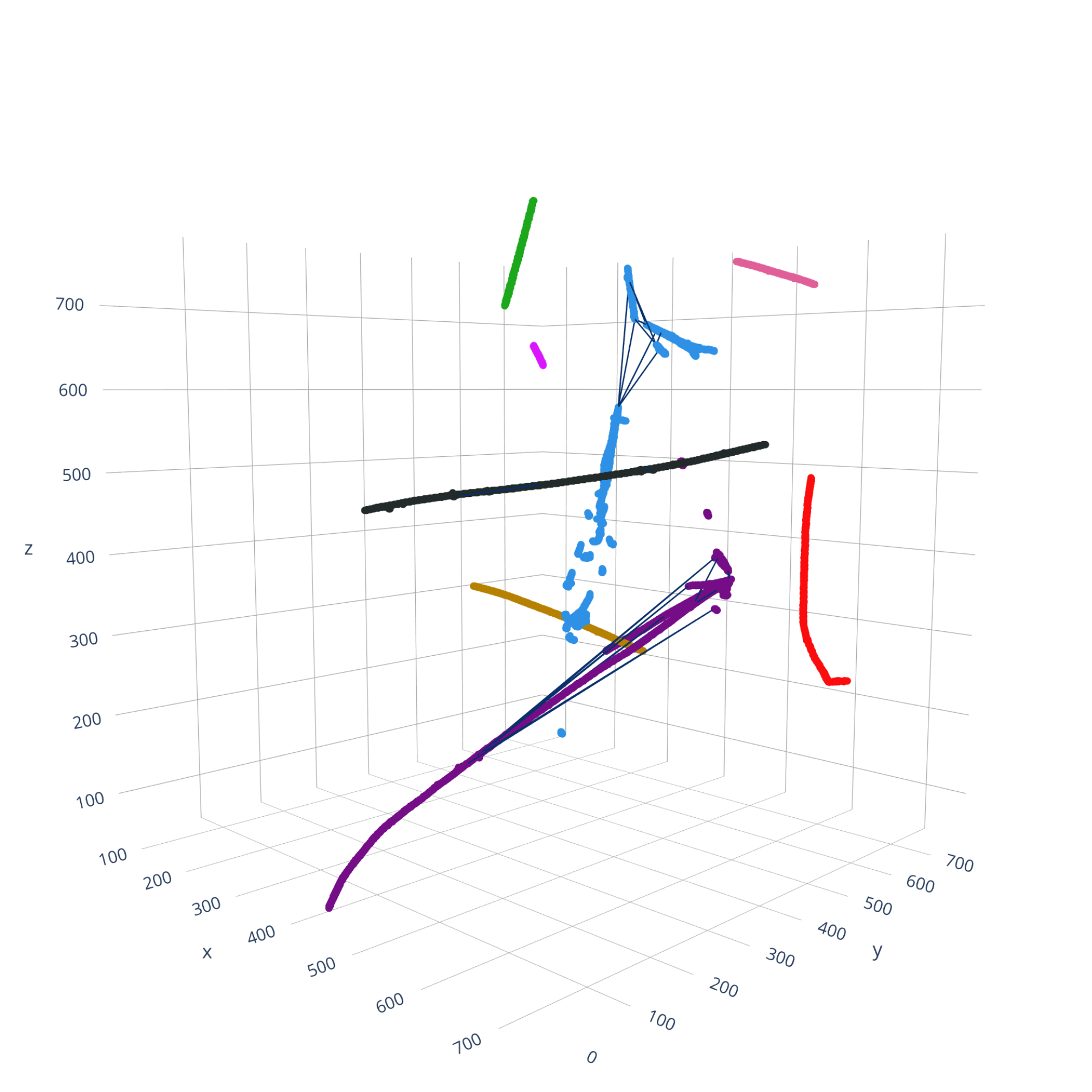}
    \hfill{}
    \includegraphics[width=0.32\textwidth, trim=1cm 0cm 1cm 2cm, clip]{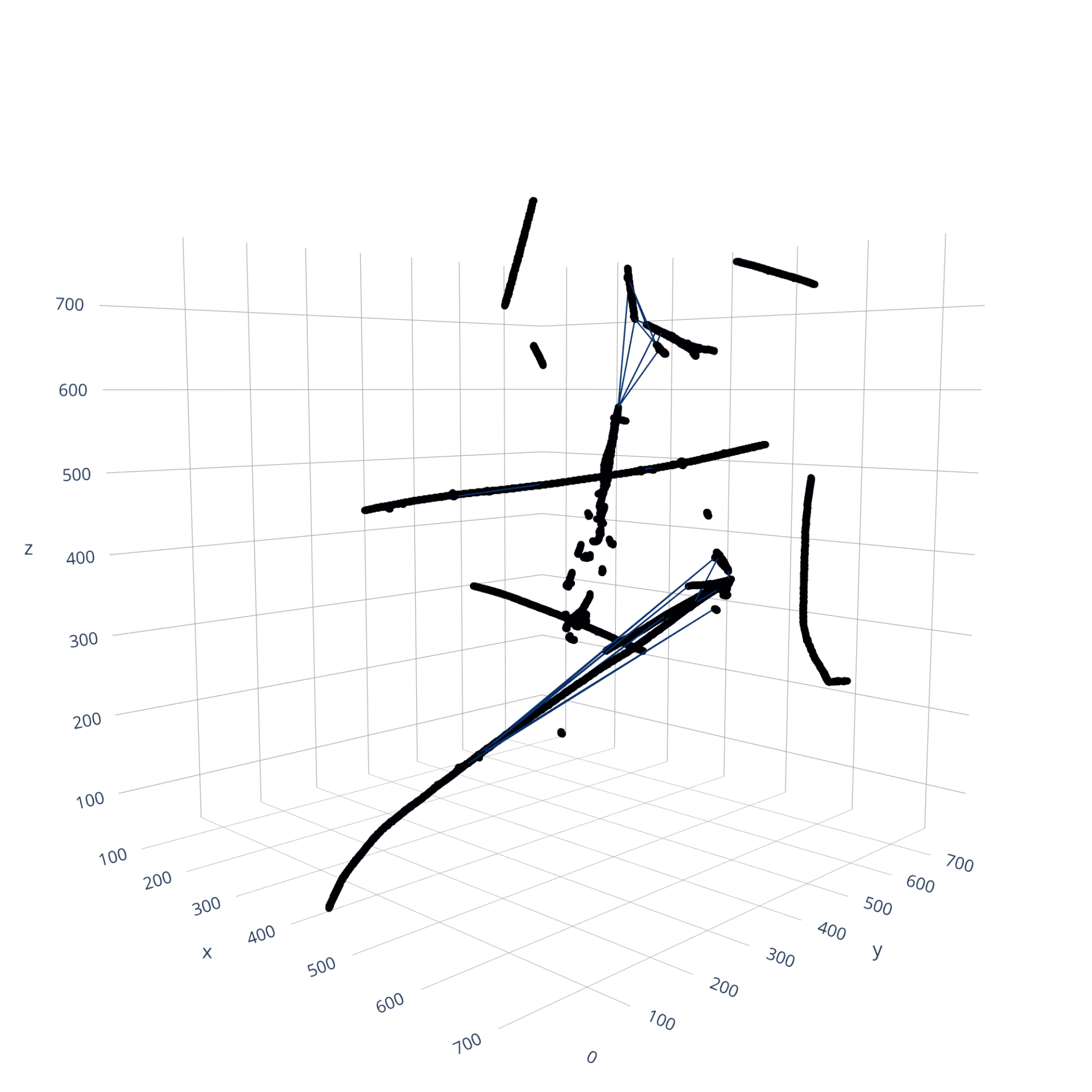}
    \hfill{}
    \begin{overpic}[width=0.32\textwidth, trim=1cm 0cm 1cm 2cm, clip]{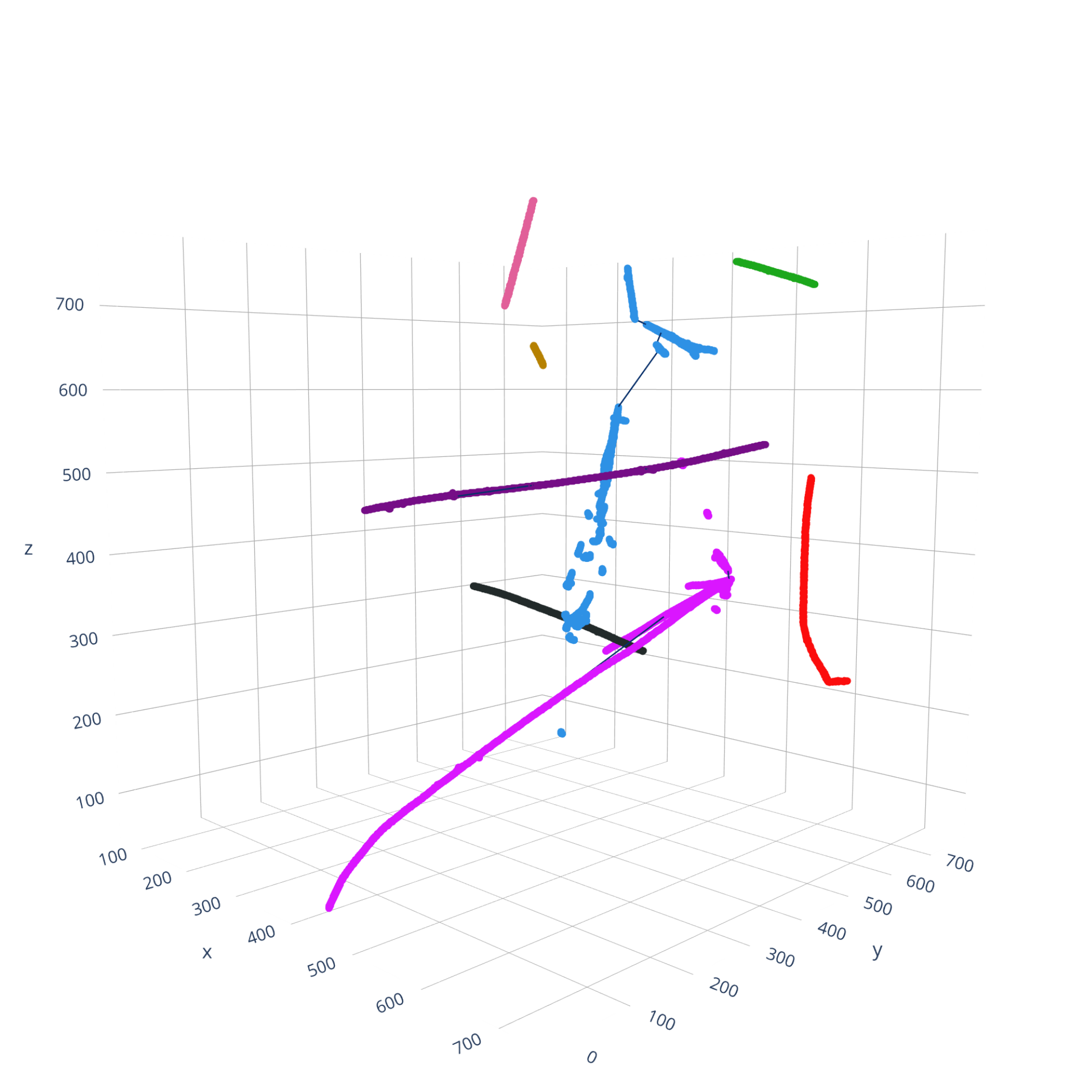}
        \put (35,15) {\fcolorbox{black}{white}{\footnotesize$\text{ARI}:\,100\,\%$}}
    \end{overpic}
    \hfill{}
    
    \hfill{}
    \includegraphics[width=0.32\textwidth, trim=1cm 0cm 1cm 2cm, clip]{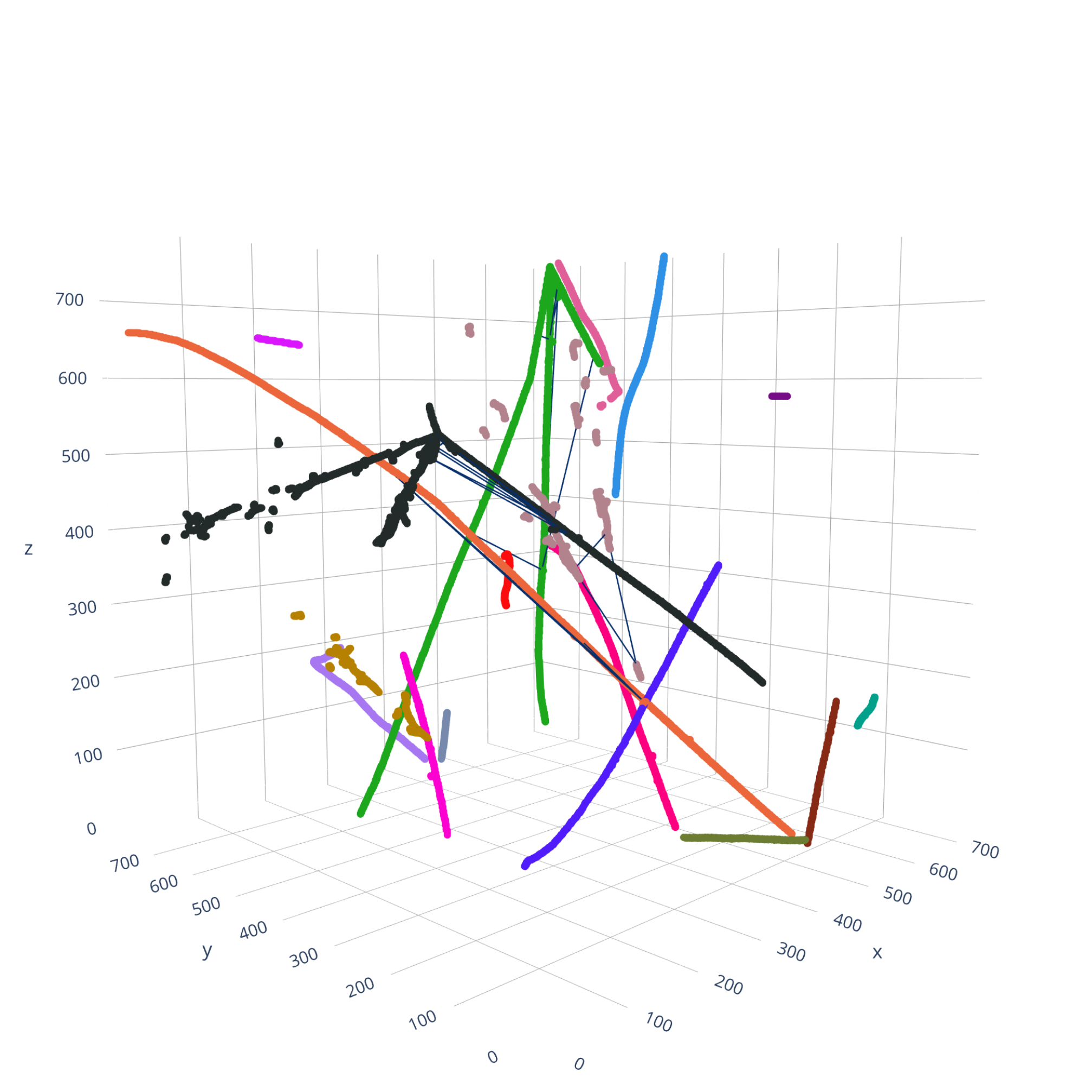}
    \hfill{}
    \includegraphics[width=0.32\textwidth, trim=1cm 0cm 1cm 2cm, clip]{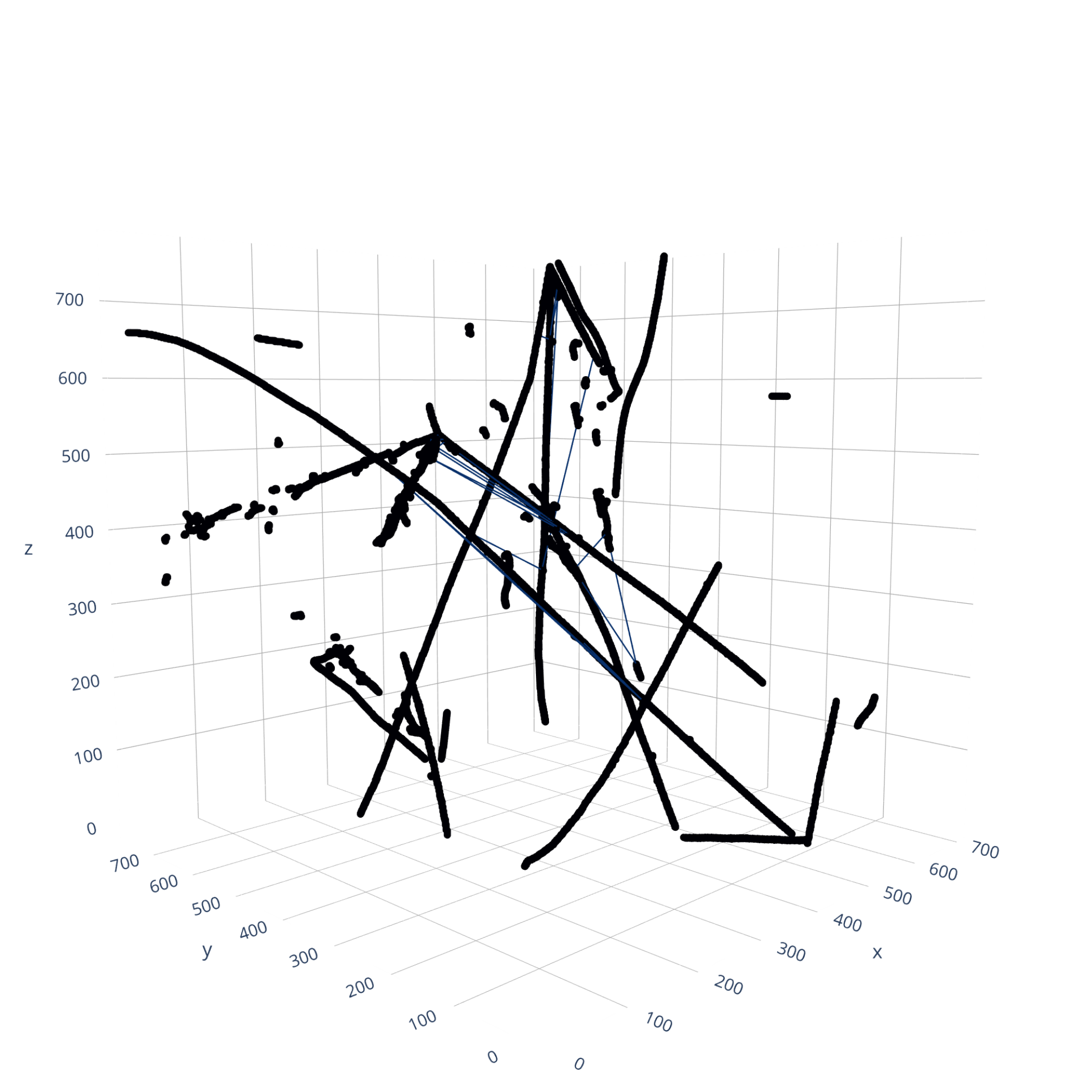}
    \hfill{}
    \begin{overpic}[width=0.32\textwidth, trim=1cm 0cm 1cm 2cm, clip]{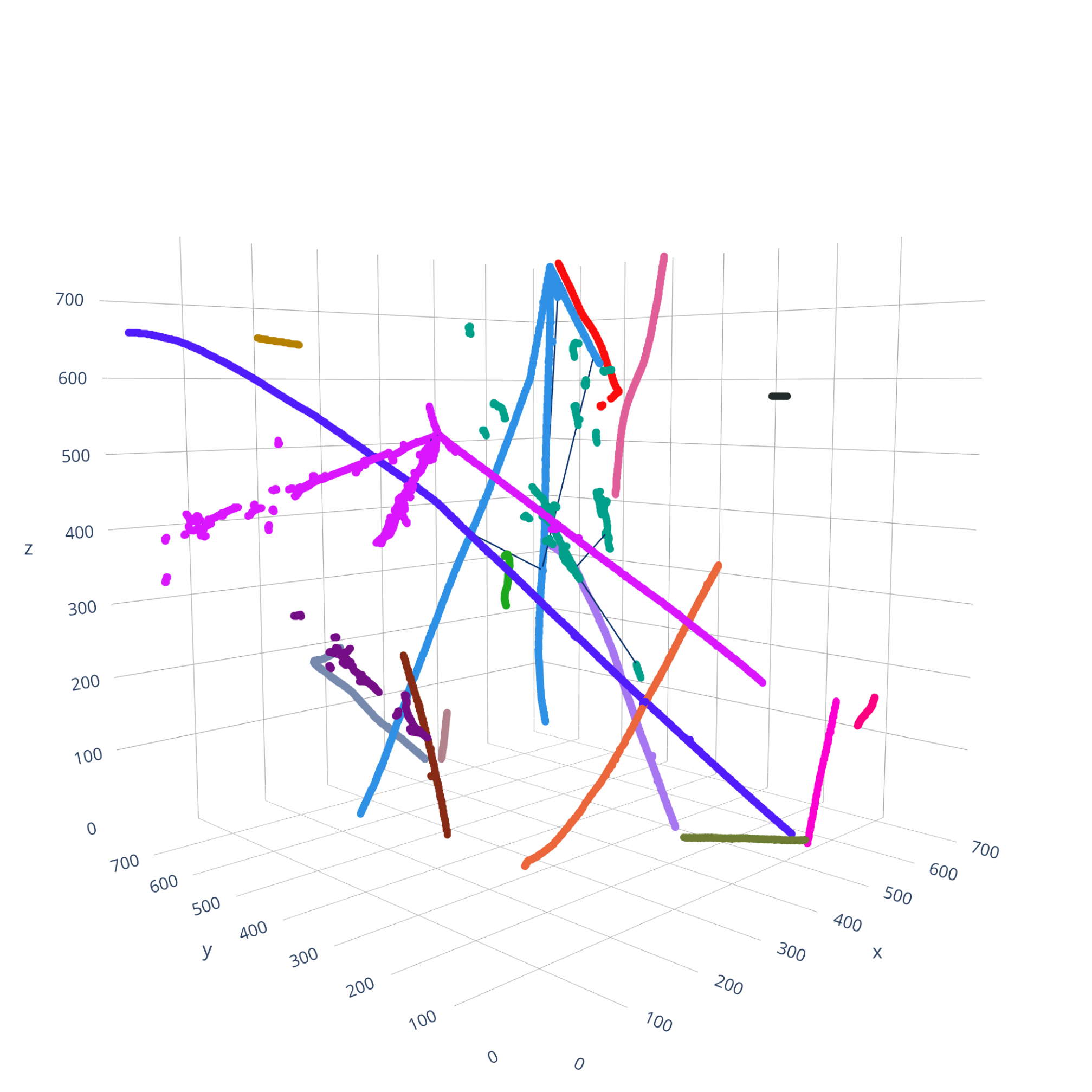}
        \put (35,15) {\fcolorbox{black}{white}{\footnotesize$\text{ARI}:\,100\,\%$}}
    \end{overpic}
    \hfill{}
    
    \hfill{}
    \includegraphics[width=0.32\textwidth, trim=1cm 0cm 1cm 2cm, clip]{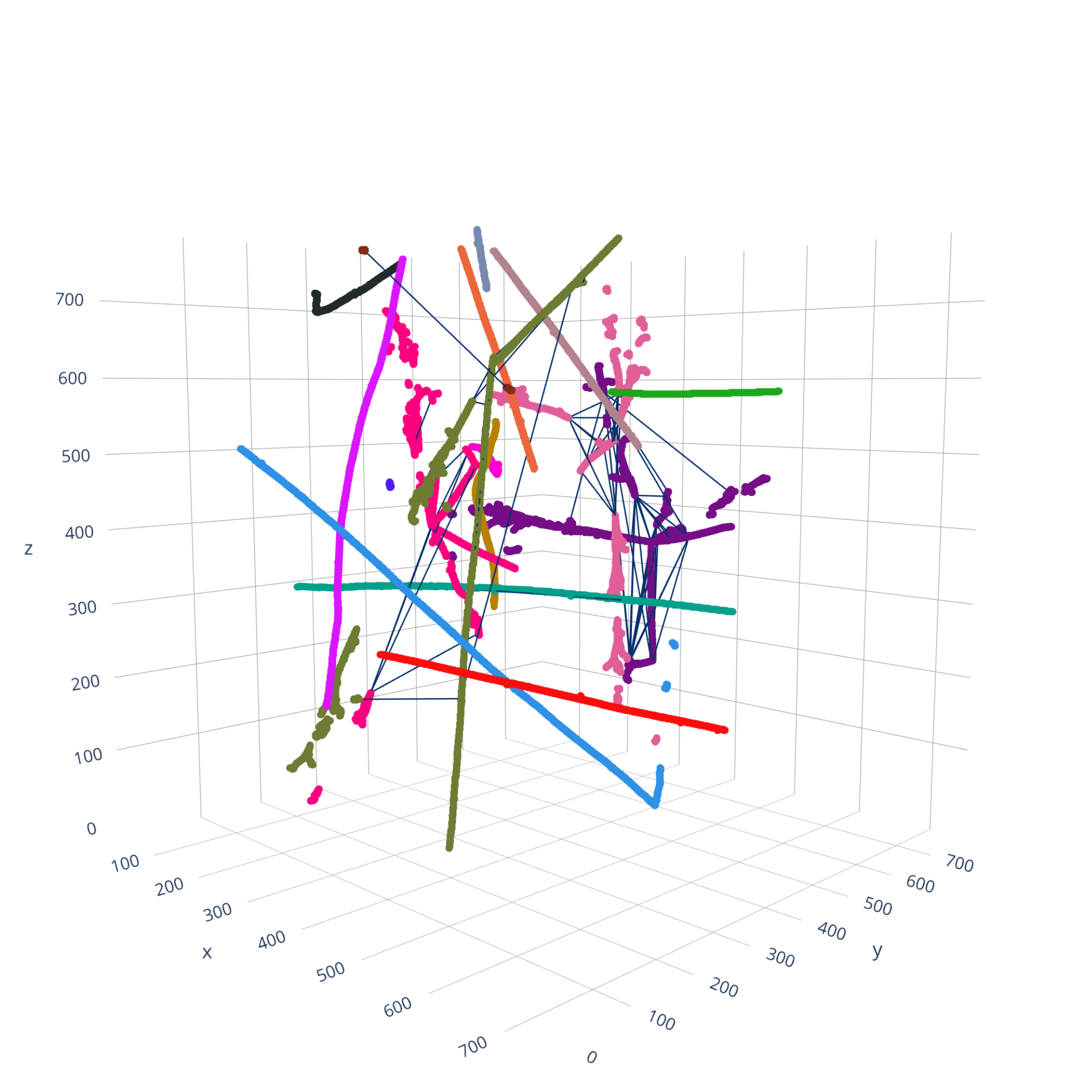}
    \hfill{}
    \includegraphics[width=0.32\textwidth, trim=1cm 0cm 1cm 2cm, clip]{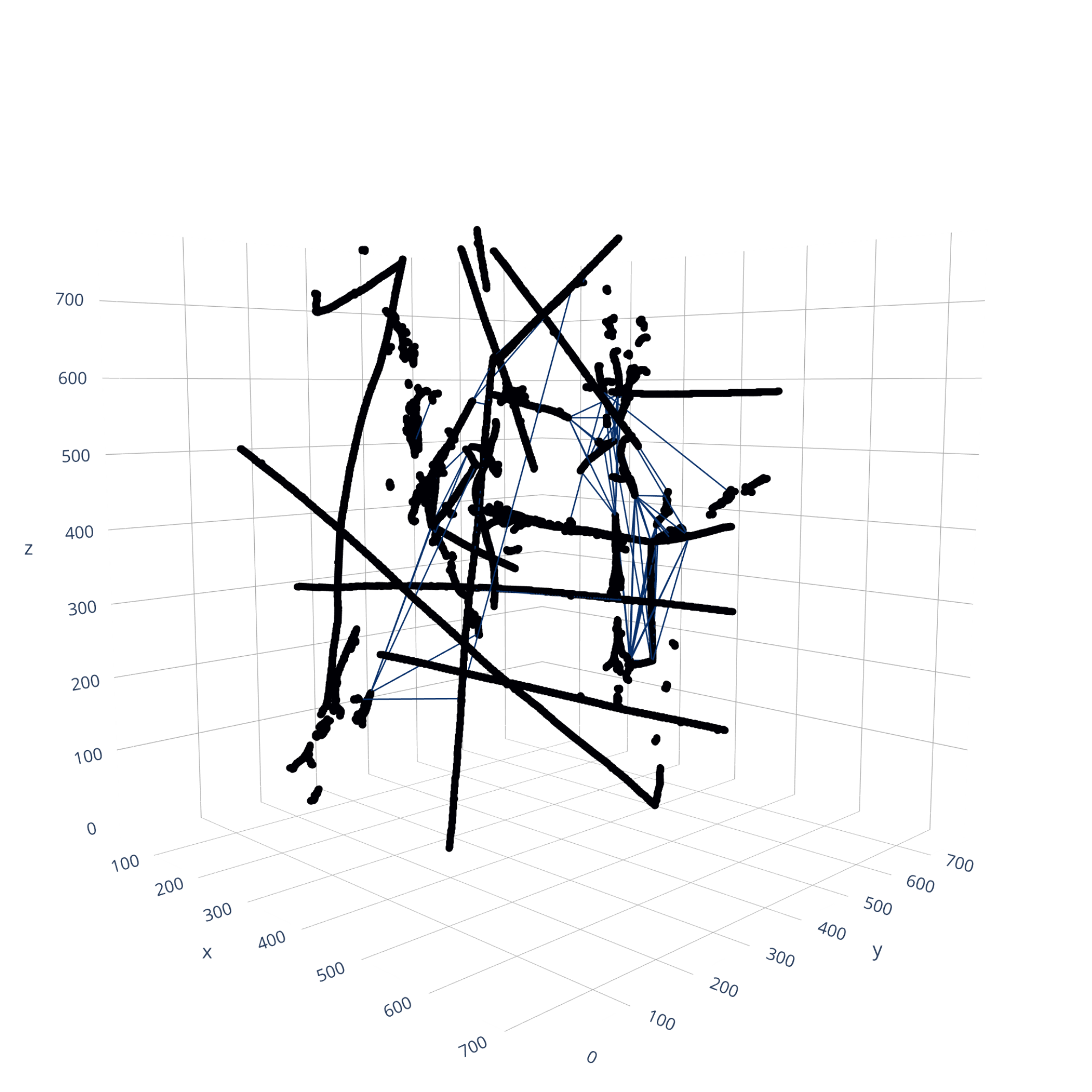}
    \hfill{}
    \begin{overpic}[width=0.32\textwidth, trim=1cm 0cm 1cm 2cm, clip]{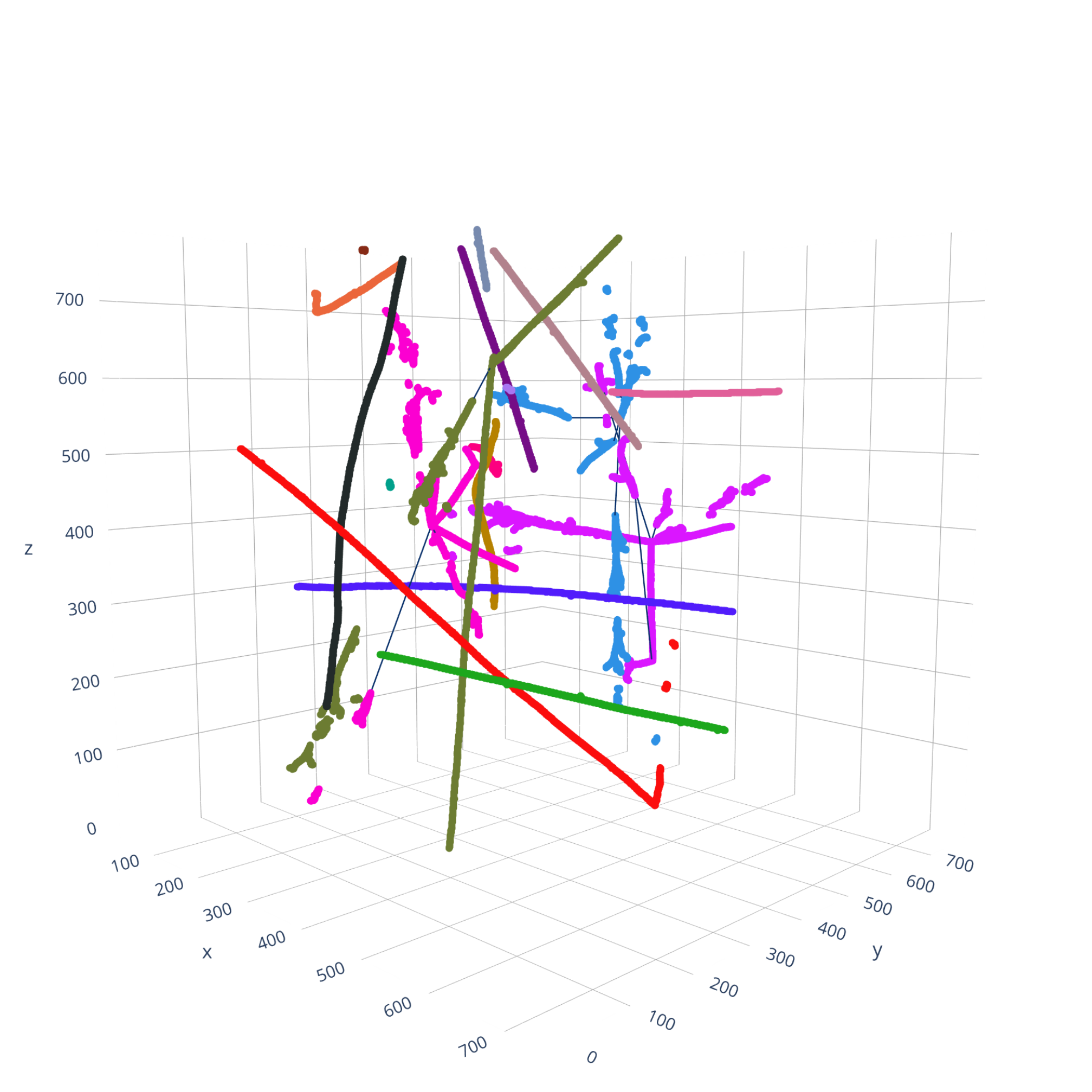}
        \put (35,15) {\fcolorbox{black}{white}{\footnotesize$\text{ARI}:\,100\,\%$}}
    \end{overpic}
    \hfill{}
    
    \caption{Interaction clustering predictions on four randomly picked events with 1, 2, 3 and 4 randomly merged images (from top to bottom). Left: ground-truth interaction labels (color) and ground-truth cluster graph edges. Middle: edges with an adjacency score $>0.5$ (the closer to 1, the darker the edge). Right: inferred interaction labels (color) and selected edges.}
    \label{fig:inter_clustering_examples}
\end{figure*}

Figure~\ref{fig:inter_clustering_examples} shows the output of the interaction clustering algorithm for four randomly selected events in the test set, with one to four interaction vertices stacked together. All four events have a high clustering accuracy, only missing or merging small particles incorrectly.

\begin{figure}[htb!]
    \centering
    \includegraphics[width=\linewidth]{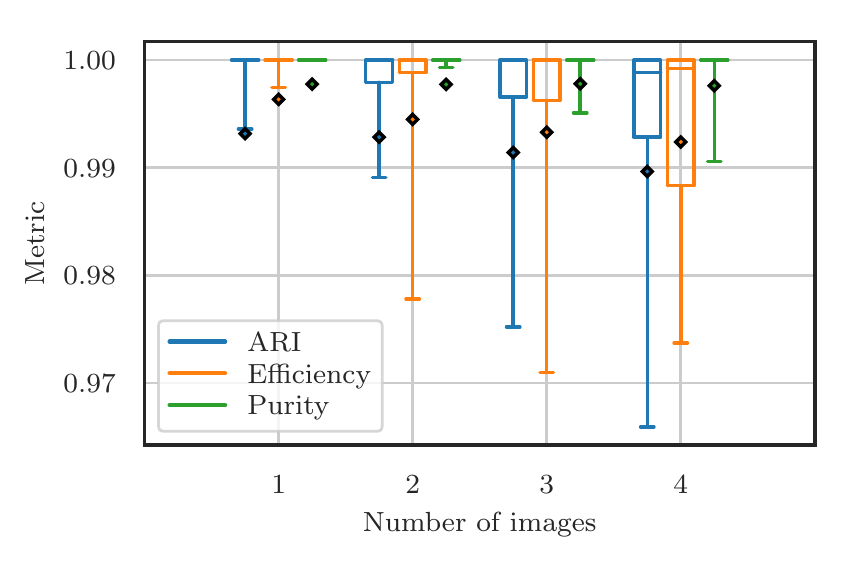}
    \caption{Interaction clustering metrics as a function of the number of interactions in the image. The diamonds represent the means, the lines the meadians, the boxes the IQRs and the whiskers span from the $10^{\text{th}}$ to the $90^{\text{th}}$ percentiles.}
    \label{fig:inter_clustering_metrics}
\end{figure}

The metrics described in section~\ref{sec:metrics} are used to systematically characterize the performance of the reconstruction chain applied to the interaction clustering task. Figure~\ref{fig:inter_clustering_metrics} shows the clustering performance as a function of the number of images that are superimposed. 

As shown in figure~\ref{fig:particle_count}, each image contains one neutrino-like interaction of $4.3\pm1.6$ particles overlayed with $3.8\pm1.7$ randomly scattered cosmic-like interactions. For an image of $\sim12\,$m$^3$, this corresponds to an interaction density of $0.40\pm0.14$\,intractions$/$m$^3$, which increases linearly with the number of superimposed images. The density observed of a single image, for instance, is equivalent to $\sim120$ interactions in a single ICARUS image, far above the expected rate~\cite{sbn}. A stack of two images contains two neutrino-like interactions, which corresponds to $\sim18$ such interactions in the DUNE-ND volume, close to the maximum expected rate~\cite{dune}, overlayed with $\sim30$ cosmic rays, which is unrealistically large. This demonstrates that this algorithm should easily deal with the expected rate of interactions in the foreseeable future.

\subsection{Mistakes analysis}
\begin{figure*}[t]
    \centering
    \hfill{}
    \includegraphics[width=0.32\textwidth, trim=1cm 0cm 1cm 2cm, clip]{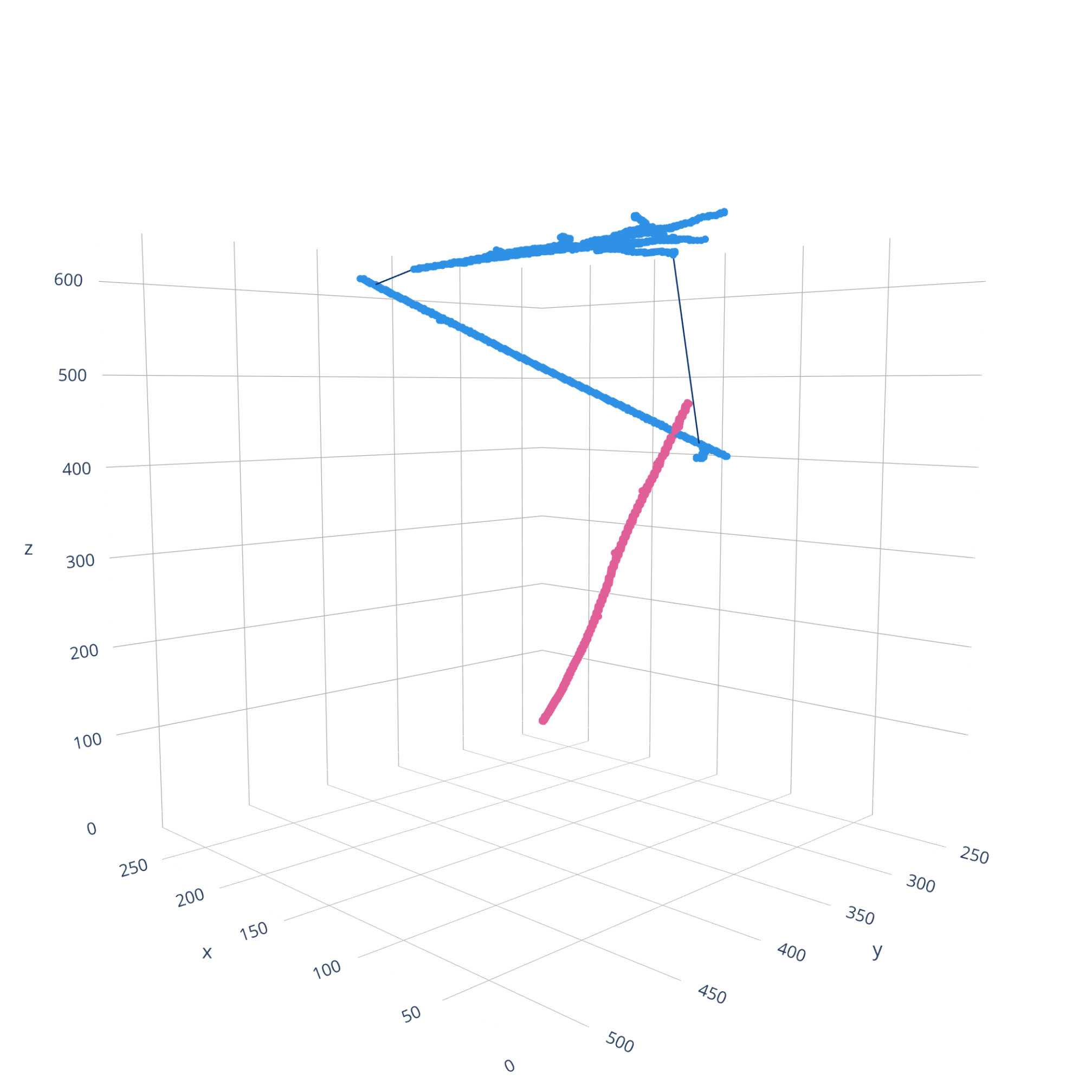}
    \hfill{}
    \includegraphics[width=0.32\textwidth, trim=1cm 0cm 1cm 2cm, clip]{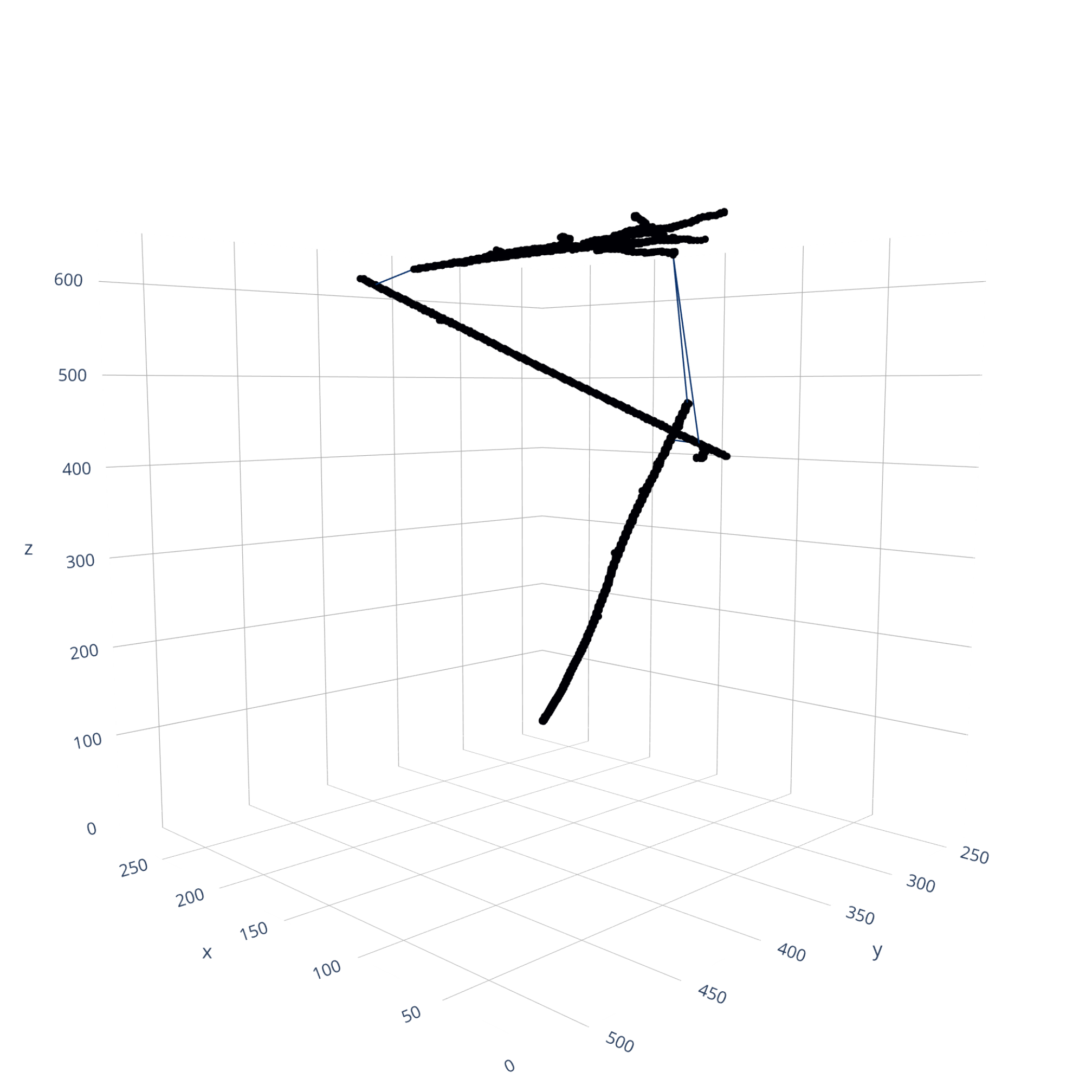}
    \hfill{}
    \begin{overpic}[width=0.32\textwidth, trim=1cm 0cm 1cm 2cm, clip]{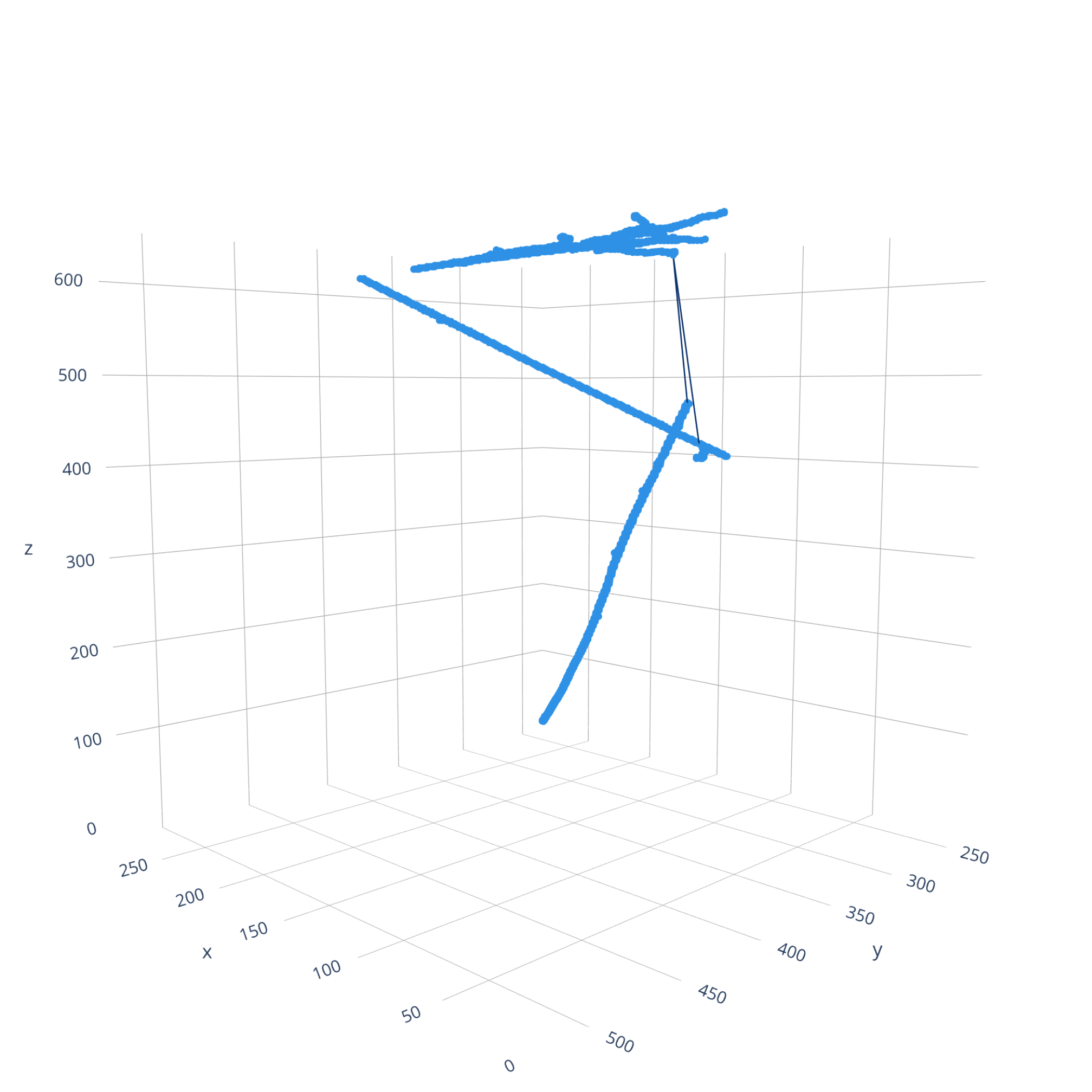}
        \put (35,15) {\fcolorbox{black}{white}{\footnotesize$\text{Pur.}:\,56.1\,\%$}}
    \end{overpic}
    \hfill{}
    
    \hfill{}
    \includegraphics[width=0.32\textwidth, trim=1cm 0cm 1cm 2cm, clip]{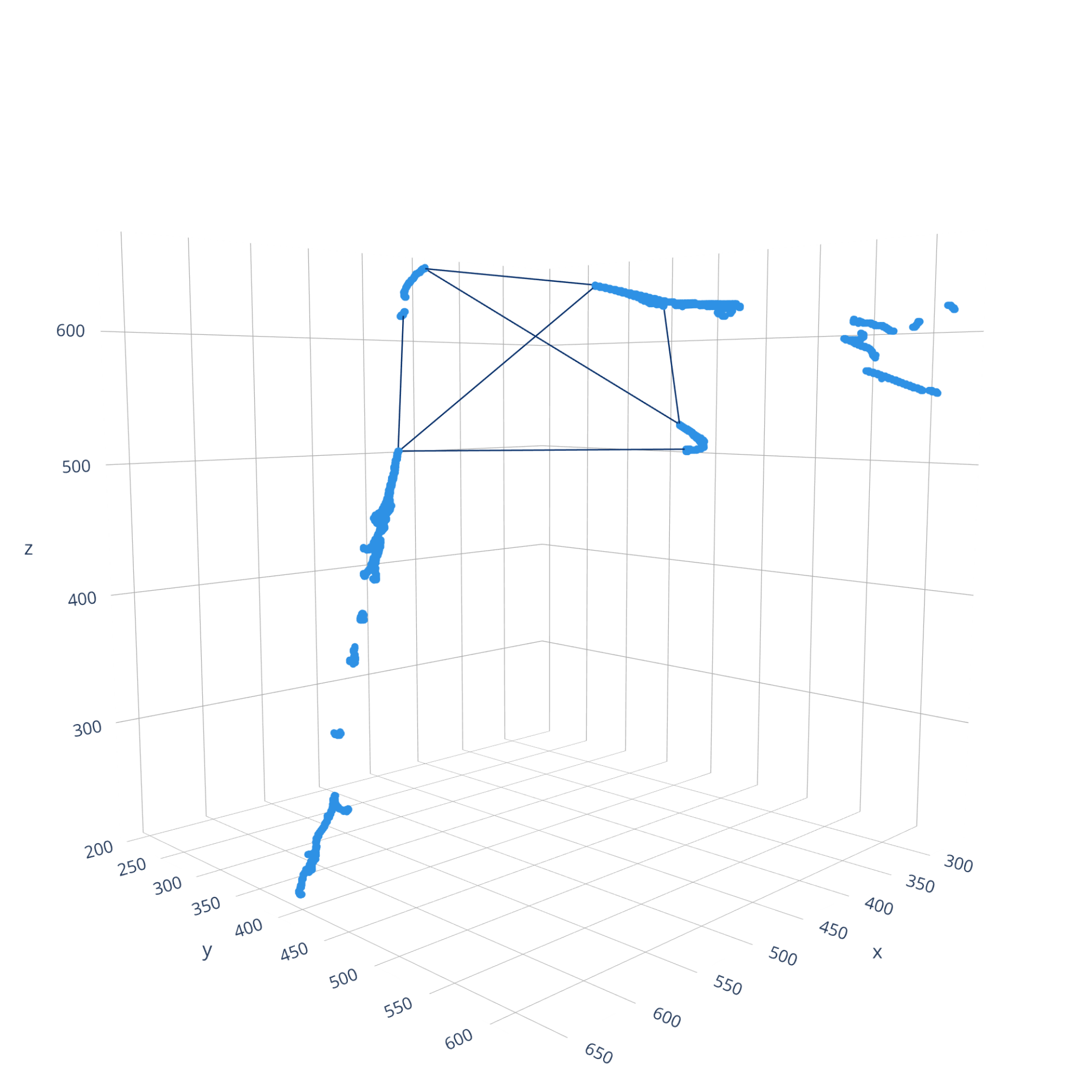}
    \hfill{}
    \includegraphics[width=0.32\textwidth, trim=1cm 0cm 1cm 2cm, clip]{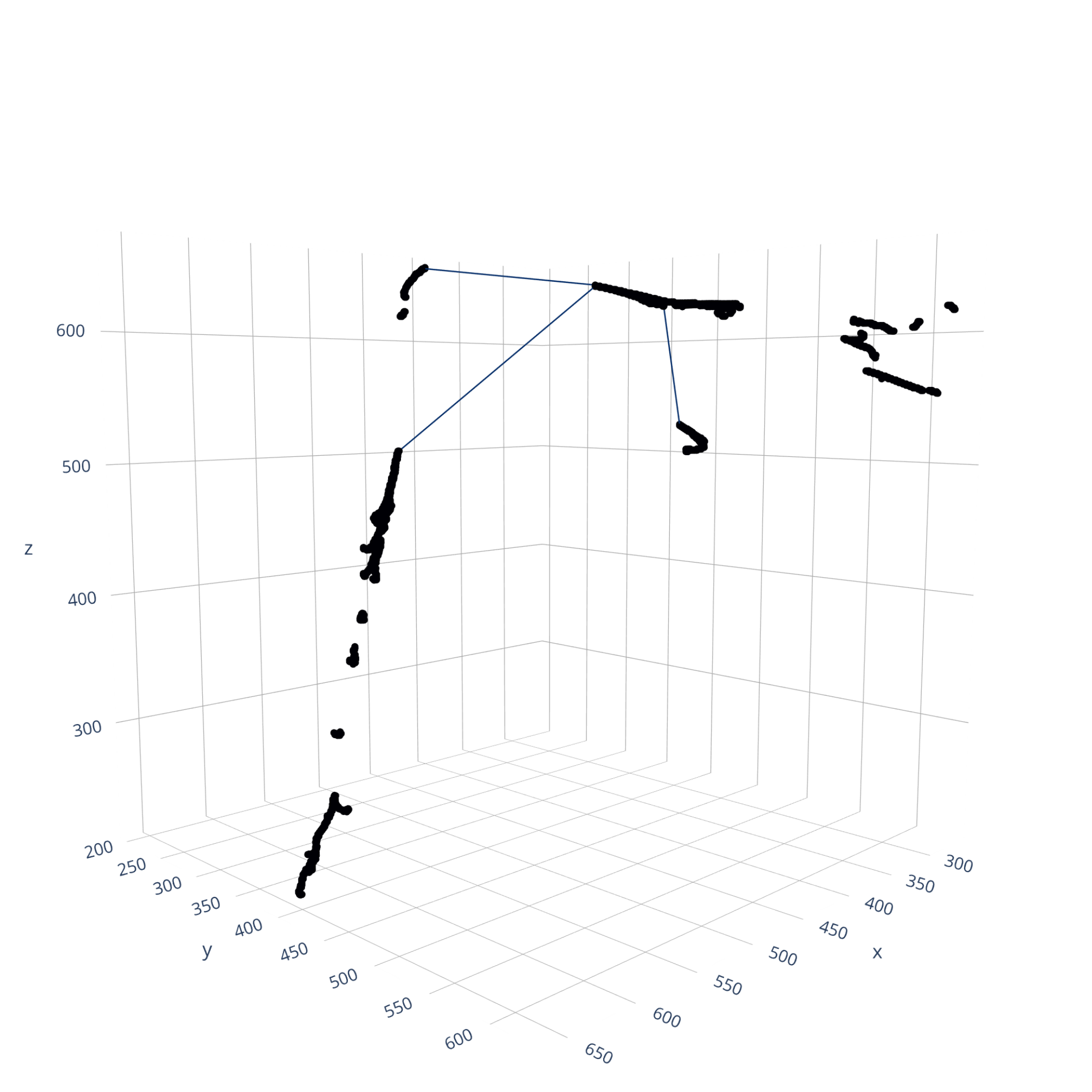}
    \hfill{}
    \begin{overpic}[width=0.32\textwidth, trim=1cm 0cm 1cm 2cm, clip]{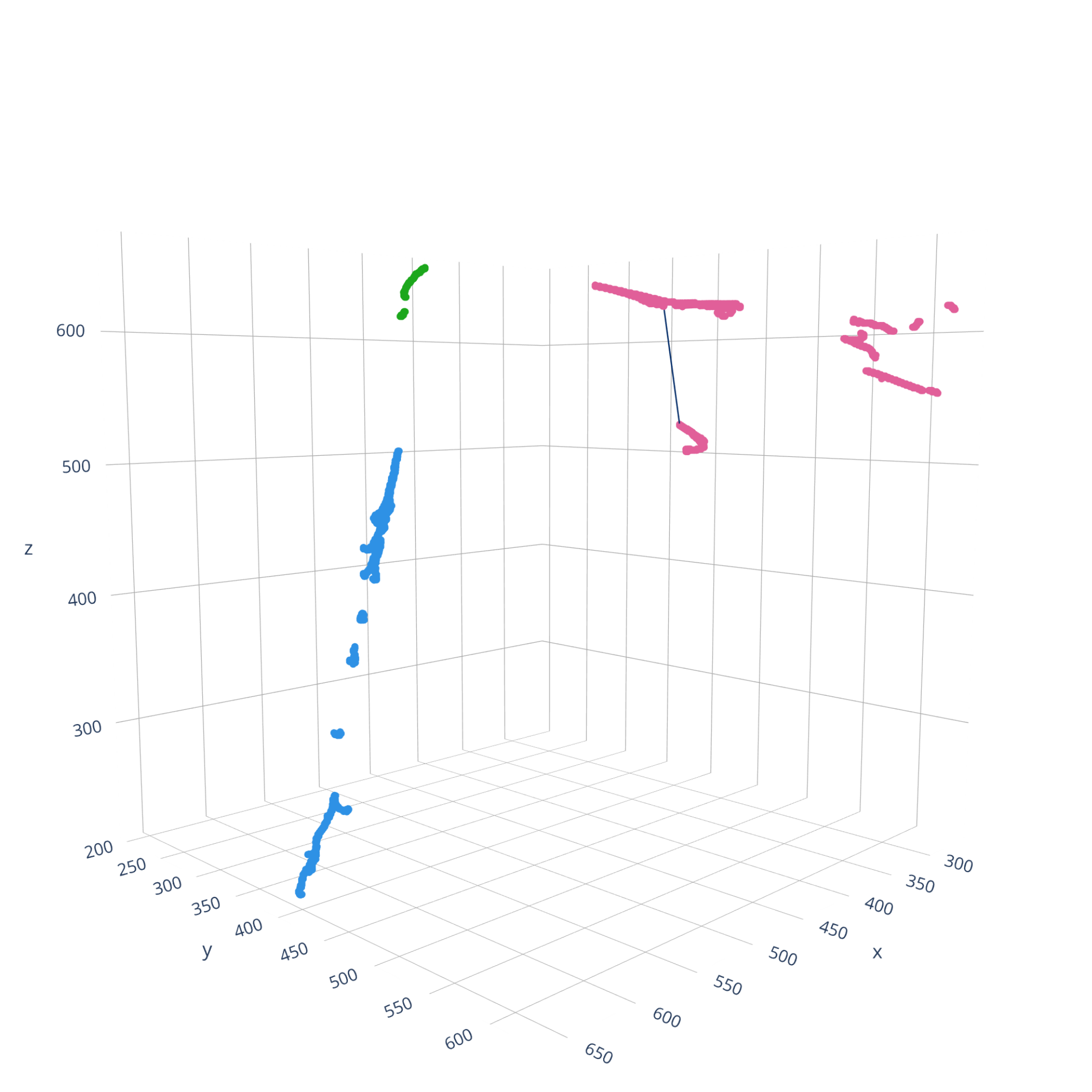}
        \put (35,15) {\fcolorbox{black}{white}{\footnotesize$\text{Eff.}:\,49.7\,\%$}}
    \end{overpic}
    \hfill{}
    
    \hfill{}
    \includegraphics[width=0.32\textwidth, trim=1cm 0cm 1cm 2cm, clip]{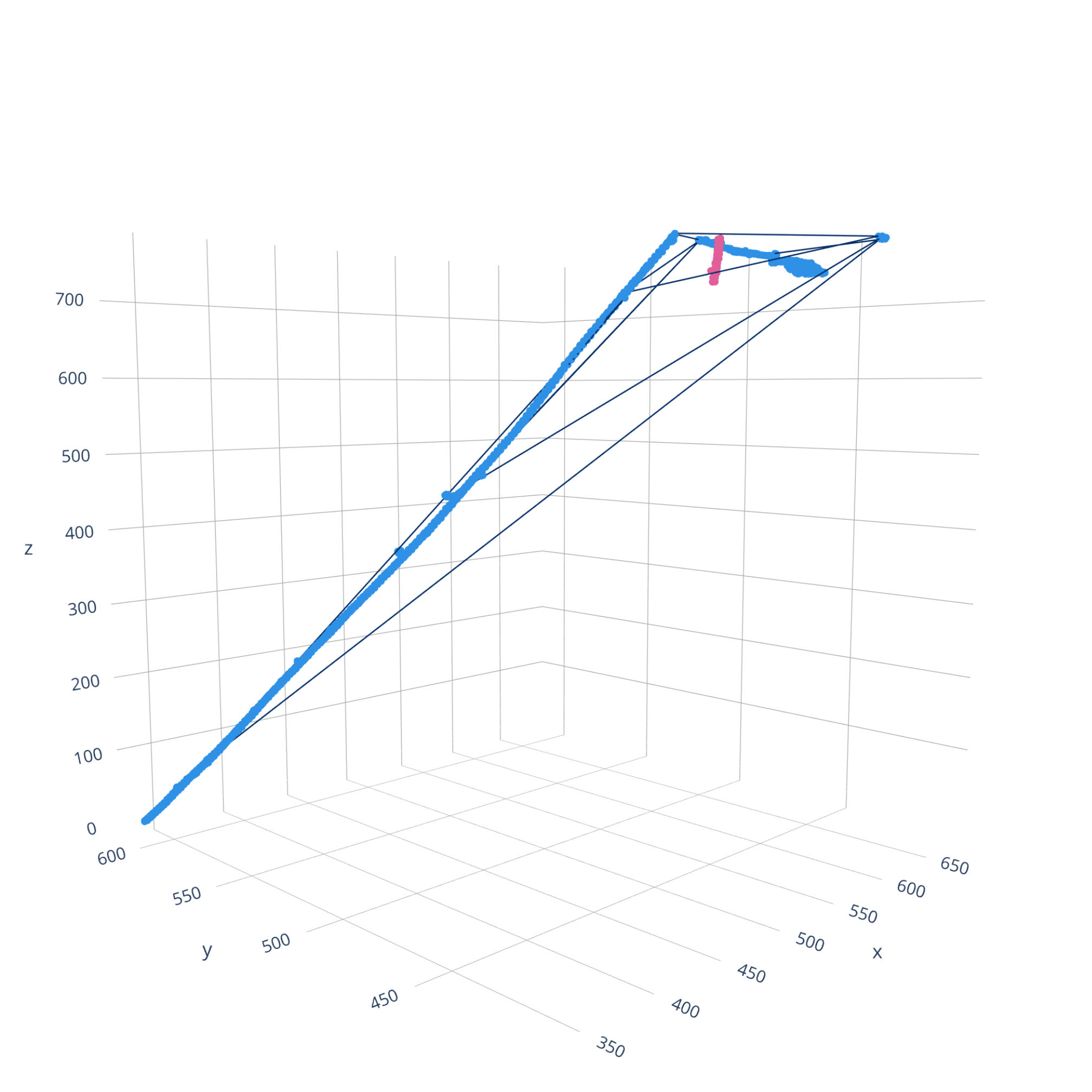}
    \hfill{}
    \includegraphics[width=0.32\textwidth, trim=1cm 0cm 1cm 2cm, clip]{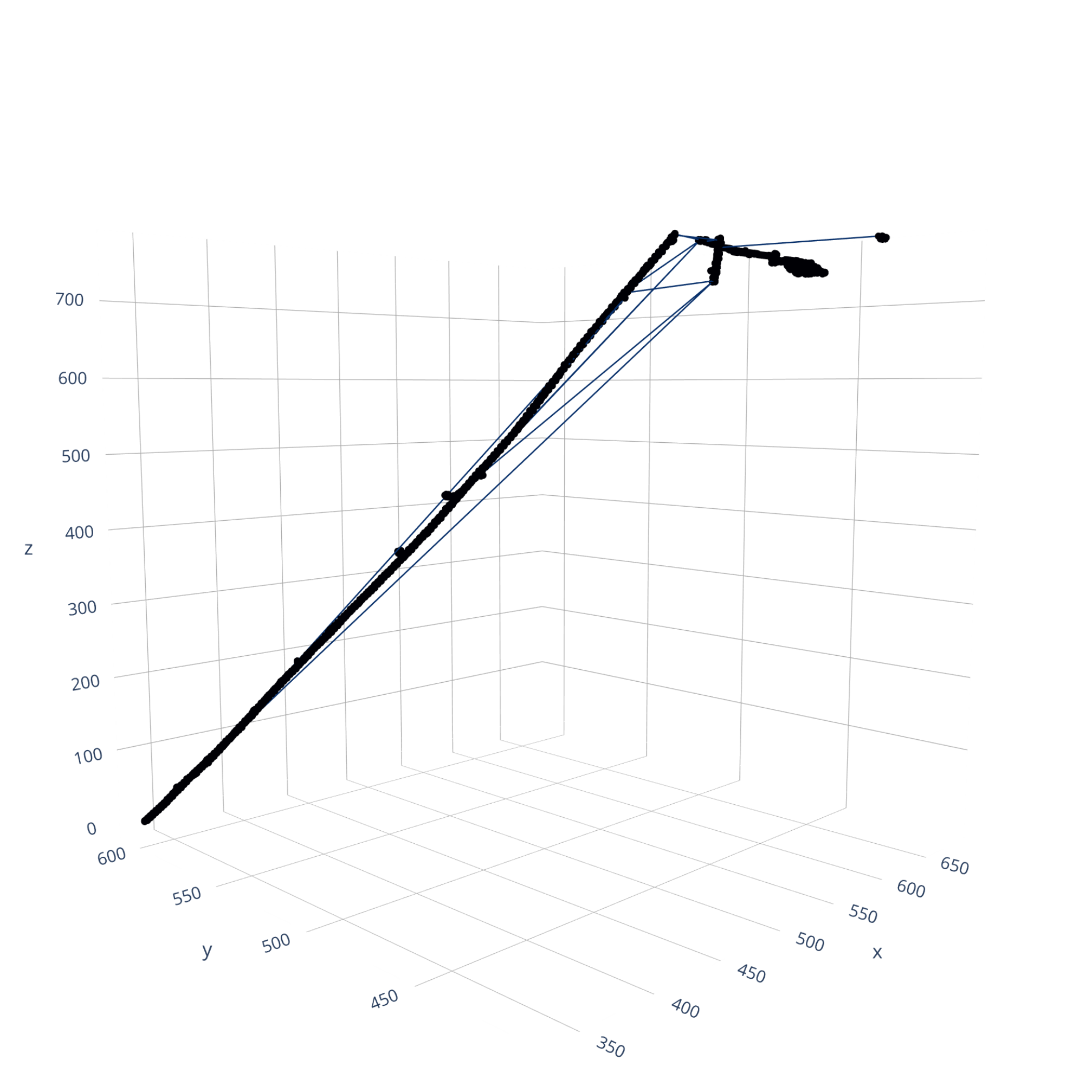}
    \hfill{}
    \begin{overpic}[width=0.32\textwidth, trim=1cm 0cm 1cm 2cm, clip]{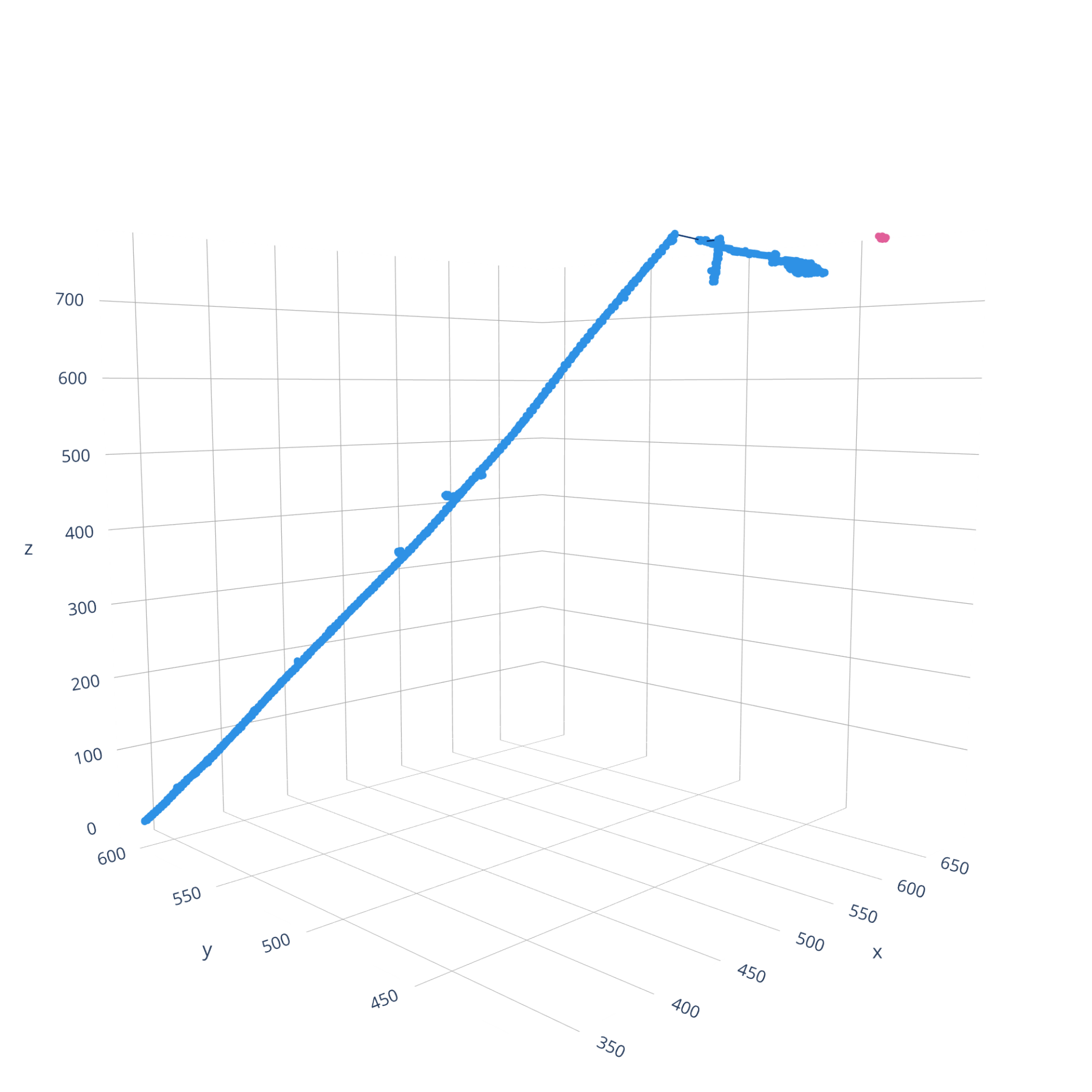}
        \put (35,15) {\fcolorbox{black}{white}{\footnotesize$\text{ARI}:\,-1.4\,\%$}}
    \end{overpic}
    \hfill{}
    
    \hfill{}
    \includegraphics[width=0.32\textwidth, trim=1cm 0cm 1cm 2cm, clip]{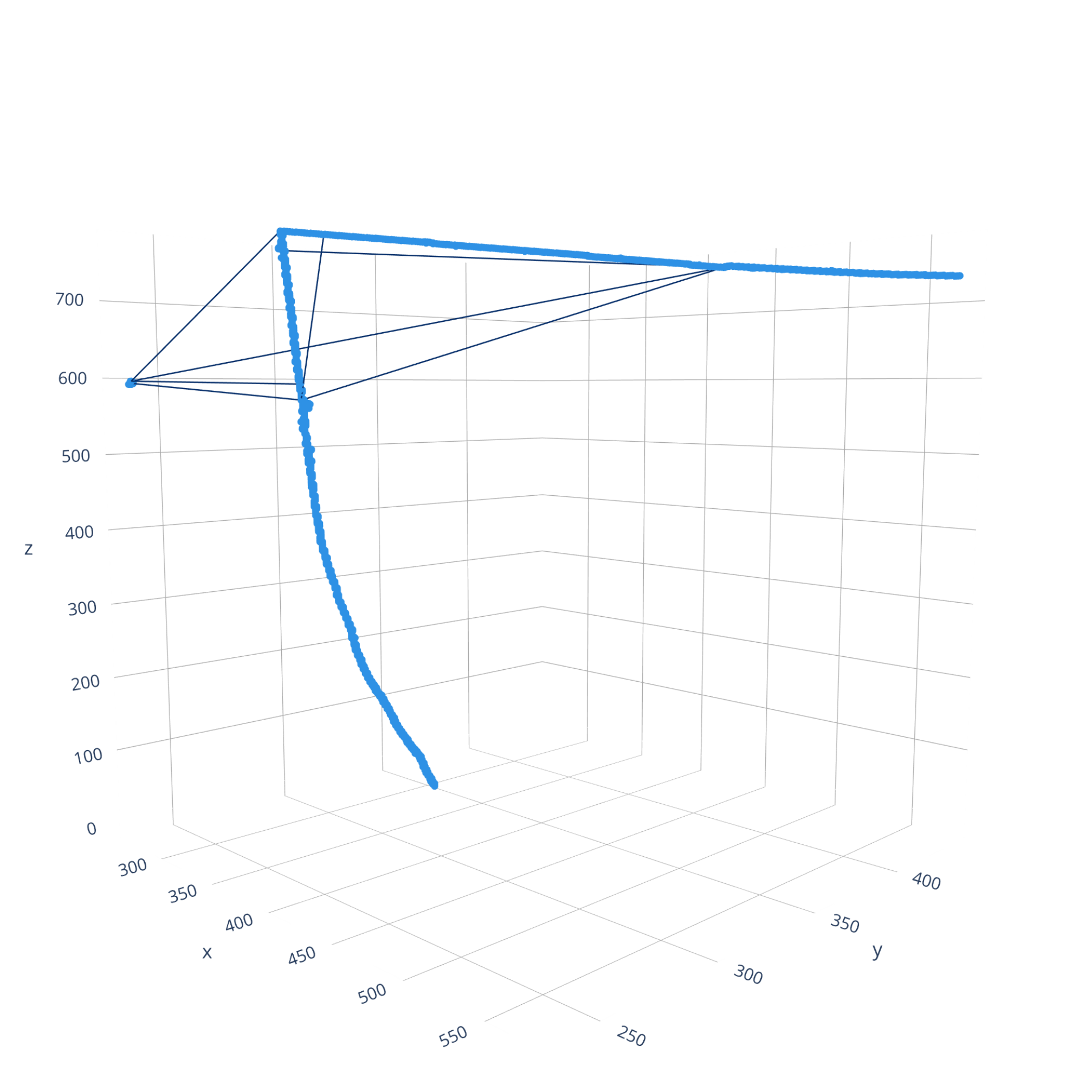}
    \hfill{}
    \includegraphics[width=0.32\textwidth, trim=1cm 0cm 1cm 2cm, clip]{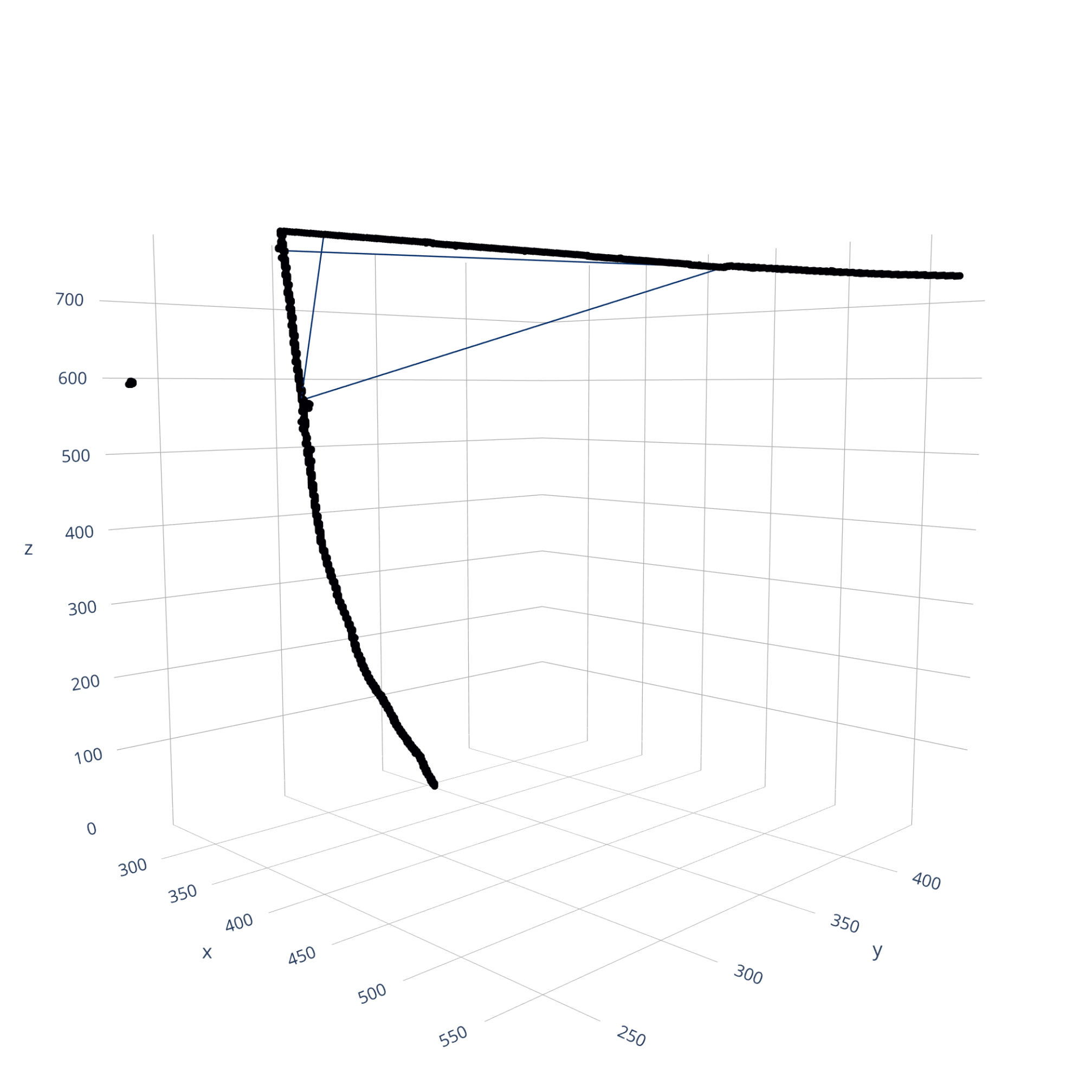}
    \hfill{}
    \begin{overpic}[width=0.32\textwidth, trim=1cm 0cm 1cm 2cm, clip]{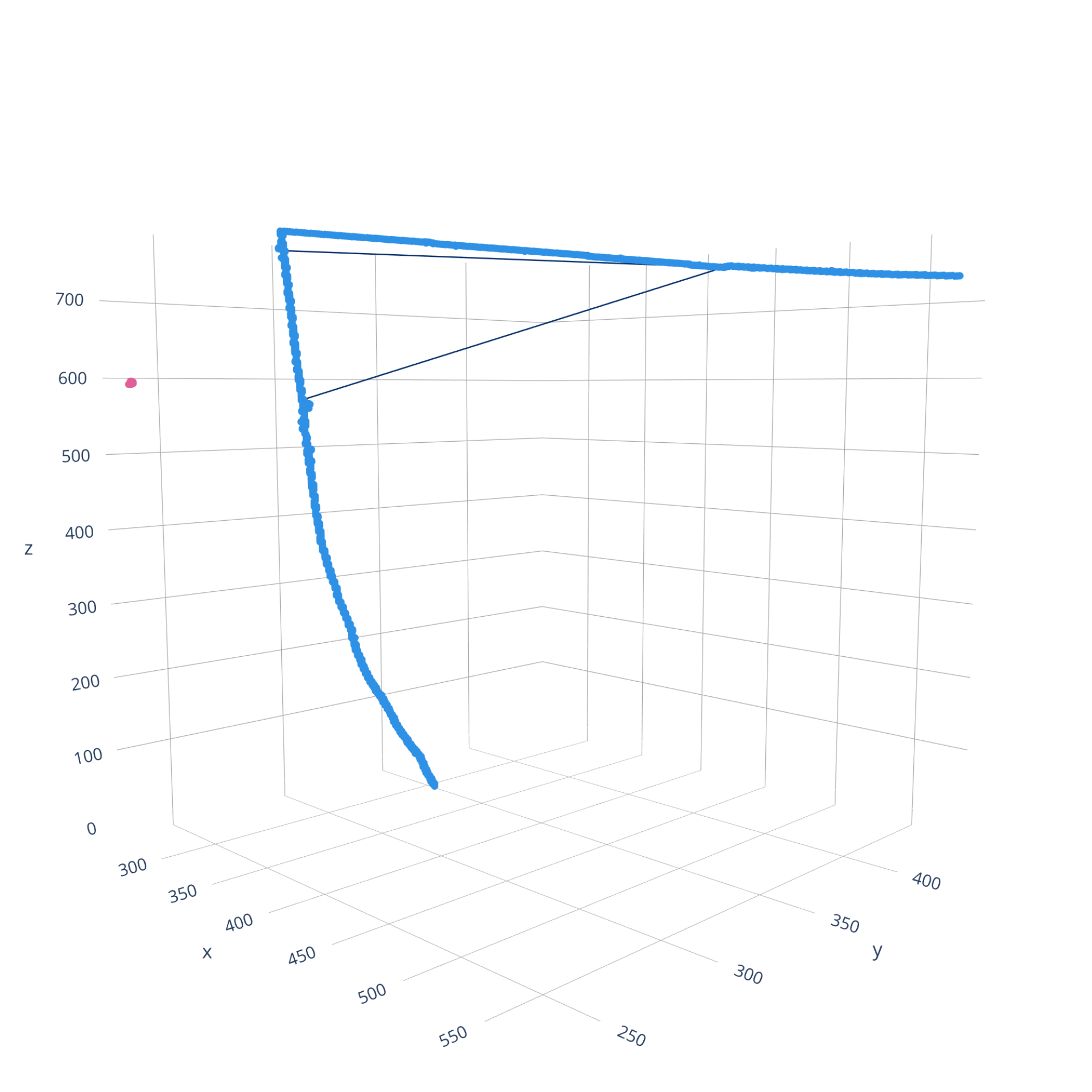}
        \put (35,15) {\fcolorbox{black}{white}{\footnotesize$\text{ARI}:\,0\,\%$}}
    \end{overpic}
    \hfill{}
    
    \caption{Interaction clustering predictions with the largest mistakes in three categories and one with an ARI of 0 (one event per row). Left: ground-truth interaction labels (color) and ground-truth cluster graph edges. Middle: edges with an adjacency score $>0.5$ (the closer to 1, the darker the edge). Right: inferred interaction labels (color) and selected edges.}
    \label{fig:inter_clustering_mistakes}
\end{figure*}

Figure~\ref{fig:inter_clustering_mistakes} shows the three events that are reconstructed with the lowest purity, efficiency and ARI on the first three rows, respectively, and an event with an ARI of 0. The first event exhibits a purity of 56.1\,\% due to a cosmic muon crossing and overlapping one of the vertex tracks. Events with purity $<0.9$ represent $\sim0.6\,\%$ of this test set. The second event has an efficiency of 49.7\,\% as the correlation between showers is not found sufficient by the network to associate them in an interaction. This may be due to a direction estimate not accurately representing the shower momentum or the vertex not being clearly defined by a track.  Events with efficiency $<0.9$ represent $\sim0.8\,\%$ of this test set. The bottom two rows show an event with an ARI of -1.3\,\% and one with an ARI of 0. These two examples have a very low ARI but only contain minor mistakes, merging a small fragment it should not while omitting another.

\section{Conclusion}
Graph Neural Networks (GNNs) are an ideally suited method to tackle the clustering of spatially detached objects in Liquid Argon Time Projection Chambers (LArTPCs).  A GNN-based reconstruction chain was developed to cluster electromagnetic showers and particle interactions.  This paper studied its performance on  a  generic  3D  sample  of  particle  interactions  in  liquid  argon  and  demonstrated  a  clustering  efficiency and  purity well above  99\,\%  for  both  tasks. A good shower energy resolution is a core requirement for the upcoming SBN program and DUNE experiment to reach their scientific goals. The reconstruction of the shower direction will be of central importance when matching neutral pion decay showers together or when back-propagating photons to a vertex. The clustering of particle into interactions will become essential for future high-rate LArTPCs and the GNN algorithm developed here shows that this can be achieved. The algorithm described in this paper will be part of an end-to-end, machine-learning-based reconstruction chain developed at SLAC for all LArTPCs.

\section{Acknowledgement}
This work is supported by the U.S. Department of Energy, Office of Science, Office of High Energy Physics, and Early Career Research Program under Contract DE-AC02-76SF00515.

\bibliography{references}

\pagebreak
\onecolumngrid

\end{document}